\makeatletter
\declare@file@substitution{revtex4-1.cls}{revtex4-2.cls}
\makeatother

\documentclass[twocolumn]{aastex631}
\usepackage[utf8x]{inputenc}
\usepackage{amsmath,amsfonts,amssymb,mathrsfs,bm,enumerate,graphicx}
\usepackage[T1]{fontenc}
\usepackage[all]{hypcap}
\hypersetup{
    colorlinks=true,
    linkcolor=magenta,
    filecolor=magenta,      
    urlcolor=blue,
}

\usepackage{ulem}
\usepackage{threeparttable}
\usepackage{ae,aecompl}
\usepackage{wasysym}
\usepackage{enumitem}

\def\red#1 {\textcolor{red}{#1}\ }   
\def\blue#1 {\textcolor{blue}{#1}\ }   
\def\purple#1 {\textcolor{purple}{#1}\ }   

\def\red#1 {\textcolor{red}{#1}\ }   
\def\blue#1 {\textcolor{blue}{#1}\ }   

\usepackage{ulem}
\usepackage{threeparttable}
\usepackage{ae,aecompl}
\usepackage{wasysym}
\usepackage{xcolor}
\definecolor{orange}{RGB}{255,127,0}
\definecolor{pinegreen}{HTML}{009B55}

\makeatletter
\newcommand{\shorteq}{\mathrel{\mkern0.2mu\mathpalette\shorteq@\relax\mkern0.2mu}}
\newcommand{\shorteq@}[2]{\scalebox{0.5}[1]{$\m@th#1=$}}
\DeclareMathSymbol{\shortminus}{\mathbin}{AMSa}{"39}
\newcommand{\compactsum}[3]{\sum_{\scriptscriptstyle ~#1\shorteq#2}^{\scriptscriptstyle #3}}

\shorttitle{ Realignment of Planetary Orbits}
\shortauthors{Mu\~noz}

\begin{document}

\title{\bf  Early Realignment of  Eccentric  Planetary Orbits without Orbital Decay}
\author[0000-0003-2186-234X]{Diego J.\ Mu\~noz}
\affiliation{Department of Astronomy and Planetary Science, Northern Arizona University, Flagstaff, AZ 86011, USA}

\begin{abstract}
Warm Jupiters (WJs) are, on average, more spin--orbit aligned than hot Jupiters, even at high eccentricities. Particularly puzzling are eccentric WJs that appear to lie along the so-called ``tidal migration track.'' These systems have been deemed candidates for ongoing high-eccentricity migration (HEM), yet their low projected obliquities are difficult to reconcile with known HEM pathways that excite large misalignments. Low obliquities suggest either that (1) dynamically cold planetary orbits can be excited to large eccentricities while remaining nearly coplanar, or (2) dynamically hot orbits can reduce their stellar obliquities faster than they circularize the orbits owing to tides on the star, despite the planets' relatively large separations. 
We explore the second possibility by calculating the tidal evolution of eccentric WJs, focusing on the $(1,0)$ component of the stellar response, forced at the tidal frequency $\tilde{\omega}_{10}=-\Omega_\star$, which dissipates stellar rotational energy while leaving the orbital energy and eccentricity unchanged, and modifies obliquity at an eccentricity-dependent rate. This frequency lies in the inertial wave range, and coincides with the natural frequency of the rigid `spin-over mode.' While this mode is known to be non-dissipative, the  dissipation strength of the full response at $\tilde{\omega}_{10}$ from non-rigid motions in rapidly rotating, inflated stars remains uncertain. We therefore enhance the dissipative response phenomenologically and find that an enhancement of $\eta_{10}\sim10^6$ may suffice for super-Jupiters and account for the observed obliquity trends. With this enhancement, stellar realignment can outpace orbital circularization if the planet is already on an eccentric warm orbit while its host is still inflated and rapidly rotating during the pre-main sequence (PMS). Realignment is most effective for super-Jupiters. If it occurs during the PMS, the obliquities of eccentric WJs need not correlate strongly with the present-day effective temperatures of their host stars. In the binary-driven HEM configurations we tested, conditions favorable to realignment can suppress secular eccentricity excitation, while on the main sequence, configurations that sustained eccentricity forcing circularized faster than they realigned. These results cast doubt on the interpretation of aligned, eccentric WJs as relics of secularly driven HEM, even when they lie along a tidal migration track.
\end{abstract}


\section{Introduction}
%
\subsection{Close-in Giants and Stellar Obliquity}

The measurement of the sky-projected stellar obliquity $\lambda$ has become a key forensic diagnostic of the dynamical past of planetary systems. Due to observational biases, most measurements have been obtained for close-in giant planets, particularly hot Jupiters (HJs; with semimajor axes $a_{\rm p}<0.1$~au). The growing sample of measured obliquities has enabled numerous statistical studies \citep[e.g.,][]{Fabrycky:2009,Winn:2010,Schlaufman:2010,Morton:2011,Albrecht:2012,Albrecht:2022,Munoz:2018,Rice:2022a,Rice:2022b,Wang:2024}, including attempts to infer the most likely origins of HJs \citep[e.g.,][]{Morton:2011,Rice:2022a}.

The broad distribution of stellar obliquities in HJ systems \citep[e.g.,][]{Albrecht:2022} is often interpreted as a relic of an earlier, highly excited dynamical state. This has been used to support a high-eccentricity migration (HEM) origin for HJs (see \citealp{Dawson:2018} for a review), since most variants of this mechanism excite substantial inclination variations alongside eccentricity growth.

Warm Jupiters (WJs; $0.1~{\rm au}\leq a_{\rm p}\lesssim0.7$~au), on the other hand, seem to be telling a different story. Among single-star systems, WJs exhibit a statistically narrower distribution of stellar obliquities than HJs \citep{Rice:2022b,Wang:2024}. Particularly striking are the remarkably low projected obliquities measured for several eccentric WJs \citep[e.g.,][]{Sedaghati:2023,Espinoza:2023,Espinoza:2025,Mireles:2025}. For example, TOI-677\,b and TOI-3362\,b have measured obliquities of $\lambda=0.3\pm1.3^\circ$ \citep{Sedaghati:2023} and $\lambda=1.2^{+2.8}_{-2.7}{}^\circ$ \citep{Espinoza:2023}, respectively, despite their substantial eccentricities.

Some of these eccentric planets lie along the so-called ``tidal migration track'' \citep{Dong:2014}, with orbital angular momenta compatible with eventual circularization into HJs along approximately constant $a_{\rm p}(1-e_{\rm p}^2)$ paths. Such systems are often dubbed ``proto-hot Jupiters'' \citep{Dawson:2012,Dawson:2015,Dong:2021,Espinoza:2023,Gupta:2024}. They are prime testbeds for the HEM hypothesis, since their eccentricities have not yet been erased by tidal circularization and their orbital architectures may still reflect the dynamical excitation that initiated migration.

Remarkably, the low obliquities of these eccentric warm Jupiters (EWJs) seem at odds with the inclination excitation expected from most HEM pathways. Nearly coplanar variants of HEM can, in principle, excite large eccentricities while maintaining small misalignments \citep[e.g.,][]{Lee:2003,Petrovich:2015b}. However, such outcomes rely on particular initial configurations, including sufficiently small mutual inclinations and finite initial eccentricities, whose prevalence in real planetary systems remains uncertain. Thus, rather than providing unambiguous evidence for ongoing HEM, these systems may represent some of its most challenging counterexamples. Alternatively, their low obliquities may not be primordial, but instead reflect subsequent realignment of their host stars. In the absence of other mechanisms (e.g., a dissipative disk), such realignment would need to arise from tidal dissipation within the star.

\subsection{Orbital Evolution Owing to Tides Acting on the Star}

Stellar tides have long been invoked to explain the evolution of close-in giant planets \citep[e.g.,][]{Lin:1996,Rasio:1996a}, including the low obliquities of HJ hosts cooler than $\sim6000$~K \citep{Winn:2010,Albrecht:2012}. However, under conventional tidal prescriptions, efficient stellar realignment is generally accompanied by orbital decay and possibly planetary destruction \citep{Winn:2010,Lai:2012}. This poses a particular problem for  EWJs, whose orbits appear to have undergone relatively little tidal circularization.

The widely used weak-friction approximation describes the stellar response through a single time lag $\Delta t$ \citep[e.g.,][]{Alexander:1973,Mignard:1979,Hut:1981}, imposing a common dissipation law on the different components of the tidal forcing. In reality, the response may depend strongly on the forcing frequency, as already recognized by \citet{Darwin:1879}, who argued that each tidal frequency should be ``retarded by its own special amount'' \citep[][§I.2]{Darwin:1879}. More general parametric models can thus assign distinct tidal quality factors $Q$ to different forcing components \citep[e.g.,][]{Lai:2012}, potentially allowing obliquity damping and orbital decay to proceed at very different rates.

There are also physical reasons to expect a highly frequency-dependent tidal response. For planetary bodies, this dependence is often modeled through viscoelastic rheologies, which determine the frequency dependence of the complex tidal Love numbers \citep[e.g.,][]{Efroimsky:2012,Efroimsky:2015}. An additional complication in stars is the resonance-rich response of their fluid interiors, where dissipation can arise from both quasi-hydrostatic tidal flows (equilibrium tides) and excited waves (dynamical tides; see \citealp{Ogilvie:2014}). These include internal gravity waves in stably stratified regions \citep{Zahn:1975,Zahn:1977,Goldreich:1989,Goodman:1998} and inertial waves (IWs) in rotating convective regions \citep{Papaloizou:1997,Savonije:1997,Ogilvie:2004,Ogilvie:2007,Ogilvie:2009,Ogilvie:2013,Barker:2009}. Recent calculations of IW dissipation and global oscillation modes \citep[e.g.,][]{Barker:2020,Barker:2022,Barker:2024,Wu:2024} further illustrate the complexity of this response, although quantitative predictions for realistic stellar interiors remain uncertain.

Of particular interest is the possibility of exciting IWs through the obliquity tide. \citet{Lai:2012} showed that spin-orbit misalignment gives rise to a tidal component with azimuthal and temporal indices $(m,m')=(1,0)$, whose frequency in the stellar rotating frame is $\tilde{\omega}_{10}=-\Omega_\star$, which lies within the inertial range \citep{Greenspan:1968}. This component is stationary in the inertial frame and therefore transfers no orbital energy, even while exerting a torque that can modify the stellar obliquity. Consequently, its dissipation can drive spin-orbit evolution without directly changing the orbital semimajor axis or eccentricity. Moreover, since its frequency lies within the inertial-wave range, its dissipation may be substantially stronger than that of other tidal components \citep{Lai:2012,Lin:2017}. However, while non-trivial inertial motions can dissipate energy at this frequency, the magnitude of the response remains poorly constrained, and existing calculations do not establish whether it is sufficient to explain the observed alignments. The pre-main-sequence (PMS) phase may offer a particularly favorable window for such realignment, as the larger stellar radii strengthen tidal forcing, while rapid rotation and extensive convection could further enhance the dissipative response \citep{Barker:2020,Barker:2022}. This possibility would require planets to acquire their eccentric, misaligned configurations sufficiently early in their evolution.

In this work, we extend the framework of \citet{Lai:2012} to orbits of arbitrary eccentricity and apply it to EWJs. We investigate whether the $(1,0)$ obliquity tide could account for their low spin-orbit misalignments, and quantify the enhanced dissipation required to do so without accompanying orbital circularization, focusing on the role of the stellar pre-main sequence phase. Because this damping efficiency remains poorly constrained, we introduce a phenomenological enhancement of the effective tidal response without prescribing its underlying mechanism. More generally, our approach allows distinct effective quality factors for the different tidal forcing components, providing a flexible framework for studying their combined effects on stellar spin and orbital evolution.

%
\begin{figure}[h!]
\includegraphics[width=0.42\textwidth]{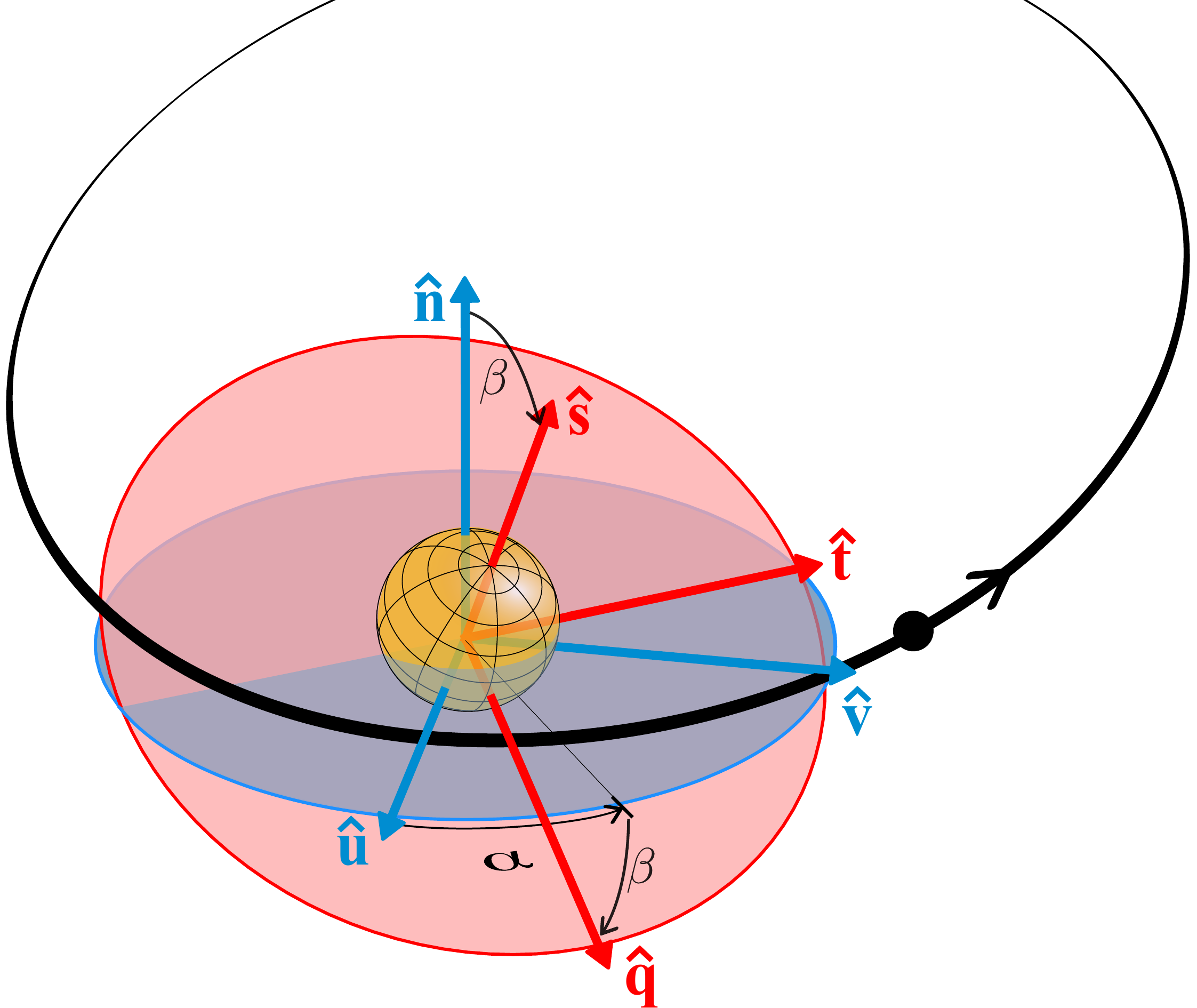}
\caption{Description of the reference frames used to decompose the tidal potential into its Fourier and multipolar components.
The orthonormal triad $(\hat{\bf q},\hat{\bf t},\hat{\bf s})$ (red vectors) is aligned with the stellar spin-axis. The orthonormal triad
$(\hat{\bf u},\hat{\bf v},\hat{\bf n})$ (blue vectors) is aligned with the orbital angular momentum and eccentricity vectors. We transform between coordinate
systems using  Euler angle rotations $(\alpha,\beta,\gamma)$ with $\gamma=0$ in the $z$-$y$-$z$ convention \citep[e.g.,][]{Rose:1957}.
\label{fig:euler_angles}}
\end{figure}

\section{A Generalized Effective Theory of Tidal Dissipation for Eccentric Orbits}
\subsection{Geometry of the Problem}
We work in two coordinate systems, an orbit-aligned frame $\mathscr{S}$ and a spin-aligned one $S$, 
as depicted in Figure~\ref{fig:euler_angles}. The cartesian unit vectors are $(\hat{\bf u},\hat{\bf v}, \hat{\bf n})$ for
the orbit-aligned system $\mathscr{S}$, where $\hat{\bf n}={\bf L}_{\rm p}/|{\bf L}_{\rm p}|$ is aligned with the orbital angular momentum,
$\hat{\bf u}={\bf e}_{\rm p}/|{\bf e}_{\rm p}|$ is  aligned with the eccentricity or Laplace vector, and $\hat{\bf v}=\hat{\bf n}\times\hat{\bf u}$
\citep[e.g.,][]{Tremaine:2009}. The cartesian unit vectors in $S$ are $(\hat{\bf q},\hat{\bf t}, \hat{\bf s})$, where
$\hat{\bf s}={\bf S}_\star/|{\bf S}_\star|$ is aligned with the stellar spin axis,
$\hat{\bf t}$ represents the line of ascending node for the equatorial plane relative to $\mathscr{S}$,
and $\hat{\bf q}=\hat{\bf t}\times\hat{\bf s}$ also lives on the equatorial plane. In addition, we use the spherical coordinates sets
$(r,\vartheta,\varphi)$ and  $(r,\theta,\phi)$ for the frames $\mathscr{S}$ and $S$, respectively.

These two coordinate systems are related via an  Euler angle rotation $\alpha{\rightarrow}\beta{\rightarrow}\gamma{=}0$ in the 
$z$-$y$-$z$ convention \citep[e.g.,][]{Rose:1957},
that is,
\begin{subequations}
\label{eq:unit_vectors_primed}
\begin{align}
\hat{\bf q}=&\cos\beta\cos\alpha\hat{\bf u}
+\cos\beta\sin\alpha \hat{\bf v}
-\sin\beta\hat{\bf n}\\
\hat{\bf t}=&-\sin\alpha \hat{\bf u}
+\cos\alpha  \hat{\bf v}
\\
 \hat{\bf s}=&\sin\beta\cos\alpha \hat{\bf u}
 +\sin\beta\sin\alpha  \hat{\bf v}
+\cos\beta  \hat{\bf n}~,
\end{align}
\end{subequations}
where $\beta$ is the stellar obliquity and $\alpha+\pi/2$ is the longitude of ascending node for the star's equatorial plane relative to $\mathscr{S}$.

\subsection{Tidal Potential and Tidal Response Ansatz}\label{sec:tidal_response}
We expand the tidal potential $V$ exerted on the star by a planet
 following a Keplerian orbit of semi-major axis $a_{\rm p}$ and eccentricity $e_{\rm p}$. Truncated to $l=2$ order, the potential
reads
\begin{equation}\label{eq:expanded_potential_rotated}
\begin{split}
V({\bf r},t)=
{-}\frac{{\cal G}M_{\rm p}r^2}{a_{\rm p}^3}\!
\!\!\compactsum{m}{\shortminus2}{2}
\!\!\compactsum{\mu}{\shortminus2}{2}
\!\!\compactsum{k}{\shortminus \infty}{\infty}&
W_{2\mu}
D^{2^{\scriptstyle *}}_{\mu m}(\alpha\beta0)
 X_{k+\mu}^{-3,\mu}(e_{\rm p})
 \\
 &
 \times
 Y_{2m}(\theta,\phi)
 {\rm e}^{-{\rm i}(k+\mu){\cal M}}~,
\end{split}
\end{equation}
where $\mu$ and $m$ are the azimuthal numbers in the $\mathscr{S}$ and $S$ systems, respectively, and where
${\cal M}$ is the orbital mean anomaly. The additional coefficients are defined as follows:
$W_{2\mu}=(4/5)\pi |Y_{2\mu}(\tfrac{\pi}{2},0)|$ \citep{Press:1977},
$X^{-3,\mu}_{k+\mu}$ represents a shifted Hansen coefficient  \citep{newc1895,plummer1918}, and
$D^2_{\mu m}(\alpha\beta0)={\rm e}^{-{\rm i}\mu\alpha}d^2_{\mu m}(\beta)$ corresponds to the Wigner D-matrix of
degree $l{=}2$, associated to the Euler angle rotation $\alpha{\rightarrow}\beta{\rightarrow}\gamma{=}0$
with $d^2_{\mu m}(\beta)$ being the so-called Wigner ``small-d'' matrix
 \citep{Wigner:1959,Rose:1957}.

In compact notation, we write the quadrupolar tidal potential as
\begin{equation}\label{eq:potential_decomposition_eccentric}
V({\bf r},t)=-\frac{{\cal G}M_{\rm p}r^2}{a_{\rm p}^3}
\sum_{\scriptscriptstyle m=-2} ^{\scriptscriptstyle+2}
\sum_{\scriptscriptstyle n=-\infty} ^{\scriptscriptstyle +\infty}
{\cal V}_{m n}
Y_{2m}(\theta,\phi){\rm e}^{-{\rm i}n \Omega t}
\end{equation}
\citep[e.g.,][]{Miranda:2015}
with $n=k+\mu$ being the time harmonic coefficient associated to the perturbation,
and where we have replaced ${\cal M}=\Omega t$ with $\Omega= [{\cal G}(M_\star{+}m_{\rm p})/a_{\rm p}^3]^{1/2}$.
 The dimensionless coefficient
${\cal V}_{m n}$
is defined as
\begin{equation}\label{eq:potential_coefficient}
\begin{split}
{\cal V}_{m n}\equiv&
\sum_{\scriptscriptstyle \mu=-2}^{\scriptscriptstyle 2}
\!\!\compactsum{k}{\shortminus \infty}{\infty}
\delta_{k+\mu,n}
W_{2\mu} D^{2^{\scriptstyle *}}_{\mu m}(\alpha\beta0)X_{k+\mu}^{-3,\mu}(e_{\rm p})
\\
=&
\sum_{\scriptscriptstyle p=-2}^{\scriptscriptstyle 2}
W_{2p}D^{2^{\scriptstyle *}}_{pm}(\alpha\beta0)X_{n}^{-3,p}(e_{\rm p})
\end{split}
\end{equation}
where we have used that $W_{2,\pm1}=0$ \citep{Press:1977}. Note that this definition
implies that
\begin{equation}
{\cal V}_{m n}^*=(-1)^m{\cal V}_{-m,-n}
\end{equation}
which is equivalent to saying that the pairs $(m,n)$ and $(-m,-n)$ are physically equivalent.

In the co-rotating frame of the star, this tidal potential is 
$\propto {\rm e}^{-{\rm i}\tilde{\omega}_{mn}t}$ with
the tidal frequency
\begin{equation}
\tilde{\omega}_{mn}\equiv n\Omega- m \Omega_\star~.
\end{equation}

In linear theory, this tidal
potential exerts a response in the star
$
\delta\rho({\bf r}, t)=\sum_{m n}\delta{\rho}_{m n}({\bf r}, t)
$,
 such that each $(m n)$ component drives a distinct fluid perturbation on the star.
 Following  \citet{Lai:2012}, we assume that the dissipative effect in the fluid response can be parametrized by a small phase lag term
 $\Delta_{m n}$, which is intrinsic to each $(m n)$ fluid perturbation:
\begin{equation}\label{eq:response_ansatz}
\delta\rho_{m n}
=\frac{M_{\rm p}}{M_\star}\left(\frac{R_\star}{a_{\rm p}}\right)^3
{\cal V}_{m n}
\delta\bar{\rho}_{m n}(r,\theta){\rm e}^{{\rm i}m\phi}
{\rm e}^{-{\rm i}n\Omega t+{\rm i}\Delta_{m n}}~.
\end{equation}
In Equation~(\ref{eq:response_ansatz}),  the  real
 function $\delta\bar{\rho}_{m n}(r,\theta)$ encapsulates the  geometric shape of the response \citep{Lai:2012},
and the characteristic response amplitude  $\simeq \nabla V|_{r=R_\star}/({\cal G}M_\star/R_\star^2)$
follows from standard static tides.

 \subsection{Tidal Response Potential and Tidal Friction Force}
 The stellar deformation in turn produces a tidal response potential, computed from solving Poisson's
 equation:
\begin{equation}
\begin{split}
\delta V({\bf r},t)
=&-{\cal G}\int {\rm d}^3x'\frac{\delta\rho({\bf r'},t)}{|{\bf r}-{\bf r}'|}
\\
=&
-\frac{{\cal G}M_{\rm p}R_\star^5}{a_{\rm p}^3r^3}
\!\sum_{\scriptscriptstyle m, n}
Y_{2m}(\theta,\phi){\cal V}_{m n}{\rm e}^{-{\rm i}n\Omega t}K_{2m n}
\end{split}
\end{equation}
where $K_{2mn}=K_{2m}(\tilde{\omega}_{mn})$ is the (complex) potential Love number \citep[e.g.,][]{Ogilvie:2013,Ogilvie:2014} with
harmonic degree $l=2$, azimuthal wavenumber $m$ and time harmonic coefficient $n$, evaluated
at the tidal frequency $\tilde{\omega}_{mn}$. In this approximation, it is
given by
\begin{equation}\label{eq:complex_love_number}
K_{2mn}=k_{2mn}{\rm e}^{{\rm i}\Delta_{m n}}
\end{equation}
where
\begin{equation}
k_{2m n}\equiv  \!\frac{4\pi}{5}\frac{1}{M_\star R_\star^2}
\int {\rm d}^3x
\delta\bar{\rho}_{m n}(r,\theta)e^{{\rm i}m\phi}
r^2Y^*_{2m}(\theta,\phi)
\end{equation}
is a generalized (real) Love number \citep[e.g.,][]{Lai:2012, Storch:2014}.

From $\delta V$, we obtain  an effective tidal friction force \citep[e.g.,][]{Mignard:1979}
by taking the gradient of the potential and evaluating it at the location of the planet, i.e.,
 $\delta{\bf F}=- M_{\rm p}\nabla\delta V|_{{\bf r}={\bf r}_{\rm p}}$. It is convenient to carry out
 this derivative in the $\mathscr{S}$ frame. Therefore, using Wigner matrices once again, we replace
\begin{equation}
Y_{2m}(\theta,\phi)=
\!\sum_{\scriptscriptstyle \nu{=}-2}^{\scriptscriptstyle 2} 
{D^{2}_{\nu m }}(\alpha\beta0)
Y_{2\nu}(\vartheta,\varphi)~.
\end{equation}
and write the gradient of $\delta V$ as
\begin{equation}
\begin{split}
\nabla\delta V({\bf r},t)
&=-\frac{{\cal G}M_{\rm p}R_\star^5}{a_{\rm p}^3r^4}\\
&~\times\!
\!\sum_{\scriptscriptstyle m n \nu}\!
{D^{2}_{\nu m }}(\alpha\beta0){\cal V}_{m n}
k_{2 m n}
{\rm e}^{-{\rm i}n\Omega t+{\rm i}\Delta_{m n}}\!
\Big\{~~
\\
& ~~~~~~ Y_{2\nu}(\vartheta,\varphi)
\Big[
{-}3\hat{\bf r}
{+}\nu\cot\vartheta \hat{\boldsymbol\vartheta}
 {+}\frac{{\rm i} \nu}{\sin\vartheta}\hat{\boldsymbol\varphi}\Big]
\\
&~~~ +Y_{2,\nu+1}(\vartheta,\varphi)\sqrt{(3+\nu)(2-\nu)}
{\rm e}^{-{\rm i}\varphi}\hat{\boldsymbol\vartheta}\Big\}\!
\end{split}
\end{equation}
where we have used  that
$\frac{\partial}{\partial \varphi}Y_{2\nu}~{=}~{\rm i} \nu Y_{2\nu}$
and that
$
\frac{\partial }{\partial \vartheta}Y_{2\nu}=
\sqrt{(3{+}\nu)(2{-}\nu)}
{\rm e}^{-{\rm i}\varphi}Y_{2,\nu+1}
{+} \nu\cot\vartheta Y_{2\nu}
$
 \citep[e.g.,][]{abramowitz+stegun}.  In the $\mathscr{S}$ frame,
 the planet's location is given by $r=r_{\rm p}$,
$\vartheta=\vartheta_{\rm p}=\tfrac{\pi}{2}$
and  $\varphi=\varphi_{\rm p}=f$, the planetary orbit's true anomaly. After writing
$\hat{\bf r}$, $\hat{\boldsymbol\vartheta}$ and $\hat{\boldsymbol\varphi}$ in
term of the cartesian vectors $\hat{\bf u}$, $\hat{\bf v}$ and $\hat{\bf n}$,
the tidal friction force evaluates to
\begin{equation}\label{eq:friction_force}
\begin{split}
\delta{\bf F}
=&
-\frac{{\cal G}M_{\rm p}^2R_\star^5}{a_{\rm p}^3r_{\rm p}^4}\\
&\times
\!\sum_{\scriptscriptstyle m n \nu}
\frac{5}{4\pi}
{D^{2}_{\nu m }}(\alpha\beta0){\cal V}_{m n}
k_{2 m n}
{\rm e}^{-{\rm i}n\Omega t+{\rm i}\Delta_{m n}}
\Big\{
\\
& ~~~~~~~~ W_{2\nu}{\rm e}^{{\rm i}\nu f}
\bigg[
(3\cos f+{{\rm i} \nu}\sin f)\hat{\bf u}
\\
&~~~~~~~~~~~~~~~~ 
+(3\sin f-{{\rm i} \nu}\cos f)\hat{\bf v}
\bigg]
\\
&~~~~ +W_{2,\nu+1}{\rm e}^{{\rm i}\nu f}
\sqrt{(3+\nu)(2-\nu)}
\hat{\bf n}\bigg\}
\end{split}
\end{equation}
This tidal friction force is accompanied by orbital energy losses given by
\begin{equation}
\dot{E}=\dot{\bf r}_{\rm p}\cdot \delta {\bf F}~,
\end{equation}
for which we may assume instantaneous Keplerian motion, i.e.,
${\bf r}_{\rm p}=r_{\rm p}[\cos f \hat{\bf u}+\sin f \hat{\bf v}]$
and thus
$\dot{\bf r}_{\rm p}=\Omega a_{\rm p}(1-e_{\rm p}^2)^{-1/2}
~[{-}\!\sin f\hat{\bf u}+(e_{\rm p}{+}\cos  f)\hat{\bf v}]$. 

Using
the Hansen coefficient expansion
\begin{equation}\label{eq:hansen_expansion}
\frac{{\rm e}^{{\rm i}\nu f}}{r_{\rm p}^\alpha}=
\frac{{\rm e}^{{\rm i}\nu \Omega t}}{a_{\rm p}^\alpha}
\sum_{s=-\infty}^\infty X_{s+\nu}^{-\alpha,\nu}{\rm e}^{{\rm i} s\Omega t}
\end{equation}
with $\alpha=4$, we write the tidal friction power as
\begin{align}
\begin{split}
\dot{E}
=&
\frac{{\cal G}M_{\rm p}^2R_\star^5\Omega}{a_{\rm p}^6\sqrt{1-e_{\rm p}^2}}\frac{5}{4\pi}\\
&\times
\!\sum_{\scriptscriptstyle \mu n \nu s}
{\rm i}
{D^{2}_{\nu \mu }}(\alpha\beta0){\cal V}_{\mu n}
k^2_{\mu n}
{\rm e}^{-{\rm i}(n-s)\Omega t+{\rm i}\Delta_{\mu n}}
\Big\{
\\
& ~~~~~~~~  \frac{1}{2}W_{2\nu}
\Big[
~2\nu  X_{s+\nu}^{-4,\nu} {\rm e}^{{\rm i}\nu \Omega t}
\\
& ~~~~~~~~~~~~~~
+(3+\nu) X_{s+\nu+1}^{-4,\nu+1}e_{\rm p} {\rm e}^{{\rm i}(\nu+1) \Omega t}
\\
& ~~~~~~~~~~~~~~ 
-(3-\nu) X_{s+\nu-1}^{-4,\nu-1}e_{\rm p} {\rm e}^{{\rm i}(\nu-1) \Omega t} 
\Big]
\bigg\}
\end{split}
\end{align}
Then, taking the time average of $\dot{E}$, we obtain,
after some rearrangement, that
\begin{equation}
\begin{split}
\langle \dot{E}\rangle
=&
\frac{{\cal G}M_{\rm p}^2R_\star^5\Omega}{a_{\rm p}^6\sqrt{1-e^2}}
\frac{5}{4\pi}
\!\sum_{\scriptscriptstyle m n  }
{\rm i}
{\cal V}_{m n}
k_{2 m n}
{\rm e}^{{\rm i}\Delta_{m n}}
\\
& ~~\times
\!\sum_{\scriptscriptstyle r}
{D^{2}_{rm }}(\alpha\beta0) W_{2r}
\big\{
r X_{n}^{-4,r} 
\\
& ~~~~~~~
+\frac{e_{\rm p}}{2}\big[
(3+r) X_{n}^{-4,r+1}
-(3-r)  X_{n}^{-4,r-1}
\big]
\big\}
\\
=&
\frac{{\cal G}M_{\rm p}^2R_\star^5\Omega}{a_{\rm p}^6}
\frac{5}{4\pi}
\\
&\times
\!\sum_{\scriptscriptstyle m n r }
{\rm i} 
n
{\cal V}_{m n}
{D^{2}_{rm }}(\alpha\beta0) W_{2r}
X_{n}^{-3,r}
k_{2 m n}
{\rm e}^{{\rm i}\Delta_{m n}}
\end{split}
\end{equation}
where we have used identity 156 in \citep{Correia:2022}
\citep[see also][]{Giacaglia:1976}. Recalling the definition of ${\cal V}_{mn}$ (Equation~\ref{eq:potential_coefficient}), we finally
write
\begin{equation}\label{eq:energy_loss}
\langle \dot{E}\rangle
=
-\frac{{\cal G}M_{\rm p}^2R_\star^5\Omega}{a_{\rm p}^6}
\frac{5}{4\pi}
\!\sum_{\scriptscriptstyle m n  }
n
|{\cal V}_{m n}|^2
k_{2 m n}
\Delta_{m n}
\end{equation}
\citep[e.g., see][eq. 25]{Lai:2012}, 
where we have additionally assumed $\Delta_{m n}\ll1$ and have only retained the real part.

Note
that these energy losses represent changes in orbital energy exclusively. Consequently,
we expect Equation~	(\ref{eq:energy_loss}) to differ from the net dissipation rate
${\cal D}$ across the system  (see Equation~\ref{eq:net_dissipation} below).

\subsection{Orbital Perturbations in Milankovitch Form}\label{sec:orbital_perturbations}
The disturbing force $\delta{\bf F}$ (Equation~\ref{eq:friction_force}) 
causes the planetary orbital elements to change over  long timescales. This secular evolution
can be expressed in terms of 
the Milankovitch state vectors 
${\bf L}_{\rm p}=\mu_{\rm p}\Omega a_{\rm p}^2({1-e^2_{\rm p}})^{1/2}\hat{\bf n}$ and 
${\bf e}_{\rm p}=e_{\rm p}\hat{\bf u}$ \citep{Milankovitch:1939}, for which the equations
of motion are:
\begin{align}
\label{eq:eom_angmom}
\dot{\bf L}_{\rm p}= & {\bf r}_{\rm p}\times\delta{\bf F}\equiv{\bf T}\\
\label{eq:eom_eccentricity}
\begin{split}\dot{\bf e}_{\rm p}
= &({\cal G}M_{\rm p}M_\star)^{-1}
\left(\mu_{\rm p}^{-1}\delta{\bf F}\times {\bf L}_{\rm p}+\dot{\bf r}_{\rm p}\times {\bf T}\right)
\end{split}
\end{align}
\citep[e.g.,][]{Eggleton:1998,Tremaine:2009, Correia:2011}.

Using Equation~(\ref{eq:friction_force}) and the series expansion of 
Equation~(\ref{eq:hansen_expansion}), we write explicitly,
\begin{equation}
\begin{split}
{\bf T}
=&{\bf r}_{\rm p}\times\delta {\bf F}
\\
=&
\frac{{\cal G}M_{\rm p}^2R_\star^5}{a_{\rm p}^6}\\
\times&
\!\sum_{\scriptscriptstyle m n \nu s}
{\rm i}\frac{5}{4\pi}
{D^{2}_{\nu m }}(\alpha\beta0){\cal V}_{m n}
k_{2 m n}
{\rm e}^{-{\rm i}(n-s)\Omega t+{\rm i}\Delta_{m n}}
\Big\{
\\
& ~~~~~ \nu W_{2\nu}X_{s+\nu}^{-3,\nu}{\rm e}^{{\rm i}\nu \Omega t}
\hat{\bf n}
\\
&~~ -{\rm i}W_{2,\nu+1}
b^{(+)}_\nu
[X_{s+\nu+1}^{-3,\nu+1}{\rm e}^{{\rm i}(\nu+1)\Omega t}\\
&~~~~~~~~~~~~~~~~~~~~ +X_{s+\nu-1}^{-3,\nu-1}{\rm e}^{{\rm i}(\nu-1)\Omega t}]\hat{\bf v}
\\
&~~ +W_{2,\nu+1}
b^{(+)}_\nu
[X_{s+\nu+1}^{-3,\nu+1}{\rm e}^{{\rm i}(\nu+1)\Omega t}\\
&~~~~~~~~~~~~~~~~~~~~ -X_{s+\nu-1}^{-3,\nu-1}{\rm e}^{{\rm i}(\nu-1)\Omega t}]
\hat{\bf u}
\Big\}~,
\end{split}
\end{equation}
\begin{align}
\begin{split}
\frac{\delta{\bf F}\times{\bf L}_{\rm p}}{\mu_{\rm p}}
=&
\frac{{\cal G}M_{\rm p}^2R_\star^5\Omega}{a_{\rm p}^5}
\sqrt{1-e_{\rm p}^2}\\
\times&
\!\sum_{\scriptscriptstyle m n \nu s}
\frac{5}{4\pi}
{D^{2}_{\nu m }}(\alpha\beta0){\cal V}_{m n}
k_{2 m n}
{\rm e}^{-{\rm i}(n-s)\Omega t+{\rm i}\Delta_{m n}}
\\
& ~\times\frac{1}{2}W_{2\nu}
\Big\{
\Big[(3+\nu) X_{s+\nu+1}^{-4,\nu+1} {\rm e}^{{\rm i}(\nu+1)\Omega t}
\\
& ~~~~~~~~~~~~
+(3-\nu) X_{s+\nu-1}^{-4,\nu-1}{\rm e}^{{\rm i}(\nu-1)\Omega t}
\Big]\hat{\bf v}
\\
&~~~~~~~~~~
+{\rm i} \Big[(3+\nu)X_{s+\nu+1}^{-4,\nu+1}{\rm e}^{{\rm i}(\nu+1)\Omega t}
\\
& ~~~~~~~~~~~~
-(3-\nu)X_{s+\nu-1}^{-4,\nu-1}{\rm e}^{{\rm i}(\nu-1)\Omega t}
\Big]\hat{\bf u}
\Big\}\!\!\!\!
\end{split}
\end{align}
and
\begin{equation}
\begin{split}
\dot{\bf r}_{\rm p}\times {\bf T}=&
\frac{{\cal G}M_{\rm p}^2R_\star^5\Omega}{a_{\rm p}^5\sqrt{1-e^2}}
\\
\times&
\!\sum_{\scriptscriptstyle m n \nu s}
{\rm i}\frac{5}{4\pi}
{D^{2}_{\nu m }}(\alpha\beta0){\cal V}_{m n}
k_{2 m n}
{\rm e}^{-{\rm i}(n-s)\Omega t+{\rm i}\Delta_{m n}}
\Big\{
\\
& ~~~ -{\rm i}\frac{\nu}{2} W_{2\nu}
[ X_{s+\nu+1}^{-3,\nu+1}{\rm e}^{{\rm i}(\nu+1)\Omega t}
\\
&~~~~~~~~~~~~~~~
- X_{s+\nu-1}^{-3,\nu-1}{\rm e}^{{\rm i}(\nu-1)\Omega t}]
\hat{\bf v}
\\
& ~~~ +\frac{\nu}{2} W_{2\nu}\big[
2 eX_{s+\nu}^{-3,\nu}{\rm e}^{{\rm i}\nu \Omega t}
+ X_{s+\nu+1}^{-3,\nu+1}{\rm e}^{{\rm i}(\nu+1)\Omega t}\\
&~~~~~~~~~~~~~~
 + X_{s+\nu-1}^{-3,\nu-1}{\rm e}^{{\rm i}(\nu-1)\Omega t}
\big]
\hat{\bf u}
\\
&~~~ -e W_{2,\nu+1}
b^{(+)}_\nu
[ X_{s+\nu+1}^{-3,\nu+1}{\rm e}^{{\rm i}(\nu+1)\Omega t}\\
&~~~~~~~~~~~~~~~
- X_{s+\nu-1}^{-3,\nu-1}{\rm e}^{{\rm i}(\nu-1)\Omega t}]
\hat{\bf n}
\Big\}
\end{split}
\end{equation}
where we have introduced the coefficients
\begin{equation}
b^{(\pm)}_{\nu}
\equiv
 \tfrac{1}{2}{\sqrt{(3\pm\nu)(2\mp \nu)}}
\end{equation}
which satisfy 
$b^{(+)}_{\nu-1}=b^{(-)}_\nu$
and
 $b^{(-)}_{\nu+1}=b^{(+)}_\nu$.

Closely following our calculation of $\langle \dot E\rangle$, we compute
the time-averages $\langle {\bf T}\rangle$, 
$\langle {\delta{\bf F}\times{\bf L}_{\rm p}}\rangle$
and $\langle \dot{\bf r}_{\rm p}\times {\bf T}\rangle$
and obtain the averaged Milankovitch equations of motion, written in
component form as:
\begin{align}
\label{eq:eom_angmom_av}
\langle \dot{\bf L}_{\rm p}\rangle
= & \langle T_u\rangle \hat{\bf u}
+\langle T_v\rangle \hat{\bf v}
+\langle T_n\rangle \hat{\bf n}
\\
\label{eq:eom_eccentricity_av}
\begin{split}
\langle\dot{\bf e}_{\rm p}\rangle
= & \langle \dot{e}_u\rangle \hat{\bf u}
+\langle \dot{e}_v\rangle \hat{\bf v}
+\langle \dot{e}_n\rangle \hat{\bf n}
\end{split}
\end{align}
where the torque terms are
\begin{equation}\label{eq:torque_u}
\begin{split}
\langle{T_u}\rangle
=&
\frac{{\cal G}M_{\rm p}^2R_\star^5}{a_{\rm p}^6}
\frac{5}{4\pi}
\!\sum_{\scriptscriptstyle m n r}
{\rm i}
k_{2 m n}
{\rm e}^{{\rm i}\Delta_{m n}}
{\cal V}_{m n}
  W_{2r}X_{n}^{-3,r}
\\
&~~
\times\!\!
\big[
{D^{2}_{r-1, m }}(\alpha\beta0)
b^{(-)}_{r}
+{D^{2}_{r+1, m }}(\alpha\beta0)
b^{(+)}_{r}
\big],\!\!\!\!
\end{split}
\end{equation}
\begin{equation}\label{eq:torque_v}
\begin{split}
\langle{T_v}\rangle
=&
-\!\frac{{\cal G}M_{\rm p}^2R_\star^5}{a_{\rm p}^6}
\frac{5}{4\pi}
\!\sum_{\scriptscriptstyle m n r}\!
k_{2 m n}
{\rm e}^{{\rm i}\Delta_{m n}}
{\cal V}_{m n}
W_{2r}
X_{n}^{-3,r}
\\
&~~~~  \times\!\!
\big[{D^{2}_{r+1, m }}(\alpha\beta0)
b^{(+)}_{r}
-{D^{2}_{r-1, m }}(\alpha\beta0)
b^{(-)}_{r}
\big]\!\!\!\!\!
\end{split}
\end{equation}
and
\begin{equation}\label{eq:torque_n}
\begin{split}
\langle{T_n}\rangle
=&
\frac{{\cal G}M_{\rm p}^2R_\star^5}{a_{\rm p}^6}
\frac{5}{4\pi}
\!\sum_{\scriptscriptstyle m n r}
{\rm i}
k_{2 m n}
{\rm e}^{{\rm i}\Delta_{m n}}
{\cal V}_{m n}
 W_{2r}X_{n}^{-3,r}
\\
&~~~~  \times
{rD^{2}_{rm }}(\alpha\beta0)~,
\end{split}
\end{equation}
and the eccentricity evolution terms are
\begin{align}
\begin{split}
\langle \dot{e}_u \rangle
&=
\frac{M_{\rm p}R_\star^5\Omega}{M_\star a_{\rm p}^5}\frac{(1-e_{\rm p}^2)^{1/2}}{e_{\rm p}}\frac{5}{4\pi}
\\
&~~\times
\!\sum_{\scriptscriptstyle m n  r}
{\rm i}
k_{2 m n}
{\rm e}^{{\rm i}\Delta_{m n}}
{\cal V}_{m n}
W_{2,r}X_{n}^{-3,r}
\\
& ~~~~~~\times
\big[n\sqrt{1-e_{\rm p}^2}-r\big]{D^{2}_{rm }}(\alpha\beta0)~,
\end{split}
\label{eq:precession_u}
\\
\begin{split}
\langle\dot{e}_v\rangle
&=
\frac{M_{\rm p}R_\star^5\Omega}{M_\star a_{\rm p}^5}
({1-e_{\rm p}^2})^{-1/2}\frac{5}{4\pi}
\\
&\times
\!\sum_{\scriptscriptstyle m n  r}
k_{2 m n}
{\rm e}^{{\rm i}\Delta_{m n}}
{\cal V}_{m n}
W_{2r}
\\
& ~\times
\Big[
\frac{3+r}{4}e_{\rm p}X^{-3,r+2}_n
+\frac{3+2r}{2}X^{-3,r+1}_n
\\
&~~~~~~ +\frac{3}{2}e_{\rm p} X^{-3,r}_n
+\frac{3-2r}{2}X^{-3,r-1}_n
\\
&~~~~~~
+\frac{3-r}{4}e_{\rm p}X^{-3,r-2}_n
\Big]
{D^{2}_{rm }}(\alpha\beta0)
\end{split}
\label{eq:precession_v}
\end{align}
and
\begin{equation}\label{eq:precession_n}
\begin{split}
\langle\dot{e}_n\rangle
&=-
\frac{M_{\rm p}R_\star^5\Omega}{M_\star a_{\rm p}^5}
({1-e_{\rm p}^2})^{-1/2}
e_{\rm p}\frac{5}{4\pi}
\\
&\times
\!\sum_{\scriptscriptstyle m n  r}
{\rm i}
k_{2 m n}
{\rm e}^{{\rm i}\Delta_{m n}}
{\cal V}_{m n}
W_{2,r}X_{n}^{-3,r}
\\
& ~\times
\big[ {D^{2}_{r-1,m }}(\alpha\beta0)b^{(+)}_{r-1} 
+{D^{2}_{r+1,m }}(\alpha\beta0)
b^{(-)}_{r+1} 
\big]~,
\end{split}
\end{equation}
where we have used standard identities  \citep{Giacaglia:1976,Correia:2022}
to ensure Hansen coefficients are of the form $X^{-3,k}_n$, and have made use of the 
relation
\begin{equation}
W_{2,m\pm2}
=- W_{2m}
\frac{b_m^{(\pm)}}{b_{m\pm1}^{(\pm)}}
\end{equation}
%
%


Before we present the final equations of the generalized effective tidal model, we clarify two further simplifications.

\paragraph{Effective Time Lag Model}
Following \citet{Lai:2012}, we assume that each phase shift $\Delta_{mn}$ depends linearly on the tidal frequency
$\tilde{\omega}_{mn}$ in the form
\begin{equation}
\frac{{\rm Im}[K_{2m}(\tilde{\omega}_{mn})]}{k_{2mn}}=\Delta_{mn} = \frac{\tilde{\omega}_{mn}\tau_{mn}}{k_{2mn}}
\end{equation}

\paragraph{Uniform Tidal Love Numbers}
Often times, the tidal Love number of second degree $k^2_{m}$ is assumed to be degenerate with the star's
azimuthal number $m$ \citep[for a detailed discussion, see e.g.,][\S 2.4]{Poisson:2014}. For rapidly spinning bodies,
however, this approximation breaks down, due to a coupling between rotational and tidal deformations, resulting
in $m$-dependent Love numbers \citep[e.g.,][]{Wahl:2017,Wahl:2020}. While these corrections can be important for
  bodies  spinning rapidly (e.g., above $\sim10\%$ of the break-up spin frequency for a model of Jupiter; \citealp{Wahl:2017}), these
  corrections are still of order unity. We thus assume that $k_{2mn}=k_2$  in the terms independent of the lag times $\tau_{mn}$ in 
  Equations~ (\ref{eq:torque_u})-(\ref{eq:torque_n}). The equations of motion finally become, for the torque terms:
\allowdisplaybreaks
\begin{equation}
\begin{split}\label{eq:torque_lag_u}
\langle{T_u}\rangle
=&
-\frac{{\cal G}M_{\rm p}^2R_\star^5}{a_{\rm p}^6}
\frac{5}{4\pi}
\!\sum_{\scriptscriptstyle m n p r}\!
W_{2p}
W_{2r}
X_{n}^{-3,p}
X_{n}^{-3,r}
\tilde{\omega}_{m n}\tau_{mn}
\\
&~~~
\times{\rm Re}\big\{
{D^{2^{\scriptstyle *}}_{pm}(\alpha\beta0)D^{2}_{r-1, m }}(\alpha\beta0)
b^{(-)}_{r}
\\
&~~~~~~ +D^{2^{\scriptstyle *}}_{pm}(\alpha\beta0){D^{2}_{r+1, m }}(\alpha\beta0)
b^{(+)}_{r}
\big\}
~~~~~~~~~~~~~~~~~~~~~
\end{split}
\end{equation}

\begin{equation}
\begin{split}\label{eq:torque_lag_v}
\langle{T_v}\rangle
=&
\frac{{\cal G}M_{\rm p}^2R_\star^5}{a_{\rm p}^6}
\frac{5}{4\pi}
\!\sum_{\scriptscriptstyle m n p r}\!
W_{2p}
W_{2r}
X_{n}^{-3,p}
X_{n}^{-3,r}
\tilde{\omega}_{m n}\tau_{m n}~~~
\\
&~~~~  
\times{\rm Im}\big\{
D^{2^{\scriptstyle *}}_{pm}(\alpha\beta0)
{D^{2}_{r+1, m }}(\alpha\beta0)
b^{(+)}_{r}
\\
&~~~~~~~
-
D^{2^{\scriptstyle *}}_{pm}(\alpha\beta0)
{D^{2}_{r-1, m }}(\alpha\beta0)
b^{(-)}_{r}
\big\}
\end{split}
\end{equation}

\begin{equation}
\begin{split}\label{eq:torque_lag_n}
\langle{T_n}\rangle
=&
-\frac{{\cal G}M_{\rm p}^2R_\star^5}{a_{\rm p}^6}
\frac{5}{4\pi}
\!\sum_{\scriptscriptstyle m n p r }
r
W_{2p}
 W_{2r}
 X_{n}^{-3,p}
 X_{n}^{-3,r}
\tilde{\omega}_{m n}\tau_{m n}
\\
&~~~~
\times{\rm Re}\big\{
D^{2^{\scriptstyle *}}_{pm}(\alpha\beta0)
D^{2}_{r m }(\alpha\beta0)
\big\}
\end{split}
\end{equation}
and for the eccentricity vector terms:
\begin{equation}
\begin{split}\label{eq:precession_lag_u}
\langle \dot{e}_u \rangle
=& -
\frac{M_{\rm p}R_\star^5\Omega}{M_\star a_{\rm p}^5}
\frac{5}{4\pi}\frac{(1-e_{\rm p}^2)^{\frac{1}{2}}}{e_{\rm p}}
\\
&\times
\!\sum_{\scriptscriptstyle m n p r}
W_{2p}
W_{2r}
X_{n}^{-3,p}
X_{n}^{-3,r}
\tilde{\omega}_{m n}\tau_{m n}
\\
&~~~~ \times
[n(1-e_{\rm p}^2)^\frac{1}{2}-r]{\rm Re}\big\{
D^{2^{\scriptstyle *}}_{pm}(\alpha\beta0)
{D^{2}_{rm }}(\alpha\beta0)
\big\}
\end{split}
\end{equation}

\begin{equation}
\begin{split}\label{eq:precession_lag_v}
\langle\dot{e}_v\rangle
=&
\frac{M_{\rm p}R_\star^5\Omega}{M_\star a_{\rm p}^5}\frac{5}{4\pi} k_2
(1-e_{\rm p}^2)^{-\frac{1}{2}}
\\
&\times
\!\sum_{\scriptscriptstyle m n  p r}
W_{2r}
W_{2p}
X_{n}^{-3,p}
\\
&~~~\times
\frac{1}{2}\big[
(3+2r)X^{-3,r+1}_n
+3e X^{-3,r}_n\\
&~~~~~~
+(3+r)\frac{e}{2}X^{-3,r+2}_n
+(3-2r)X^{-3,r-1}_n
\\
&~~~~~~
+(3-r)\frac{e}{2}X^{-3,r-2}_n
\big]
\\
&\times
{\rm Re}\big\{
D^{2^{\scriptstyle *}}_{pm}(\alpha\beta0)
{D^{2}_{rm }}(\alpha\beta0)
\big\}
\\
-&
\frac{M_{\rm p}R_\star^5\Omega}{M_\star a_{\rm p}^5}\frac{5}{4\pi} 
(1-e_{\rm p}^2)^{-\frac{1}{2}}
\\
&\times
\!\sum_{\scriptscriptstyle m n  p r}
W_{2r}
W_{2p}
X_{n}^{-3,p}
\\
& ~~~\times
\frac{1}{2}\big[
(3+2r)X^{-3,r+1}_n
+3e X^{-3,r}_n\\
&~~~~~~
+(3+r)\frac{e}{2}X^{-3,r+2}_n
+(3-2r)X^{-3,r-1}_n
\\
&~~~~~~
+(3-r)\frac{e}{2}X^{-3,r-2}_n
\big]
\tilde{\omega}_{m n}\tau_{m n}
\\
&~~~~\times
{\rm Im}\big\{
D^{2^{\scriptstyle *}}_{pm}(\alpha\beta0)
{D^{2}_{rm }}(\alpha\beta0)
\big\}
\end{split}
\end{equation}

\begin{equation}
\begin{split}\label{eq:precession_lag_n}
\langle\dot{e}_n\rangle
=&
\frac{M_{\rm p}R_\star^5\Omega}{M_\star a_{\rm p}^5}
 \frac{5}{4\pi}
e_{\rm p}(1-e_{\rm p}^2)^{-\frac{1}{2}}
\\ 
&~~~~\times
\!\sum_{\scriptscriptstyle m n  p r}
W_{2p}
W_{2r}
X_{n}^{-3,p}
X_{n}^{-3,r}
\tilde{\omega}_{m n}\tau_{m n}
\\
& ~~~~~~\times{\rm Re}\big\{
 D^{2^{\scriptstyle *}}_{pm}(\alpha\beta0){D^{2}_{r-1,m }}(\alpha\beta0)b^{(-)}_{r} 
\\
&~~~~~~~~~~~
+D^{2^{\scriptstyle *}}_{pm}(\alpha\beta0){D^{2}_{r+1,m }}(\alpha\beta0)
b^{(+)}_{r} 
\big\}
\end{split}
\end{equation}
These equations are valid for arbitrary eccentricity and inclination. Moreover, and except for our simplifying assumption that
${\rm Re}[K_{2m}(\tilde{\omega}_{mn})]=k^2_{mn}=k_2$=constant, these equations are general
and can be applied to dissipation models other than constant time lag model. Recalling the equivalence
 ${\rm Im}[K_{2m}(\tilde{\omega}_{mn})]= k_2\tilde{\omega}_{mn}\tau_{mn}$, one can implement these equations
 of motion for prescribed $K_{2m}(\tilde{\omega})$ functions for  arbitrary rheologies \citep{Correia:2022}.

%
\subsubsection{Equations of Motion in Orbital Element Form}
The equations of motion in Milankovitch form (\ref{eq:eom_angmom}-\ref{eq:eom_eccentricity})
are amenable for numerical integration; they are also particularly useful when adding complexity to the system (e.g., additional bodies,
short-range forces; see, for example, \citealp{Liu:2015}). But it is also true that these equations can obscure the simple physical interpretation of the dynamics.
For the simple interaction of two bodies, it is instructive to focus on just four equations of motion in orbital element form: $\dot{a}_{\rm p}$,  $\dot{e}_{\rm p}$,
$\dot{\Omega}_\star$ and $\dot{\beta}$.

The equation of motion for the semi-major axis $a_{\rm p}$ follows directly from Equation~(\ref{eq:energy_loss}) and from the relation
$\dot{a}_{\rm p}/a_{\rm p}=-\dot{E}/{E}$, where $E=-\tfrac{1}{2}{\cal G}M_\star M_{\rm p}/a_{\rm p}$:
\begin{equation}\label{eq:orb_semimajor}
\begin{split}
\frac{\langle \dot{a}_{\rm p}\rangle}{a_{\rm p}}
=
-\frac{2M_{\rm p}R_\star^5\Omega}{M_\star a_{\rm p}^5}\frac{5}{4\pi}&
\!\sum_{\scriptscriptstyle m n  p r}
n
W_{2p}
W_{2r}
X_{n}^{-3,p}
X_{n}^{-3,r}\tilde{\omega}_{m n}\tau_{m n}
\\
&~~\times
{\rm Re}\{D^{2^{\scriptstyle *}}_{pm}(\alpha\beta0)D^{2}_{rm}(\alpha\beta0)\}
\end{split}
\end{equation}
The eccentricity equation follows directly from Equation~(\ref{eq:precession_u}) since
$ \dot{e}_{\rm p}=\dot{\bf e}_{\rm p}\cdot{\bf e}_{\rm p}/|{\bf e}_{\rm p}|=\dot{e}_u$, i.e.,
\begin{equation}\label{eq:orb_eccentricity}
\begin{split}
\frac{\langle \dot{e}_{\rm p} \rangle}{e_{\rm p}}
=& -
\frac{M_{\rm p}R_\star^5\Omega}{M_\star a_{\rm p}^5}
\frac{(1-e_{\rm p}^2)^{\frac{1}{2}}}{e_{\rm p}^2}
\frac{5}{4\pi}
\\
&\times
\!\sum_{\scriptscriptstyle m n p r}
W_{2p}
W_{2r}
X_{n}^{-3,p}
X_{n}^{-3,r}
\tilde{\omega}_{m n}\tau_{m n}
\\
&~~~~ \times
[n(1-e_{\rm p}^2)^\frac{1}{2}-r]{\rm Re}\big\{
D^{2^{\scriptstyle *}}_{pm}(\alpha\beta0)
{D^{2}_{rm }}(\alpha\beta0)
\big\}
\end{split}
\end{equation}
For the rate of change of the spin rate ${\Omega}_\star=I_\star^{-1}{ S}_\star$, we have
$\dot{\Omega}_\star=-({\bf T}\cdot\hat{\bf s})/{I_\star}$, and thus,
$\langle\dot{\Omega}_\star\rangle
= -\frac{1}{I_\star}\big[
\sin\beta\cos\alpha \langle T_u\rangle
 +\sin\beta\sin\alpha  \langle T_v\rangle
+\cos\beta  \langle T_n\rangle
\bigg]$

\begin{equation}\label{eq:orb_spinrate}
\begin{split}
\langle\dot{\Omega}_\star\rangle
=&
\frac{{\cal G}M_{\rm p}^2R_\star^5}{I_\star a_{\rm p}^6}
\frac{5}{4\pi}
\!\sum_{\scriptscriptstyle m n p r}
W_{2p}
  W_{2r}
  X_{n}^{-3,p}
  X_{n}^{-3,r}
  \tilde{\omega}_{m n}\tau_{m n}
\\
&
~~~~~~~~\times\!\!
{\rm Re}\big\{
\!\sin\beta{\rm e}^{{\rm i}\alpha}
  D^{2^{\scriptstyle *}}_{pm}(\alpha\beta0){D^{2}_{r+1, m }}(\alpha\beta0)
b^{(+)}_{r}
\\
&~~~~~~~~~~~~
+\sin\beta {\rm e}^{-{\rm i}\alpha}
  D^{2^{\scriptstyle *}}_{pm}(\alpha\beta0) {D^{2}_{r-1, m }}(\alpha\beta0)
b^{(-)}_{r}
\\
&~~~~~~~~~~~~
+\cos\beta{r
  D^{2^{\scriptstyle *}}_{pm}(\alpha\beta0) D^{2}_{rm }}(\alpha\beta0)
\big\}
\\
=&
\frac{{\cal G}M_{\rm p}^2R_\star^5}{I_\star a_{\rm p}^6}
\frac{5}{4\pi}
\!\sum_{\scriptscriptstyle m n p r}
mW_{2p}
  W_{2r}
  X_{n}^{-3,p}
  X_{n}^{-3,r}
  \tilde{\omega}_{m n}\tau_{m n}
\\
&
~~~~~~~~\times\!\!
D^{2^{\scriptstyle *}}_{pm}(\alpha\beta0) 
 D^2_{rm}(\alpha\beta0)
\end{split}
\end{equation}
where we have used Equation~(\ref{eq:varshalovich_identity_1}) in
the last equality \citep[see][]{Varshalovich:1988}.

Next, we obtain an explicit equation of motion for the stellar obliquity angle $\beta$
from expanding 
$\tfrac{{\rm d}}{{\rm d}t}( {\bf S}_\star\cdot{\bf L}_{\rm p})
=\tfrac{{\rm d}}{{\rm d}t}( S_\star  L_{\rm p}\cos\beta)$, and using that
$\dot{\bf S}_\star={-}\dot{\bf L}_{\rm p}={-}\bf{T}$, such that
\begin{equation}
\begin{split}
\nonumber
\dot{\beta}
=&\frac{({\bf T}\cdot\hat{\bf n})}{\sin\beta}\left(\frac{\cos\beta}{L_{\rm p}}+\frac{1}{S_\star}\right)
-\frac{({\bf T}\cdot\hat{\bf s})}{\sin\beta}\left(\frac{\cos\beta}{S_\star}+\frac{1}{L_{\rm p}}\right)\\
=
&-\left(
{T_u}\cos\alpha
+{T_v}\sin\alpha\right)
\left(\frac{\cos\beta}{S_\star}+\frac{1}{L_{\rm p}}\right)
+\frac{T_n}{S_\star} \sin\beta
\end{split}
\end{equation}
Replacing the expressions for $\langle T_u\rangle$,
 $\langle T_v\rangle $ and $\langle T_n\rangle$
and using identity (4.8.17) in \citet{Varshalovich:1988}, we have
\begin{equation}\label{eq:orb_obliquity}
\begin{split}
\langle{\dot{\beta}}\rangle
=&
-\frac{{\cal G}M_{\rm p}^2R_\star^5}{a_{\rm p}^6}
\frac{5}{4\pi}
\!\sum_{\scriptscriptstyle m n p r}
W_{2p}
  W_{2r}
  X_{n}^{-3,p}
  X_{n}^{-3,r}
  \tilde{\omega}_{m n}\tau_{m n}
  \\
&~~
\times\!\!
\bigg[
r\frac{\sin\beta}{S_\star}
-
\frac{m-r\cos\beta}{\sin\beta}
\left(\frac{\cos\beta}{S_\star}+\frac{1}{L_{\rm p}}\right)
\bigg]
\\
&~~
\times
{\rm Re}\{
D^{2^{\scriptstyle *}}_{pm}(\alpha\beta0)
D^{2}_{rm }(\alpha\beta0)
\}
\end{split}
\end{equation}

These expressions allow us to compute the {\it net} dissipation rate, which differs from
Equation~(\ref{eq:energy_loss}).
Since the net dissipation rate ${\cal D}$ is defined
as the change in the total energy budget 
$E_{\rm tot}=E+\tfrac{1}{2}I_\star\Omega_\star^2$, then
 \begin{equation}
 {\cal D}=-\dot{E}_{\rm tot}=-\dot{E}-I_\star\Omega_\star\dot{\Omega}_\star~.
 \end{equation}
 Replacing Equations~(\ref{eq:energy_loss}) and~(\ref{eq:orb_spinrate}), we obtain
\begin{equation}\label{eq:net_dissipation} 
\begin{split}
\langle {\cal D}\rangle
=
\frac{{\cal G}M_{\rm p}^2R_\star^5}{a_{\rm p}^6}
&\frac{5}{4\pi}
\!\sum_{\scriptscriptstyle m n pr }
\left(n\Omega-m\Omega_\star \right)
W_{2p}
W_{2r}
X_{n}^{-3,p}
X_{n}^{-3,r}
\\
&\times
D^{2^{\scriptstyle *}}_{pm}(\alpha\beta0)
D^{2}_{rm}(\alpha\beta0)
  \tilde{\omega}_{m n}\tau_{m n}
  \\
  =
  \frac{{\cal G}M_{\rm p}^2R_\star^5}{a_{\rm p}^6}
&\frac{5}{4\pi}
\!\sum_{\scriptscriptstyle m n  }
|{\cal V}_{mn}|^2
  \tilde{\omega}_{m n}^2\tau_{m n}
  \end{split}
\end{equation}
and the per-component dissipation rate is written as
\begin{equation}\label{eq:specific_dissipation}
\begin{split}
 {\cal D}_{2mn}
  =&
  \frac{{\cal G}M_{\rm p}^2R_\star^5}{a_{\rm p}^6}
\frac{5}{4\pi}
|{\cal V}_{mn}|^2
  \tilde{\omega}_{m n}^2\tau_{m n}
  \\
    =&
  \frac{{\cal G}M_{\rm p}^2R_\star^5}{a_{\rm p}^6}
\frac{5}{4\pi}
|{\cal V}_{mn}|^2
  \tilde{\omega}_{mn}{\rm Im}(K_{2mn})~,
  \end{split}
\end{equation}
which is identical to the usual relation between dissipation and ${\rm Im}[K_{lm}]$
\citep[][eq. 41]{Ogilvie:2013} associated to a tidal potential
of the form $V({\bf r},t)=A_{lmn}(r/R_\star)^lY_{lm}(\theta,\phi){\rm e}^{-{\rm i}\tilde{\omega}_{mn}t}$
after identifying  $A_{2mn}=-2({{\cal G}M_{\rm p}R_\star^2}/a_{\rm p}^3)~{\cal V}_{m n}$, where
 the extra factor of 2 comes from the combination of the physically equivalent pairs $(m,n)$ and $(-m,-n)$ in
 Equation~(\ref{eq:net_dissipation}).

Note that Equations~(\ref{eq:orb_semimajor})-(\ref{eq:orb_obliquity}) describe a system of 5 unknowns
with only 4 equations, having omitted the equation of motion for the angle $\alpha$. While there exist  situations
in which this angle cannot be ignored,
in cases where the apsidal precession of the orbit is fast, one can average again over this precession period \citep{Correia:2009,Correia:2011}.

This additional averaging step is easily carried out in our current notation, since
\begin{equation}
{\rm e}^{{\rm i}s \alpha}
D^{l^{\scriptstyle *}}_{\mu m}(\alpha\beta\gamma)
D^l_{\nu m}(\alpha\beta\gamma)
={\rm e}^{{\rm i}(s+\mu-\nu) \alpha}
d^l_{\nu m}(\beta)
d^l_{\mu m}(\beta)
\end{equation}
is non-oscillatory only for $s+\mu-\nu=0$. Thus, after carrying the sum over $r$, we have
\begin{equation}\label{eq:orb_semimajor_doubleav}
\begin{split}
\frac{\langle\langle \dot{a}_{\rm p}\rangle\rangle}{a_{\rm p}}
=&
-\frac{2M_{\rm p}R_\star^5\Omega}{M_\star a_{\rm p}^5}
\\
&\times
\frac{5}{4\pi}
\!\sum_{\scriptscriptstyle m n  p}
n
\left(W_{2p}\right)^2
[X_{n}^{-3,p}]^2
[d^2_{pm}(\beta)]^2
\tilde{\omega}_{m n}\tau_{m n}
\end{split}
\end{equation}
\begin{equation}\label{eq:orb_eccentricity_doubleav}
\begin{split}
\frac{\langle\langle \dot{e}_{\rm p} \rangle\rangle}{e_{\rm p}}
=& -
\frac{M_{\rm p}R_\star^5\Omega}{M_\star a_{\rm p}^5}
\frac{(1-e_{\rm p}^2)^{\frac{1}{2}}}{e_{\rm p}^2}
\\
&\times
\frac{5}{4\pi}
\!\sum_{\scriptscriptstyle m n p}
\Big\{
[n(1-e_{\rm p}^2)^\frac{1}{2}-p]
\left(W_{2p}\right)^2
[X_{n}^{-3,p}]^2
\\
&~~~~~~~~~~~~~
\times
[d^{2}_{pm}(\beta)]^2
\tilde{\omega}_{m n}\tau_{m n}
\Big\}
\end{split}
\end{equation}
\begin{equation}\label{eq:orb_spinrate_doubleav}
\begin{split}
\langle\langle\dot{\Omega}_\star\rangle\rangle
=&
\frac{{\cal G}M_{\rm p}^2R_\star^5}{I_\star a_{\rm p}^6}
\\
&\times
\frac{5}{4\pi}
\!\sum_{\scriptscriptstyle m n p}
m
\left(  W_{2p}\right)^2
[  X_{n}^{-3,p}]^2
[  d^{2}_{pm}(\beta)]^2
  \tilde{\omega}_{m n}\tau_{m n}
\end{split}
\end{equation}
and
\begin{equation}\label{eq:orb_obliquity_doubleav}
\begin{split}
\langle\langle{\dot{\beta}}\rangle\rangle
=&
-\frac{{\cal G}M_{\rm p}^2R_\star^5}{a_{\rm p}^6}
\frac{1}{\sin\beta}
\\
&\times
\frac{5}{4\pi}
\!\sum_{\scriptscriptstyle m n p }
\Big\{
\left(W_{2p}\right)^2
[  X_{n}^{-3,p}]^2
[d^{2}_{pm}(\beta)]^2
  \tilde{\omega}_{m n}\tau_{m n}
  \\
&~~~~~~~~~~~
\times\!\!
\Big[
\frac{p-m\cos\beta}{S_\star}
+\frac{p\cos\beta-m}{L_{\rm p}}
\Big]
\Big\}
\end{split}
\end{equation}
where we have used identity (4.8.17) in \citet{Varshalovich:1988} to simplify 
Equation~(\ref{eq:orb_spinrate_doubleav}).

\subsection{Equations of Motion in the Weak Friction Limit}\label{sec:weak_friction}
In the standard weak friction limit, we assume a constant time lag $\tau_{mn}=\tau$ \citep[e.g.,][]{Alexander:1973,Mignard:1979,Hut:1981}.
 In this limit, expressions (\ref{eq:torque_u})-(\ref{eq:precession_n}) become greatly simplified by virtue of the orthogonality properties of
the Wigner D-matrices (see Appendix~\ref{app:weak_friction}). The torque  terms reduce to
\begin{equation}\label{eq:torque_u_equi}
\langle{T_u}\rangle_{\!\rm \scriptscriptstyle eq}
=
\frac{3{\cal G}M_{\rm p}^2R_\star^5}{a_{\rm p}^6}
\tau
\Omega_\star
\sin\beta\cos\alpha
\frac{1}{2}
\left(X_{0}^{-6,0}-X_{0}^{-6,2}\right)~,
\end{equation}
\begin{equation}\label{eq:torque_v_equi}
\langle{T_v}\rangle_{\!\rm  \scriptscriptstyle eq}
=
\frac{3{\cal G}M_{\rm p}^2R_\star^5}{a_{\rm p}^6}
\tau
\Omega_\star
\sin\beta\sin\alpha
\frac{1}{2}
\left(X_{0}^{-6,0}+X_{0}^{-6,2}\right)~,
\end{equation}
and
\begin{equation}\label{eq:torque_n_equi}
\langle{T_n}\rangle_{\!\rm \scriptscriptstyle eq}
=3\frac{{\cal G}M_{\rm p}^2R_\star^5\Omega}{a_{\rm p}^6}\tau
\!
\left[
\frac{\Omega_\star}{\Omega} \cos\beta
X_0^{-6,0}
\!
{-}
({1-e_{\rm p}^2})^{\frac{1}{2}}X_0^{-8,0}
\!
\right],
\end{equation}
and the eccentricity vector terms reduce to
\begin{align}\label{eq:precession_u_equi}
\begin{split}
\langle \dot{e}_u \rangle_{\!\rm  \scriptscriptstyle eq}
=& 
\frac{3M_{\rm p}R_\star^5\Omega}{M_\star a_{\rm p}^5}
\tau \Omega
e_{\rm p}^{-1}(1-e_{\rm p}^2)
\\
&\times 2\big[
-3X_0^{-9,0}+2X_0^{-8,0}+(1-e^2)X_0^{-10,0}
\big]
\\
 &
 +
\frac{3M_{\rm p}R_\star^5\Omega}{M_\star a_{\rm p}^5}
\tau \Omega_\star \cos\beta
\\
&~~~\times
\left[(1-e_{\rm p}^2)X_0^{-8,0}-X_0^{-6,0}\right]e_{\rm p}^{-1}(1-e_{\rm p}^2)^{\frac{1}{2}}
\end{split}
\end{align}
\begin{align}\label{eq:precession_v_equi}
\begin{split}
\langle\dot{e}_v\rangle_{\!\rm  \scriptscriptstyle eq}
&=
\frac{3M_{\rm p}R_\star^5\Omega}{M_\star a_{\rm p}^5} k_2
(1-e_{\rm p}^2)^{-\frac{1}{2}}
\\
&~~\times
\frac{1}{2}\left(
2X_0^{-6,1}
+eX_0^{-6,0}
+eX_0^{-6,2}
\right)
\end{split}
\end{align}
and
\begin{equation}\label{eq:precession_n_equi}
\begin{split}
\langle\dot{e}_n\rangle_{\!\rm  \scriptscriptstyle eq}
&=-
\frac{M_{\rm p}R_\star^5\Omega}{M_\star a_{\rm p}^5}
 \frac{3}{2}
e_{\rm p}(1-e_{\rm p}^2)^{-\frac{1}{2}} \tau \Omega_\star
\sin\beta\cos\alpha
\\ 
&~~~~\times \left(X_{0}^{-6,0}-X_{0}^{-6,2}\right)
\end{split}
\end{equation}
Using the closed-form definition of the $X_0^{l,m}$ Hansen coefficients \citep{Hughes:1981}, it is straightforward
to confirm that Equation~(\ref{eq:torque_n_equi})-(\ref{eq:precession_n_equi}) are equivalent to 
Equations~22 and 21 of \citet{Correia:2011} after replacing $\tau = k_2 \Delta t$, where $\Delta t$
is the conventional time lag \citep{Mignard:1979,Hut:1981}. Namely, we have
\begin{equation}\label{eq:torque_equi}
\begin{split}
\langle{{\bf T}}\rangle_{\!\rm \scriptscriptstyle eq}
=3\frac{{\cal G}M_{\rm p}^2R_\star^5}{a_{\rm p}^6(1-e_{\rm p}^2)^{9/2}}\tau
\Omega 
\Bigg\{
\frac{1}{2}
f_4(e_{\rm p}^2)
\frac{\Omega_\star}{\Omega}(\hat{\bf s}\cdot\hat{\bf u} )
\hat{\bf u}~~~\\
+\frac{1}{2}
f_6(e_{\rm p}^2)
\frac{\Omega_\star}{\Omega}(\hat{\bf s}\cdot\hat{\bf v} )
\hat{\bf v}~~~\\
\!
+\Big[
(\hat{\bf s}\cdot\hat{\bf n} )\frac{\Omega_\star}{\Omega}
f_5(e_{\rm p}^2)
-
(1-e_{\rm p}^2)^{-3/2}f_2(e_{\rm p}^2)\!
\Big]
\hat{\bf n}
\Bigg\}
\end{split}
\end{equation}
and
\begin{equation}\label{eq:eccentricity_equi}
\begin{split}
\langle{\dot{\bf e}}\rangle_{\!\rm \scriptscriptstyle eq}
=&e_{\rm p}
\frac{3M_{\rm p}R_\star^5\Omega^2}{M_\star a_{\rm p}^5}
\tau 
\bigg\{
\Big[
\\
&-9(1-e_{\rm p}^2)^{-13/2}f_3(e_{\rm p}^2)
\\
 &
 +\frac{11}{2}(1-e_{\rm p}^2)^{-5}
 \frac{\Omega_\star}{\Omega} (\hat{\bf s}\cdot\hat{\bf n} )
f_4(e_{\rm p}^2)
\Big]
\hat{\bf u}
\\
&
+\Big[-\frac{1}{2}
(1-e_{\rm p}^2)^{-5} \frac{ \Omega_\star}{\Omega}
(\hat{\bf s}\cdot\hat{\bf u} )
 f_4(e_{\rm p}^2)
\Big]\hat{\bf n}\bigg\}
\\
&+e_{\rm p}\frac{15M_{\rm p}R_\star^5\Omega}{2M_\star a_{\rm p}^5} k_2
(1-e_{\rm p}^2)^{-5}f_4(e_{\rm p}^2)
\hat{\bf v}
\end{split}
\end{equation}
where we have used the $f_i(e^2)$ functions  ($i=1,..,5$) defined in
\citet{Hut:1981}, plus the additional definition
$f_6(e^2)=1+\tfrac{9}{2}e^2+\tfrac{5}{8}e^4$.


Similarly, the weak friction theory of equilibrium tides can be recovered
 in orbital element form
 if we set $\tau_{mn}=\tau$ in Equations~(\ref{eq:orb_semimajor_doubleav})-(\ref{eq:orb_obliquity_doubleav}).
 For completeness,
we list these familiar expressions here (see  also Appendix~\ref{app:weak_friction}):
\begin{equation}\label{eq:orb_semimajor_equi}
\begin{split}
\frac{\langle \dot{a}_{\rm p}\rangle_{\!\rm \scriptscriptstyle eq}}{a_{\rm p}}
=
-\frac{6M_{\rm p}R_\star^5\Omega^2}{M_\star a_{\rm p}^5}\tau\bigg[&
f_1(e_{\rm p}^2)(1-e_{\rm p}^2)^{-15/2}
\\
&
-\frac{\Omega_\star}{\Omega}
\cos\beta
f_2(e_{\rm p}^2)(1-e_{\rm p}^2)^{-6}
\bigg]
\end{split}
\end{equation}
\begin{equation}\label{eq:orb_eccentricity_equi}
\begin{split}
\frac{\langle \dot{e}_{\rm p} \rangle_{\!\rm \scriptscriptstyle eq}}{e_{\rm p}}
&=
-\frac{27M_{\rm p}R_\star^5\Omega^2}{M_\star a_{\rm p}^5(1-e_{\rm p}^2)^{13/2}}
\tau
\Bigg\{
\\
&~~
  f_3(e_{\rm p}^2)
-\frac{\Omega_\star}{\Omega}
\cos\beta
\frac{11}{18}(1-e_{\rm p}^2)^{3/2}
 f_4(e_{\rm p}^2)
\Bigg\}
\end{split}
\end{equation}
\begin{equation}\label{eq:orb_spinrate_equi}
\begin{split}
\langle\dot{\Omega}_\star\rangle_{\!\rm \scriptscriptstyle eq}
=&
\frac{3{\cal G}M_{\rm p}^2R_\star^5\Omega}{I_\star a_{\rm p}^6(1-e_{\rm p}^2)^{6}}
\tau
\bigg\{
\cos\beta
f_2(e_{\rm p}^2)
\\
-&
\frac{1}{2}\frac{\Omega_\star}{\Omega}(1-e_{\rm p}^2)^{3/2}
\big[
(1+\cos^2\beta)f_5(e_{\rm p}^2)
\\
&~~~~~~
-\sin^2\beta\cos2\alpha\left(f_5(e_{\rm p}^2)-f_4(e_{\rm p}^2)\right)
\big]
\bigg\}
\end{split}
\end{equation}
\begin{equation}\label{eq:orb_obliquity_equi}
\begin{split}
\langle{\dot{\beta}}\rangle_{\!\rm \scriptscriptstyle eq}
=&
-\frac{3{\cal G}M_{\rm p}^2R_\star^5\Omega}{a_{\rm p}^6(1-e_{\rm p}^2)^{6}}
\frac{\sin\beta}{S_\star}
\tau
\Bigg\{f_2(e_{\rm p}^2)
\\
& 
-\frac{\Omega_\star}{2\Omega}
\left(\cos\beta{-}\frac{S_\star}{L_{\rm p}}\right)
 f_5(e_{\rm p}^2)(1-e_{\rm p}^2)^{3/2}
\\
&-\frac{\Omega_\star}{2\Omega}
\cos 2\alpha
\left(\cos\beta{+}\frac{S_\star}{L_{\rm p}}\right)\\
&~~~~~~~~~~~~~~\times
   [ f_5(e_{\rm p}^2)- f_4(e_{\rm p}^2)](1-e_{\rm p}^2)^{3/2}
\Bigg\}
\end{split}
\end{equation}
where, as before, double brackets denote averaging over orbital and apsidal precession periods.
These equations are in agreement with equations 42-44 in \citet{Lai:2012} when one takes the limit $e_{\rm p}=0$.
 With the replacement
 $\tau = k_2\Delta t$, where $\Delta t$ is the conventional 'constant time lag' \citep{Mignard:1979,Hut:1981,Correia:2011},
 Equations~(\ref{eq:orb_semimajor_equi})-(\ref{eq:orb_obliquity_equi}) are also in agreement with equations
 9-11 and 13 of \citet{Hut:1981} to first order in $\beta$.

\section{Obliquity Damping from Dissipation of Inertial Waves}\label{sec:inertial_waves}
\subsection{The Uniqueness of the (1,0) Component}\label{sec:10_component}
In this work, we focus our attention on the orbital evolution due to an enhanced dissipative response
to the (1,0) component of the tidal potential. Direct inspection of Equations~(\ref{eq:orb_semimajor_doubleav}) and~(\ref{eq:orb_eccentricity_doubleav}) shows that,
for $n=0$,
\begin{equation}\label{eq:no_orbital_decay}
\langle{\dot{a}}_{\rm p}\rangle_{\scriptscriptstyle 10}=
\langle{\dot{e}}_{\rm p}\rangle_{\scriptscriptstyle 10}=0
\end{equation}
i.e., the rates of orbital decay and circularization are  identically zero.

The obliquity damping rate, on the other hand, is finite. \citet{Lai:2012} has  computed the obliquity damping rate of the (1,0) component for circular orbits.  Their results can be generalized to eccentric orbits straightforwardly from Equation~(\ref{eq:orb_obliquity_doubleav})
after directly replacing $(m,n)=(\pm1,0)$.  Knowing that $X_0^{-3,p}=X_0^{-3,-p}=0$ for 
$-|p|-1\leq-3$ \citep[e.g.,][]{Brumberg:2012}, then $p=0$ becomes the only non-vanishing term in the sum.
 Thus, after replacing  
$d^2_{01}(\beta)=\sqrt{{3}/{2}}\cos\beta\sin\beta$ 
and $X_{0}^{-3,0}(e)=(1-e^2)^{-3/2}$,  we have
\begin{equation}\label{eq:obliquity_damping_10}
\begin{split}
\langle{\dot{\beta}}\rangle_{\scriptscriptstyle 10}
=&-
\frac{3}{4}
\frac{{\cal G}M_{\rm p}^2R_\star^5}{a_{\rm p}^6I_\star(1-e_{\rm p}^2)^{3}}
\sin\beta\cos^2\beta
\left(\cos\beta+\frac{S_\star}{L_{\rm p}}\right)
 \tau_{1 0}
\end{split}
\end{equation}
where we have omitted the double brackets for simplicity and the extra factor of 2 comes from summing the 
physically equivalent $m=1$ and $m=-1$ components. This expression is
identical to eq.(52) in \citet{Lai:2012}, except for the eccentricity-dependent factor, which
 further accelerates  the obliquity damping effect at high $e_{\rm p}$.

As  discussed by \citet{Lai:2012}, Equation~(\ref{eq:no_orbital_decay}) indicates that
the energy losses due to the (10) component cannot come from losses in orbital energy.
In this case, energy
loss comes at the expense of the star's rotational kinetic energy. 
Replacing  $(m,n)=(\pm1,0)$  into Equation~(\ref{eq:orb_spinrate_doubleav}) results in
\begin{equation}\label{eq:spin_damping_10}
\begin{split}
\langle\dot{\Omega}_\star\rangle_{\scriptscriptstyle 10}
=&
-\frac{3{\cal G}M_{\rm p}^2R_\star^5}{4I_\star a_{\rm p}^6(1-e_{\rm p}^2)^{3}}
\sin^2\beta\cos^2\beta
\Omega_\star \tau_{1 0}~.
\end{split}
\end{equation}
i.e., the star slows down as a result of  dissipation of the (1,0) component.

A physically motivated value for $\tau_{10}$ can be obtained  from the relation ${\rm Im}[K_{2mn}]={\rm Im}[K_2(\tilde{\omega}_{mn})]=\tilde{\omega}_{mn}\tau_{mn}$ 
(Equation~\ref{eq:complex_love_number}), which
allows to write
the modified lag time $\tau_{mn}$ in terms of a known response function $K_{2m}(\tilde{\omega})$ evaluated at nominal resonance,
i.e.,
\begin{equation}\label{eq:10_lagtime}
\tau_{10}= \frac{{\rm Im}[K_{21}(\tilde{\omega}_{10})]}{\tilde{\omega}_{10}}
\end{equation}
where  $K_{2m}(\tilde{\omega})$  must be obtained by direct solution of the linearized equations of fluid dynamics \citep[e.g.,][]{Ogilvie:2013,Dewberry:2023} or via a modal decomposition \citep{Wu:2024}. 

\subsection{Inertial Wave Dissipation at the $\tilde{\omega}_{10}$ Tidal Frequency}\label{sec:iw_dissipation}
A resonant response at tidal frequency  $\tilde{\omega}_{10}$ guarantees enhanced obliquity damping. Such behavior was originally
reported by \citet{Barker:2009} in their modeling of IW dissipation in Solar type stars under tidal forcing  $\propto Y_{21}$. But this enhancement was later identified by
\citet{Ogilvie:2013} to be a `formal resonance' with the so-called spin-over global mode  \citep{Poincare:1910, Greenspan:1968},
which only appeared to provide enhanced dissipation due to an oversimplified treatment of wave damping. If instead of an ad hoc damping force, 
one assumes viscous damping with a dynamical viscosity $\mu$, then the energy dissipation rate is
\begin{equation}\label{eq:viscous_dissipation}
{\cal D}=\int_{V_{\rm conv}}\mu
\delta{\bf S}^*:\delta{\bf S}
dV
\end{equation}
with $\delta{\bf S}$  being the traceless strain tensor associated to the velocity perturbations ${\bf v}$:
\begin{equation}\label{eq:strain_tensor}
\delta{\bf S}=\frac{1}{2}\left[\nabla {\bf v}+(\nabla{\bf v})^T-\tfrac{2}{3}(\nabla\cdot{\bf v})\mathbb{I}\right]~,
\end{equation}
 \citep[e.g.,][]{Landau_Fluids,KunduCohen:2001}.  The tensor $\delta{\bf S}$ vanishes exactly when ${\bf v}$ is in rigid rotation,
 which is precisely what the spin-over mode is: ${\bf v}=\delta\boldsymbol{\Omega}_\perp \times {\bf r}$, 
where $\delta\boldsymbol{\Omega}_\perp=\epsilon\Omega_\star\left[\cos t\hat{\bf x}+\sin t\hat{\bf y}\right]$ is an arbitrary angular velocity vector
perpendicular to the background uniformly-rotating flow \citep{Zhang-Liao:2017}. Thus, excitation of the global mode is expected
to provide exactly zero dissipation at the forcing frequency $\tilde{\omega}_{10}$.

At the same time, the spin-over mode is not the only dynamical response that emerges from tidal forcing at this frequency. Work by
\citet{Lin:2017} has shown that the traveling waves excited at this frequency can provide substantial viscous dissipation, although its magnitude depends strongly on the internal structure of the fluid body, particularly the presence of a rigid core. More recently, \citet{Dewberry:2022} have explored the nature of the spin-over mode
in direct 2D calculations of tidal forcing in rotationally deformed stars. Although these authors find that the spin-over mode retains its zero inertial-frame frequency even under strong centrifugal deformation, their  seminal work on modified spectrum of inertial oscillations reveals that the
 effects of mode mixing in rapidly rotating stars remain incompletely explored, potentially opening additional dissipation channels.

In this work, we take an agnostic approach to IW dissipation, while aiming to infer the amount of dissipation needed to explain the observational trends
with WJs. We  thus defer further discussion on the nature of the spin-over mode to Section~\ref{sec:discussion_spin-over} below.
We proceed with some simple scaling arguments and parametrize the enhancement
of the linear response at the tidal frequency $\tilde{\omega}_{10}$ by introducing an enhancement coefficient  $\eta_{10}$, via
\begin{equation}
\eta_{10}\equiv 
\frac{{\rm Im}[K_{2m}(\tilde{\omega}_{10})]}
{{\rm Im}[K_{2m}^{({\rm bg})}(\tilde{\omega}_{10})]}
\end{equation}
and where $K_{2m}^{({\rm bg})}(\tilde{\omega})$ represents the `background' response function. This background is well represented by
\begin{equation}
{\rm Im}\{K_{lm}^{({\rm bg})}(\tilde{\omega})\}\frac{\omega_{\rm dyn}^2}{\Omega_\star\tilde{\omega}} \approx {\cal K}
\end{equation}
where empirically, ${\cal K}\sim 10^{-6}$-$10^{-5}$ \citep[e.g.][]{Wu:2024,Dewberry:2024,Dewberry:2026} and is the result of the contribution of multiple broad resonances, such
as large-scale f-modes \citep{Dewberry:2024}.  Replacing these scaling into Equation~(\ref{eq:10_lagtime}), we then have
\begin{equation}\label{eq:10_lagtime_alt}
\begin{split}
\tau_{10}=&
\eta_{10}{\cal K}\frac{\Omega_\star}{\omega_{\rm dyn}^2}
\\
\approx &
1.84\times10^{-4}{\rm s}\,\,
\eta_{10}\!
\left(\frac{{\cal K}}{10^{-6}}\right)\!
\left(\frac{P_\star}{{\rm d}}\right)^{\!\!-1}\!\!\!
 \left(\frac{R_\star}{R_\odot}\right)^{3}\!
 \left(\frac{M_\star}{M_\odot}\right)^{\!\!-1}
\end{split}
\end{equation}

The strength $\eta_{mn}$ associated to any resonance peak in the inertial range is commonly found to be in the range $10^4$-$10^9$ \citep{Ogilvie:2004,Ogilvie:2007,Ogilvie:2013,Wu:2024,Dewberry:2026}. In most cases, however, this is for coplanar orbits (i.e., without obliquity tides),
while the few studies focused specifically on the (1,0) response conclude that its contribution to obliquity damping is modest at best \citep{Lin:2017}. 
As stated above, in order to explain the observational trends, we speculate that the response at $\tilde{\omega}_{10}$ could be enhanced
 (see discussion in Section~\ref{sec:discussion_spin-over}), and thus leave $\eta_{10}$ as a tunable free parameter.

As a last step in constructing our toy model, we tie the dimensionless coefficient ${\cal K}$ to stellar structure. \citet{Dewberry:2026} shows that ${\cal K}$
depends on the extent of the stellar convective envelope. Since the dissipation is inherently viscous, we thus assume that  ${\cal D}$ in Equation~(\ref{eq:viscous_dissipation})
scales with the product of the convective envelope's mass $M_{\rm conv}$ and the kinematic viscosity coefficient for turbulent eddies $\nu$. Following the argument made by \citet{Zahn:1977}, we equate the eddy turnover time-scale $\tau_{\rm conv}\sim (M_\star R_\star^2/L_\star)^{1/3}$
to the viscous time-scale $\tau_\nu\sim R_\star^2/\nu$ to solve for the kinematic eddy viscosity
$\nu\sim R_\star^{4/3}L_\star^{1/3}M_\star^{-1/3}$ and thus, applying this dimensional analysis only to the star's convective
envelope \citep[e.g.,][]{Vick:2020}, we write $M_{\rm conv}\nu \sim l_{\rm conv}^{4/3}L_\star^{1/3}M_{\rm conv}^{2/3}$, where
$l_{\rm conv}$ is the length-scale of the convective region. We thus  derive a heuristic scaling relation:
\begin{equation}\label{eq:linear_response_bg_coeff}
{\cal K}=10^{-6}\left(\frac{M_{\rm conv}}{M_\odot}\right)^{2/3}\left(\frac{l_{\rm conv}}{l_{\rm conv,\odot}}\right)^{4/3}\left(\frac{L_\star}{L_\odot}\right)^{1/3}
\end{equation}
which will prove useful later on when we consider evolving stars.

\subsection{Equations of Motion: Enhanced (1,0)-Response  Coexisting with Weak Friction Theory}
\label{sec:equations_of_motion}
A convenient way to minimally modify the weak friction equations of motion and enhance 
dissipation due a specific $(m,n)$ components is that laid out by \citet{Lai:2012}, who propose to assume 
$\tau_{mn}=\tau_{\!\rm  \scriptscriptstyle eq}$ for all modes {\it except} for $(m,n)=(\pm1,0)$. Thus the equations of motion become
those of weak friction (Equation~\ref{eq:torque_equi}-\ref{eq:eccentricity_equi})
but where the $(\pm1,0)$ terms evaluated with $\tau_{10}=\tau_{\!\rm  \scriptscriptstyle eq}$
are removed, and then reinstated with a different $\tau_{10}$. We here take the same approach, now In Milankovitch vector form.

 The $(\pm1,0)$ terms of $\langle{T_u}\rangle$ (Equation~\ref{eq:torque_lag_u}) are
\begin{equation}
\begin{split}\label{eq:torque_u_10}
\langle{T_u}\rangle_{\scriptscriptstyle 10}
=&
2\frac{{\cal G}M_{\rm p}^2R_\star^5}{a_{\rm p}^6}
\Omega_\star\tau_{10}
\frac{5}{4\pi}
\left(W_{20}\right)^2
[X_{0}^{-3,0}]^2
\\
&~~~
\times
\sqrt{\frac{3}{2}}
~d^{2}_{01}(\beta)\left[
d^{2}_{-11 }(\beta)
+
d^2_{11}(\beta)\right]\cos\alpha
~~~~~~~~~~~~~~~~~~~~~
\\
=&
\frac{3}{4}\frac{{\cal G}M_{\rm p}^2R_\star^5}{a_{\rm p}^6(1-e_{\rm p}^2)^{3}}
\Omega_\star\tau_{10}
\cos\alpha
\cos^2\beta\sin\beta
\end{split}
\end{equation}
where we have again used that $X_0^{-3,p}{=}0$ for  $-|p|{-}1\leq-3$,
$d^2_{01}(\beta)=\sqrt{{3}/{2}}\cos\beta\sin\beta$ and $X_{0}^{-3,0}(e)=(1-e^2)^{-3/2}$,  alongside
that that  $d^l_{\pm11}(\beta){=}\mp\tfrac{1}{2}(1\mp2\cos\beta)(1\pm\cos\beta)$
(Appendix~\ref{app:wigner_matrices}). As before, the extra factor of 2 comes from combining
the $(1,0)$ and $(-1,0)$ components. 

Similarly, we obtain
\allowdisplaybreaks
\begin{subequations}
\begin{align}
\begin{split}\label{eq:torque_v_10}
\langle{T_v}\rangle_{\scriptscriptstyle 10}
=&
\frac{3{\cal G}M_{\rm p}^2R_\star^5}{4a_{\rm p}^6(1-e_{\rm p}^2)^{3}}
\Omega_\star\tau_{10}
 \sin\alpha
\cos^2\beta
\sin\beta
\end{split}
\\
\begin{split}\label{eq:torque_n_10}
\langle{T_n}\rangle_{\scriptscriptstyle 10}
=&0
\end{split}
\end{align}
\end{subequations}
and
\begin{subequations}
\begin{align}
\begin{split}\label{eq:precession_u_10}
\langle \dot{e}_u \rangle_{\scriptscriptstyle 10}
=& 0
\end{split}
\\
\begin{split}\label{eq:precession_v_10}
\langle\dot{e}_v\rangle_{\scriptscriptstyle 10}
=&
\frac{9M_{\rm p}R_\star^5\Omega e_{\rm p}}{4M_\star a_{\rm p}^5(1-e_{\rm p}^2)^{7/2}}k_2
\cos^2\beta\sin^2\beta
\end{split}
\\
\begin{split}\label{eq:precession_n_10}
\langle\dot{e}_n\rangle_{\scriptscriptstyle 10}
=&
\frac{-3M_{\rm p}R_\star^5\Omega e_{\rm p}}{4M_\star a_{\rm p}^5(1-e_{\rm p}^2)^{7/2}}
\Omega_\star\tau_{10}
\cos^2\beta\sin\beta
\cos\alpha
\end{split}
\end{align}
\end{subequations}
Note that $\langle\dot{e}_v\rangle_{\scriptscriptstyle 10}$ just reproduces the global precession rate as it does not depend explicitly 
on $\tau_{10}$.

Then, the evolution equations of  modified weak friction end up being remarkably simple:
\begin{equation}\label{eq:torque_enhanced}
\begin{split}
\dot{\bf L}_{\rm p}\Big|_{\star}=& \langle{T_u}\rangle_{\!\rm \scriptscriptstyle eq,\star}\hat{\bf u}
 +\langle{T_v}\rangle_{\!\rm \scriptscriptstyle eq,\star}\hat{\bf v}
  +\langle{T_n}\rangle_{\!\rm \scriptscriptstyle eq,\star}\hat{\bf n}
\\
&
+
\frac{T_0}
{(1-e_{\rm p}^2)^{3}}
(\hat{\bf s}\cdot\hat{\bf u})
(\hat{\bf s}\cdot\hat{\bf n})^2
\Omega_\star (\tau_{1 0}-\tau_{{\rm eq},\star})
\hat{\bf u}
\\
&
+\frac{T_0}
{(1-e_{\rm p}^2)^{3}}
(\hat{\bf s}\cdot\hat{\bf v})
(\hat{\bf s}\cdot\hat{\bf n})^2
\Omega_\star (\tau_{1 0}-\tau_{{\rm eq},\star})
\hat{\bf v}
\end{split}
\end{equation}
\begin{equation}\label{eq:precession_enhanced}
\begin{split}
\frac{\dot{\bf e}_{\rm p}}{e_{\rm p}}\bigg|_\star=& \frac{\langle{\dot{e}_u}\rangle_{\!\rm \scriptscriptstyle eq,\star}}{e_{\rm p}}\hat{\bf u}
+ \frac{\langle{\dot{e}_v}\rangle_{\!\rm \scriptscriptstyle eq,\star}}{e_{\rm p}}\hat{\bf v}
+ \frac{\langle{\dot{e}_n}\rangle_{\!\rm \scriptscriptstyle eq,\star}}{e_{\rm p}}\hat{\bf n}
\\
&
-\frac{T_0({1-e_{\rm p}^2})^{-\frac{7}{2}}}{\mu_{\rm p}\Omega a_{\rm p}^2}
(\hat{\bf s}\cdot\hat{\bf u})
(\hat{\bf s}\cdot\hat{\bf n})^2
\Omega_\star( \tau_{10}-\tau_{{\rm eq},\star})
\hat{\bf n}~~~~~
\end{split}
\end{equation}
with
\begin{equation}
T_0\equiv \frac{3}{4}\frac{{\cal G}M_{\rm p}^2R_\star^5}{a_{\rm p}^6}
=\frac{3}{4}\frac{\mu_{\rm p}M_{\rm p}}{M_\star} \frac{R_\star^5}{a_{\rm p}^3}\Omega^2 
\end{equation}
and where quantities denoted by $\langle \cdot \rangle_{\!\rm \scriptscriptstyle eq,\star}$ refer to
those in Equations~(\ref{eq:torque_u_equi}-\ref{eq:precession_n_equi}), now explicitly labeling them as
tidal dissipation in the star.

\subsection{Realignment and Circularization Timescales}\label{sec:timescales}

If the weak friction model and the (1,0) enhancement coexist,
then $\langle \dot{a}_{\rm p}\rangle \neq 0$
and $\langle \dot{e}_{\rm p}\rangle \neq 0$. Efficient realignment thus requires that
the obliquity damping time-scale be shorter than the circularization timescale. We rewrite
Equation~(\ref{eq:spin_damping_10}) introducing a damping time-scale $t_{\beta}^{({\rm 10,\star})}$
\begin{equation}
\begin{split}
\frac{\langle\dot{\Omega}_\star\rangle_{\scriptscriptstyle 10}}{\Omega_\star}
=&
-\frac{1}{t_{\beta}^{({\rm 10,\star})}}
\sin^2\beta\cos^2\beta
\end{split}
\end{equation}
where
\begin{equation}\label{eq:realignment_timescale}
\begin{split}
t_{\beta}^{({\rm 10,\star})}
=&
\frac{4I_\star a_{\rm p}^6(1-e_{\rm p}^2)^{3}}{3{\cal G}M_{\rm p}^2R_\star^5 \tau_{1 0}}
\end{split}
\end{equation}
Analogously, the circularization time-scale due to weak friction tides in the star(cf. Equation~\ref{eq:orb_eccentricity_equi}) is
\begin{equation}\label{eq:circularization_timescale_star}
t_{e}^{({\rm eq,\star})}=
\frac{M_\star a_{\rm p}^5(1-e_{\rm p}^2)^{13/2}}{27M_{\rm p}R_{\star }^5\Omega^2\tau_{\rm eq,\star}}
\end{equation}
while the circularization timescale due to weak friction in the planet is 
\begin{equation}\label{eq:circularization_timescale_planet}
t_{e}^{({\rm eq,p})}=
\frac{M_{\rm p} a_{\rm p}^5(1-e_{\rm p}^2)^{13/2}}{27M_\star R_{\rm p}^5\Omega^2\tau_{\rm eq,p}}
\end{equation}
which is generally $\ll t_{e}^{({\rm eq,\star})}$ as we typically have that
$\tau_{\rm eq,p}\gg \tau_{\rm eq,\star}$. Since
the main driver of circularization is oftentimes tidal dissipation  within the planet, 
the condition for efficient realignment is given
 by ${t_{\beta}^{({\rm 10,\star})}}/{t_{e}^{({\rm eq,p})}}\ll1$. If we write
\begin{equation}\label{eq:damping_timescale_ratio}
\begin{split}
\frac{t_{\beta}^{(10,\star)}}{t_{e}^{({\rm eq,p})}}
=&
\frac{36  }{(1-e_{\rm p}^2)^{7/2} }
\frac{I_\star }
{\mu_{\rm p} a_{\rm p}^2}
\left(\frac{M_\star}{M_{\rm p}}\right)^2
\left(\frac{R_{\rm p}}{R_\star}\right)^5
\frac{\tau_{\rm eq,p}}{\tau_{1 0}}
\\
 \approx &
\frac{40}{(1-e_{\rm p}^2)^{7/2}}
 \frac{\tau_{\rm eq, p}}{\tau_{1 0}}
\left( \frac{k_\star}{0.1}\right)
\!\left( \frac{M_\star}{1{\rm M}_\odot}\right)^{3}
\!\left( \frac{M_{\rm p}}{1{\rm M}_{\rm J}}\right)^{-3}
\\
&~~~~
\times\left( \frac{R_\star}{1{\rm R}_\odot}\right)^{-3}
\left( \frac{R_{\rm p}}{1{\rm R}_{\rm J}}\right)^5
\!\left( \frac{a_{\rm p}}{0.15\,{\rm au}}\right)^{\!-2}~,
\end{split}
\end{equation}
then efficient realignment of eccentric orbits requires that
\begin{equation}
 \tau_{10}\gtrsim (1-e_{\rm p}^2)^{-7/2}\tau_{10,{\rm min}} 
\end{equation}
where we have defined
\begin{equation}\label{eq:minimum_damping_timescale}
\begin{split}
 \tau_{10,{\rm min}} \equiv  &
 20~{\rm s}
\left(\frac{ \tau_{\rm eq, p}}{0.5~{\rm s}}\right)
\left( \frac{k_\star}{0.1}\right)
\!\left( \frac{M_\star}{1{\rm M}_\odot}\right)^{3}
\!\left( \frac{M_{\rm p}}{1{\rm M}_{\rm J}}\right)^{-3}
\\
&~~~~
\times\left( \frac{R_\star}{1{\rm R}_\odot}\right)^{-3}
\left( \frac{R_{\rm p}}{1{\rm R}_{\rm J}}\right)^5
\!\left( \frac{a_{\rm p}}{0.12\,{\rm au}}\right)^{\!-2}
\end{split}
\end{equation}
which is orders of magnitude larger than $\tau_{10}$ in 
Equation~(\ref{eq:10_lagtime_alt}) if one assumes $\eta_{10}$ of order unity.

Thus, dissipation in the planet imposes a significant constraint for this mechanism
to operate effectively.  However, the strong dependence on stellar radius suggests that conditions could be much more
 amiable for realignment with minimal circularization while stars are still contracting. We explore this possibility below.


\section{Spin Realignment During the Pre-Main Sequence}
We now focus on three key properties of PMS stars.
First, during this phase the stellar radius is larger,  thus enhancing the damping rate since $\langle{\dot{\beta}}\rangle_{\scriptscriptstyle 10}\propto R_\star^3$ (Equation~\ref{eq:obliquity_damping_10}). Second, during a substantial portion of this phase, the star is fully convective, thus making it more effective in dissipating IWs \citep{Barker:2022,Wu:2024}.  And third, young stars can be rapidly rotating, thus further enhancing the IW dissipation according to the scaling of Equation~(\ref{eq:10_lagtime_alt}).

But the PMS phase is short-lived, and thus obliquity damping only has a finite amount of time to operate before the mechanism is essentially shut off. In the following sections,
we develop a toy model of PMS evolution.

\subsection{Toy Model for a Contracting Star}\label{sec:toy_model}

\subsubsection{Stellar Structure}
We adopt a simple analytical model of a contracting pre-main sequence star. We assume
that while the star is fully convective, it contracts at constant effective temperature as an $n{=}\tfrac{3}{2}$ polytrope and
that  the luminosity is solely due to gravitational contraction. A straightforward calculation  \citep[e.g.,][]{Hansen:2004,Spalding:2014}
 gives, for an approximate Hayashi track,
\begin{equation}\label{eq:hayashi_track}
R^{{\rm Hay}}_\star(t)
=R_\odot
\left[\frac{t-t_0}{\tau_{\rm Hay}}
+\left(\frac{R_{\star,{\rm init}}}{R_\odot}\right)^{-3}
\right]^{-1/3}~,
\end{equation}
where
\begin{equation}\label{eq:hayashi_timescale}
\tau_{\rm Hay}
\equiv 0.423
\tau_{\rm KH}
\left(\frac{M_\star}{M_\odot}\right)^{74/51}
\end{equation}
is the characteristic Hayashi time-scale, and where
$\tau_{\rm KH}\equiv {\cal G}M_\odot^2/(R_\odot L_\odot)\approx3.142\times10^7$~yr is 
the Sun's Kelvin-Helmholtz time. In Equation~(\ref{eq:hayashi_track}), we choose
$R_{\star,{\rm init}}=10 R_\odot (M/M_\odot)$ and 
$t_0=5\times10^4$~yr in order to match the PMS models of \citet{Baraffe:2015}.

Similarly, once the star has developed a substantial radiative core, 
we assume it contracts as a polytrope of effective index $n_{\rm eff}\lesssim3$, following the relation
$ L_\star \sim T_{\rm eff}^m$ with $m\approx2$. This approximate Henyey track
results in the contraction law
\begin{equation}
R^{\rm Hen}_\star(t)
=
R_\odot
\left[
-\frac{t-t_{\rm min}}{\tau_{\rm Hen}}
+\left(\frac{R_{\star,{\rm min}}}{R_\odot}\right)
\right] 
\end{equation}
where $R_{\rm min}$ and $t_{\rm min}$  are the radius and time, respectively,
 at which the Hayashi luminosity reaches the PMS track turnover point
 $L_{\rm min}/L_\odot \approx 0.52 (M_\star/M_\odot)^3$, and where
\begin{equation}
\tau_{\rm Hen}\equiv
1.2348(5-n_{\rm eff})
\tau_{\rm KH}
\left(\frac{M_\star}{M_\odot}\right)^{-3.5172}
\end{equation}
 is the characteristic Henyey time-scale. We set $n_{\rm eff}=2.6$,  
for which $\tau_{\rm Hen}\approx10^8$~yr for a $1M_\odot$ star.

The Hayashi and Henyey tracks must intersect at $L_\star{=}L_{\rm min}$ on the HR diagram. From this requirement, 
the effective temperatures  for these two phases are, respectively
$T_{\rm eff}^{\rm Hay}=T_{\rm eff,0}\approx 4400~{\rm K} \left({M_\star}/{M_\odot}\right)^{7/51}$ and $T_{\rm eff}^{\rm Hen}=0.946\,T_{\rm eff,\odot}\,(R_\star^{\rm Hen}/R_\odot)^{-1}$.

We construct a single PMS contraction law by simply taking the
weighted average of $R^{\rm Hen}_\star$ and $R^{{\rm Hay}}_\star$:
\begin{equation}\label{eq:premainsequence_contraction}
R_\star(t)
=[1-s(t)]R^{{\rm Hay}}_\star(t)
+s(t)R^{{\rm Hen}}_\star(t)
\end{equation}
where $s(t)$ is a ramp-up function that we choose as
\begin{equation}
s(t)=1-\exp\left[-1.3\frac{t-t_{\rm rad}}{t_{\rm contr}}\right]
\end{equation}
where $t_{\rm rc}{\approx}\tau_{\rm Hay}\left({R_{\star,{\rm rc}}}/{R_\odot}\right)^{-3} $
marks the time at which the radiative core first appears during the Hayashi phase,
with $R_{\star,{\rm rc}}{\approx}2.1R_\odot\, \left({M_\star}/{M_\odot}\right)^{1.38}$
\citep[e.g.,][]{Kippenhahn:2012},
and $t_{\rm contr}\approx \tau_{\rm Hay}\left({R_{\star,{\rm final}}}/{R_\odot}\right)^{-3} $ is the  contraction time,
with $R_{\star,{\rm final}}=R_{\star,{\rm ZAMS}}=0.9(M_\star/M_\odot)^{1.2}$
marking the end of the contraction phase.  
The effective temperature $T_{\rm eff}$ is likewise a weighted composite of $T_{\rm eff}^{\rm Hay}$
and $T_{\rm eff}^{\rm Hen}$, from which we write the PMS luminosity $L_\star/L_\odot=(R_\star/R_\odot)^2 (T_{\rm eff}/T_{\rm eff,\odot})^4$,
which we will use in Equation~(\ref{eq:pms_convective_time}), below
For more details, see Appendix~\ref{app:pms_model}. 

\begin{figure}
\includegraphics[width=0.47\textwidth]{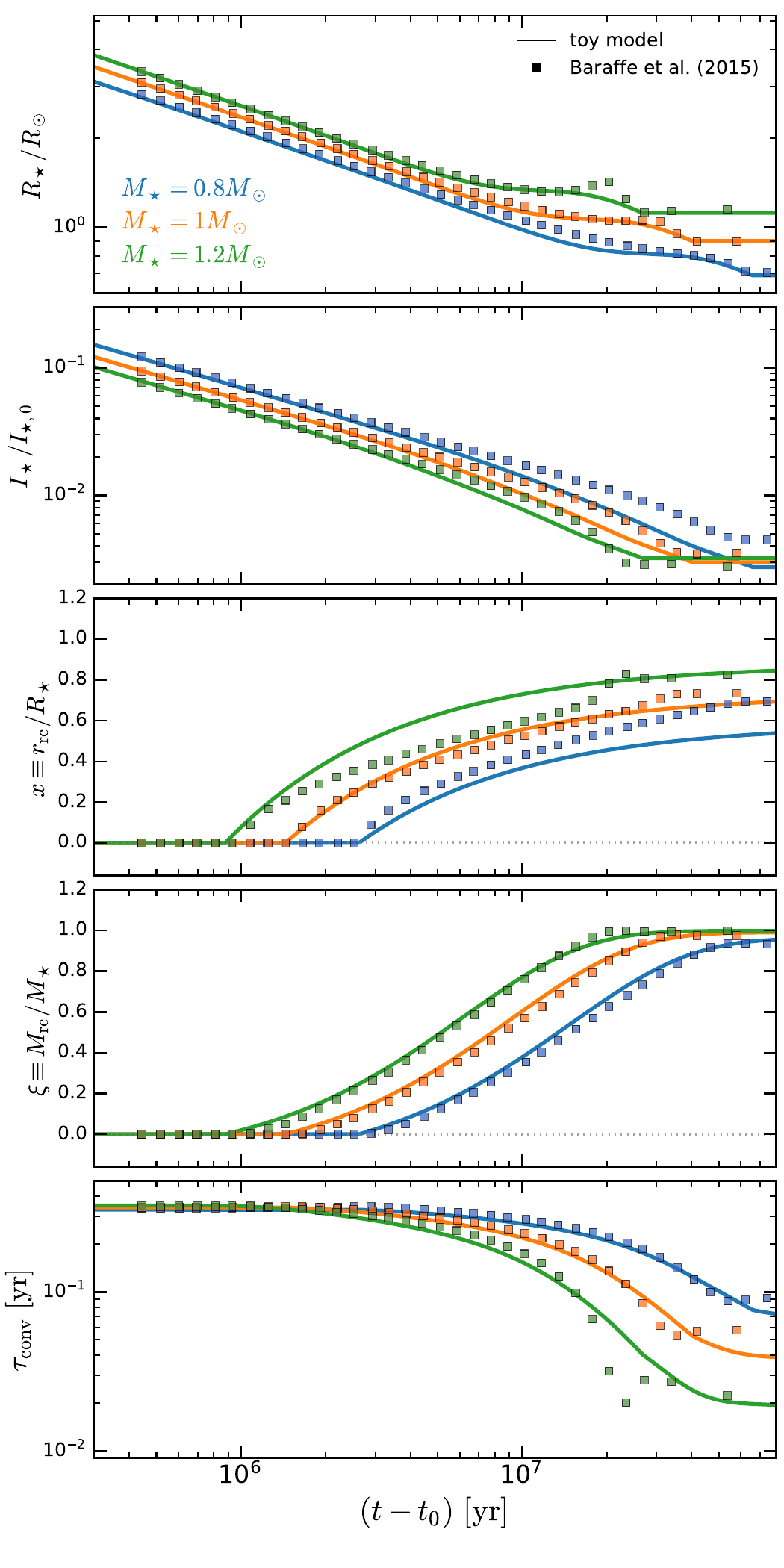}
\caption{Toy model of PMS contraction (solid lines; see Section~\ref{sec:toy_model}) compared to the detailed evolution models of \citet{Baraffe:2015} (square markers).
 {\it Top panel:} Stellar radius $R_\star$ (Equation~\ref{eq:premainsequence_contraction}) as a function
of time for $M_\star=0.8M_\odot$ (blue), $1M_\odot$ (orange) and $1.2M_\odot$ (green).   
{\it Second panel from top:} Stellar moment of inertia $I_\star=k_\star M_\star R_\star^2$.
 {\it Third panel}: relative extent  of the radiative-convective boundary $x$.
{\it Fourth panel}: mass fraction of the radiative core $\xi$.
{\it Fifth panel}:  eddy turnover time computed from $x$ and $\xi$ from (Equation~\ref{eq:pms_convective_time}).
\label{fig:stellar_contraction}}
\end{figure}

The evolving PMS radius (Equation~\ref{eq:premainsequence_contraction})
is shown in the top panel of Figure~\ref{fig:stellar_contraction} alongside the models by \citet{Baraffe:2015}, showing that we achieve a good
approximation to both the Hayashi and Henyey phases. Including the Henyey phase is important as it delays contraction onto the main sequence
for another $\sim20$~Myr, thus extending the period during which the (1,0) dissipation can be effective due to enlarged stellar radii.

To further model the structure constants of this contracting model, we assume that these
are given by
$k_\star=\frac{2}{5}\kappa_n~$ (radius of gyration)
$k_{q,\star}\simeq \frac{\kappa_n^2}{2}\left(1-\frac{n}{5}\right)$ (rotational distortion)
and $k_{2,\star}=3k_{q,\star}$ (tidal Love number)
where $\kappa_n$ is solely determined by the polytropic index $n$ \citep[see][]{Lai:1993,Lai:1994a},
and takes the value of  $0.5115$ for $n=3/2$
and $0.1884$ for $n=3$ \citep[e.g., see Table 1 in][]{Lai:1993}. We interpolate between the structure constants
of $n=3/2$ and $n=3$ functions using the same weighted average as in Equation~(\ref{eq:premainsequence_contraction}).
We compare the evolution of the stellar moment of inertia $I_\star=k_\star M_\star R_\star^2$ according to this toy model
to the values from  \citet{Baraffe:2015} in the second panel of Figure~\ref{fig:stellar_contraction}.

Finally, we adopt phenomenological laws for the radius of the radiative-convective boundary $r_{\rm rc}=x R_\star$
and for the mass of the radiative core $M_{\rm rc}=\xi M_\star$ to match the results of \citet{Baraffe:2015} (third and fourth panels
of Figure~\ref{fig:stellar_contraction}). We use these to
write the eddy turnover time-scale following \citet{Hurley:2002}
\begin{equation}\label{eq:pms_convective_time}
\begin{split}
\tau_{\rm conv}=&
\left[\frac{M_{\rm conv}(R_\star-R_{\rm rc})(R_\star+R_{\rm rc})}{6L_\star}\right]^{1/3}
\\
\approx &
0.237~{\rm yr}\!
\left(\frac{M_\star}{M_\odot}\right)^\frac{1}{3}\!\!
\left(\frac{T_{\rm eff}}{T_{\rm eff,\odot}}\right)^{\!-\frac{4}{3}} \!\!\!
\left[(1-\xi)(1-x^2)\right]^{1/3}
\end{split}
\end{equation}
\citep[see also][]{Rasio:1996a,Belczynski:2008}. We compare it to  $\tau_{\rm conv}$ as derived from
the quantities reported by \citet{Baraffe:2015} in the fifth panel of Figure~\ref{fig:stellar_contraction}.

\subsection{Evolving Tidal Dissipation in a Contracting Star}\label{sec:evolving_dissipation}
During the PMS, $\tau_{\rm eq,\star}$ and $\tau_{10}$ become functions of time. We estimate these as follows:

\paragraph{Modified lag time for weak friction}
As a PMS star contracts and lands on the MS the modified timelag $\tau_{\rm eq,\star}$ transitions from
  $\tau_{\rm eq,\star}^{\rm (PMS)}\sim0.5$~s to  $\tau_{\rm eq,\star}^{\rm (ZAMS)}\sim10^{-4}$~s. 
Following the picture of \citet{Zahn:1977}, 
we write the modified lag time, in Hut's notation,  as $\tau_{\rm eq,\star}=k_{2,\star} [{\omega_{\rm dyn}^2}T]^{-1}$, where
 $\omega_{\rm dyn}=({\cal G}M_\star/R_\star^3)^{1/2}$ is  the dynamical frequency and $T$ is the ``typical''
tidal time-scale. Thus, when $T\approx\tau_{\rm conv}M_{\rm conv}/M_\star$, we write
\begin{equation}\label{eq:modified_lagtime_equi}
\tau_{\rm eq,\star} =k_{2,\star}  [{\omega_{\rm dyn}^2}{\tau_{\rm conv}}]^{-1}\frac{M_{\rm conv}}{M_\star }  {\cal F}
\end{equation}
where the extra factor ${\cal F}$ represents the reduction
in the effectiveness of eddy viscosity when the forcing period is less than the turnover period of the largest eddies \citep[e.g.,][]{Zahn:1977, Goldreich:1977a, Goldreich:1977b, Goodman:1997}. This reduction factor is generally a function of tidal frequency $\tilde{\omega}$, but in this work we will simple assume that ${\cal F}=0.1$, noting that this assumption
overestimates the amount of tidal dissipation.

A fully convective $1M_\odot$ PMS star has $\xi=x=0$ and $\tau_{\rm conv}\approx 0.35$~yr, while at
a  $1M_\odot$  ZAMS star has $\xi\approx0.97$,  $x=0.65$ and $\tau_{\rm conv}\approx 0.05$~yr \citep[e.g.,][]{Moe:2018}.
Replacing these values into
Equation~(\ref{eq:modified_lagtime_equi}) assuming a $k_{2,\star}\approx 0.2747$ for the PMS
and $k_{2,\star}\approx 0.0213$ for the ZAMS, allows us to write
\begin{equation}
\tau_{\rm eq,\star}^{\rm (PMS)}\approx 0.8~{\rm s}
\left(\frac{M_\star}{M_\odot}\right)^{-1}
\left(\frac{R_\star}{5R_\odot}\right)^{3}
\left(\frac{\tau_{\rm conv}}{0.35~{\rm yr}}\right)^{-1}
\end{equation}
and
\begin{equation}
\tau_{\rm eq,\star}^{\rm (ZAMS)}\approx 10^{-4}~{\rm s}
\left(\frac{M_\star}{M_\odot}\right)^{-1}
\left(\frac{R_\star}{R_\odot}\right)^{3}
\left(\frac{\tau_{\rm conv}}{0.05~{\rm yr}}\right)^{-1}
\end{equation}
A  value for the star's modified time lag of 
$\tau_{{\rm eq},\star}\approx 10^{-4}$~s is $\sim5000$ times smaller 
 than the value usually assumed
for gas giants \citep[e.g.,][]{Anderson:2016}.  In the often-used parametrization of 
\citet{Eggleton:2001}, $\tau_{{\rm eq},\star}$
corresponds to a viscous time of $t_v\approx 2500$~yr.\footnote{
The viscous time $t_v$ is related to the modified lag time via $\tau=3R_\star^2/({\cal G}M_\star)(1+k_2)^2/t_v$. For $t_v\approx2500$~yr 
then $\tau\approx 10^{-4}$s.
} 
In turn, this lag can be related to a modified tidal quality
factor $Q_\star'$ via $\tau_{{\rm eq},\star}=\tfrac{2}{3}[ Q_\star' \tilde{\omega}]^{-1}$ 
for a tidal forcing frequency $\tilde{\omega}$. For the problem at hand,
$\tilde{\omega}=\Omega_\star|1-\frac{\Omega}{\Omega_\star}|$ and the forcing period is $\sim1\textup{--10}$~days, 
and thus $Q_\star'\sim10^8\textup{--}10^9$, in rough agreement with estimates of the 
tidal quality factor from turbulent dissipation in the star’s convective envelope \citep{Penev:2011}.

\paragraph{Modified lag time for the (1,0) component}
Writing Equation~(\ref{eq:linear_response_bg_coeff}) in terms of $\xi$ and $x$ and replacing into
Equation~(\ref{eq:10_lagtime_alt}), we have
\begin{equation}
\begin{split}
\tau_{10}=&
3~{\rm m}\,\,
\left(\frac{\eta_{10}}{10^6}\right)
\left(1 - \xi\right)^{2/3} \left(1-x\right)^{4/3} 
\\
&\times
\left(\frac{ L_\star}{L_\odot}\right)^{1/3}
\left(\frac{P_\star}{{\rm d}}\right)^{\!\!-1}\!\!\!
 \left(\frac{R_\star}{R_\odot}\right)^{3}\!
 \left(\frac{M_\star}{M_\odot}\right)^{\!\!-1}
\end{split}
\end{equation}
which is the time-evolving expression for $\tau_{10}$ that we adopt from here onward in this work.

\subsection{Short-Term Evolution}\label{sec:short_term}

We now proceed to numerically integrate the secular 2-body system governed by
\begin{align}\label{eq:2body_system_L}
\dot{\bf L}_{\rm p}=&
 \dot{\bf L}_{\rm p}\Big|_{\star}
 +\dot{\bf L}_{\rm p}\Big|_{\rm p}
+ \dot{\bf L}_{\rm p}\Big|_{\rm CSRF}
  \\
  \label{eq:2body_system_e}
  \frac{\dot{\bf e}_{\rm p}}{e_{\rm p}}=&
   \dot{\bf e}_{\rm p}\Big|_{\star}
 +\dot{\bf e}_{\rm p}\Big|_{\rm p}
+ \dot{\bf e}_{\rm p}\Big|_{\rm CSRF}
\end{align}
where $ \dot{\bf L}_{\rm p}\Big|_{\star}$ and $\dot{\bf e}_{\rm p}\Big|_{\star}$
denote the equations of motion due to the combined effects of  viscous damping of the star's equilibrium tide
and dissipation due to its $(1,0)$ mode, and where we also include dissipation within the planet  due to weak friction only, i.e.,
\begin{align}\label{eq:}
\dot{\bf L}_{\rm p}\Big|_{\rm p}=& \langle{T_u}\rangle_{\!\rm \scriptscriptstyle eq,p}\hat{\bf u}
 +\langle{T_v}\rangle_{\!\rm \scriptscriptstyle eq,p}\hat{\bf v}
  +\langle{T_n}\rangle_{\!\rm \scriptscriptstyle eq,p}\hat{\bf n}
  \\
  \frac{\dot{\bf e}_{\rm p}}{e_{\rm p}}\bigg|_{\rm p}=& \frac{\langle{\dot{e}_u}\rangle_{\!\rm \scriptscriptstyle eq,p}}{e_{\rm p}}\hat{\bf u}
+ \frac{\langle{\dot{e}_v}\rangle_{\!\rm \scriptscriptstyle eq,p}}{e_{\rm p}}\hat{\bf v}
+ \frac{\langle{\dot{e}_n}\rangle_{\!\rm \scriptscriptstyle eq,p}}{e_{\rm p}}\hat{\bf n}
\end{align}
where now the quantities $\langle \cdot \rangle_{\!\rm \scriptscriptstyle eq,p}$ also refer to
 Equations~(\ref{eq:torque_u_equi}-\ref{eq:precession_n_equi}), but with the stellar
 quantities replaced by the planetary ones.

\begin{figure*}[ht]
\begin{center}
\includegraphics[width=0.3\textwidth]{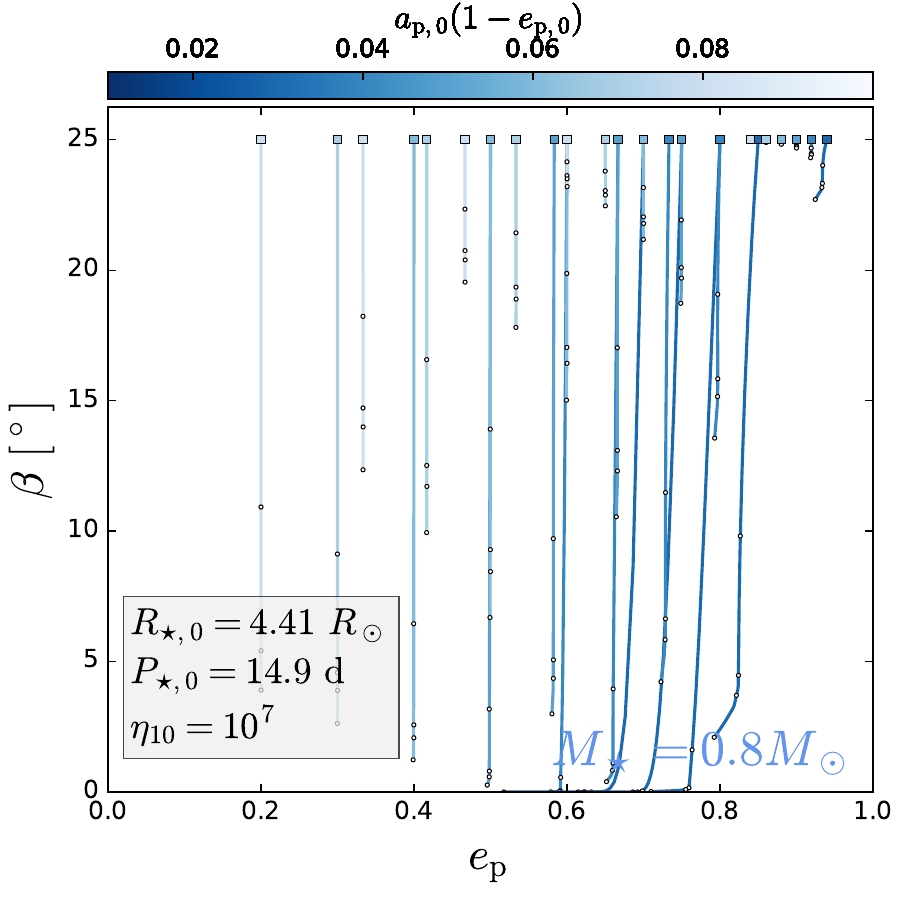}
\includegraphics[width=0.3\textwidth]{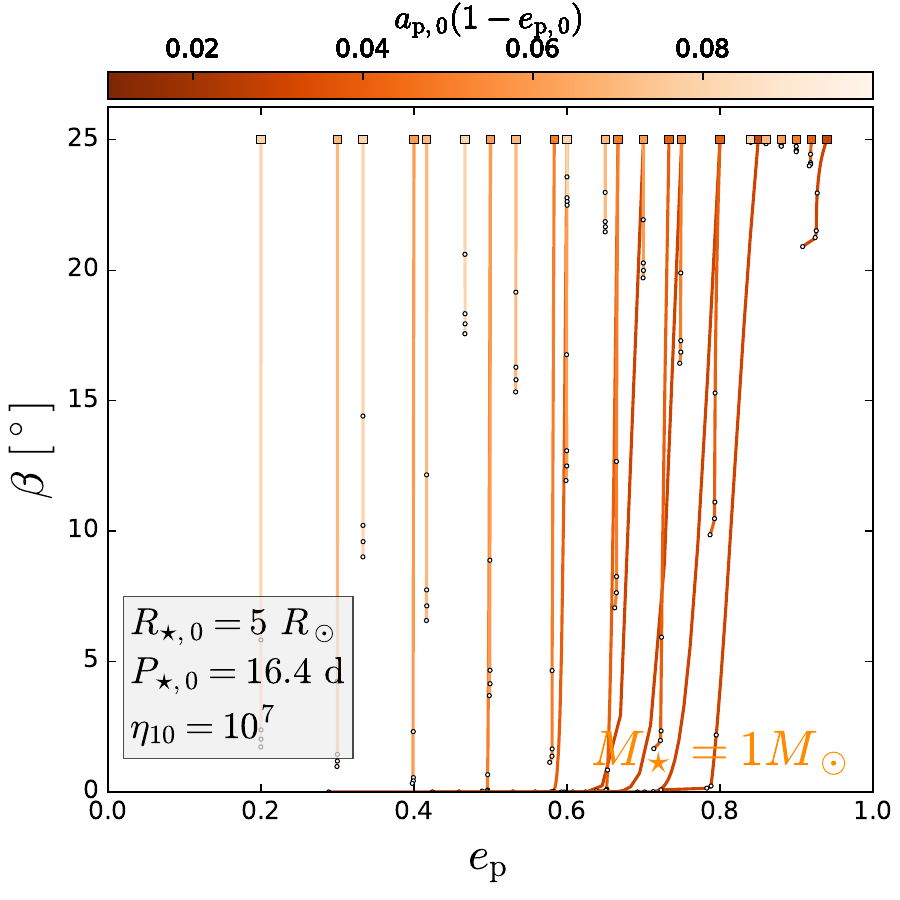}
\includegraphics[width=0.3\textwidth]{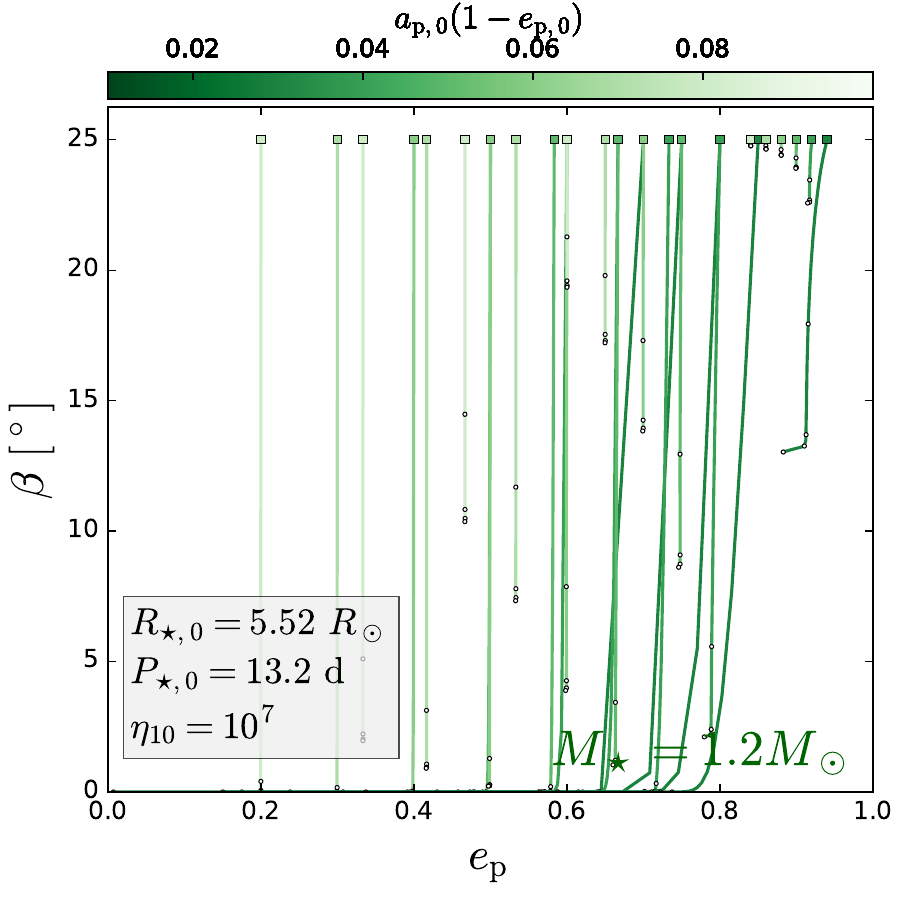}
\includegraphics[width=0.3\textwidth]{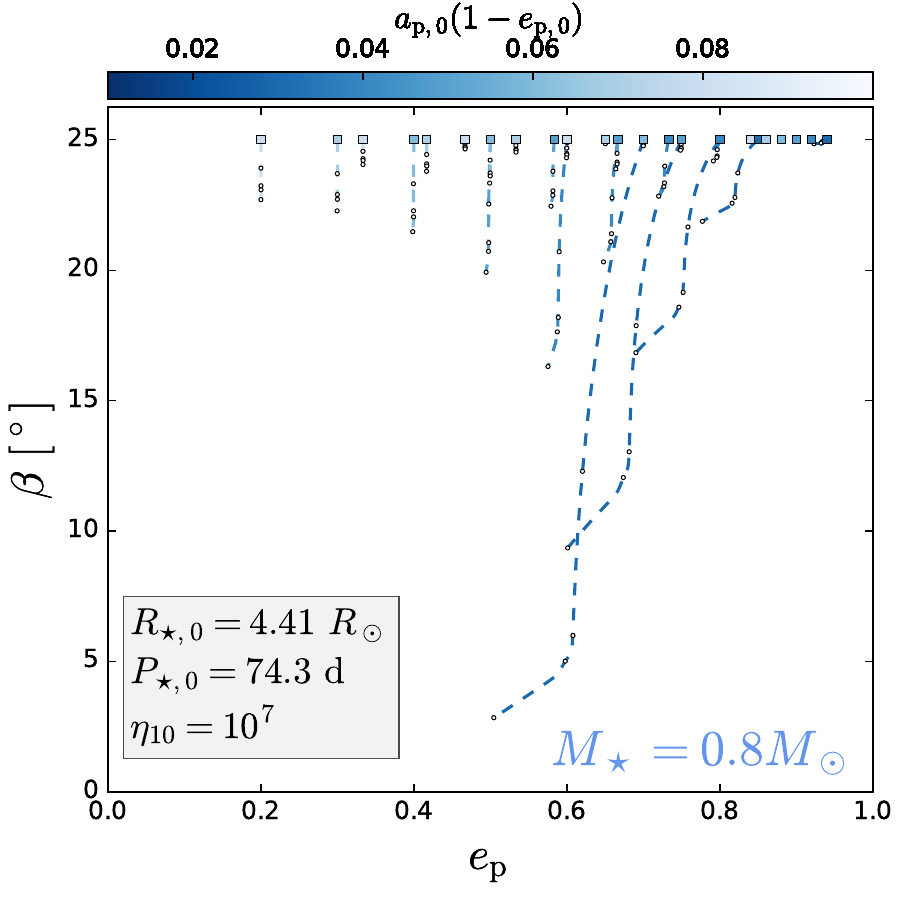}
\includegraphics[width=0.3\textwidth]{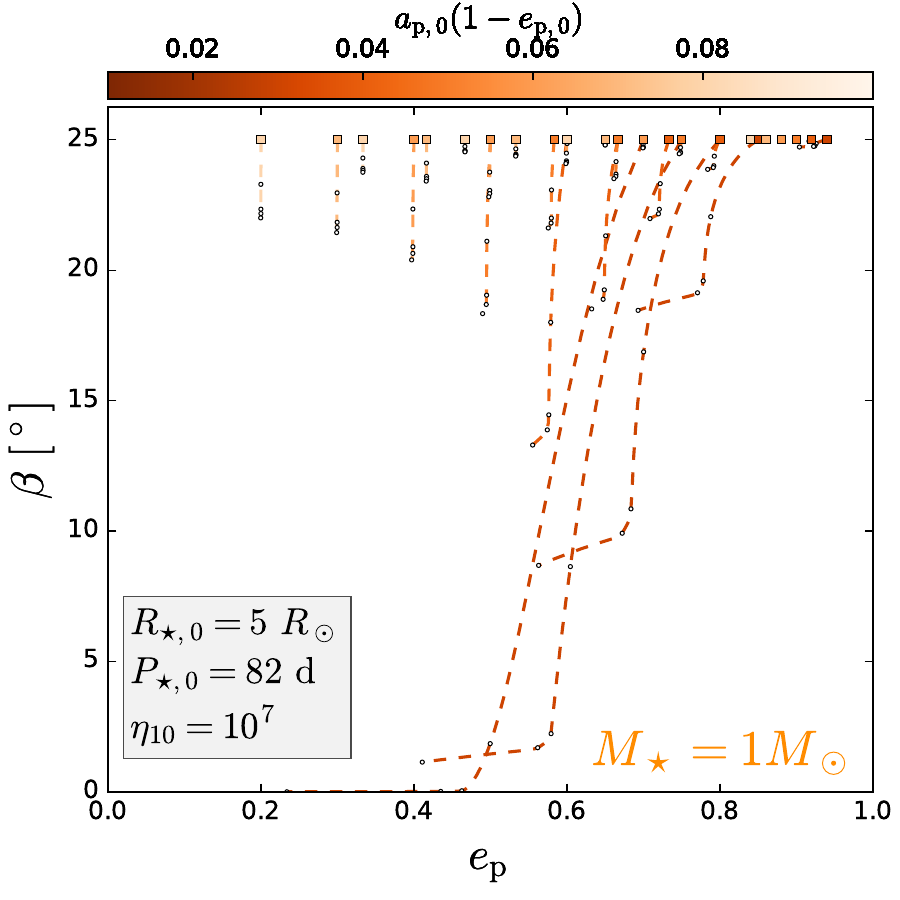}
\includegraphics[width=0.3\textwidth]{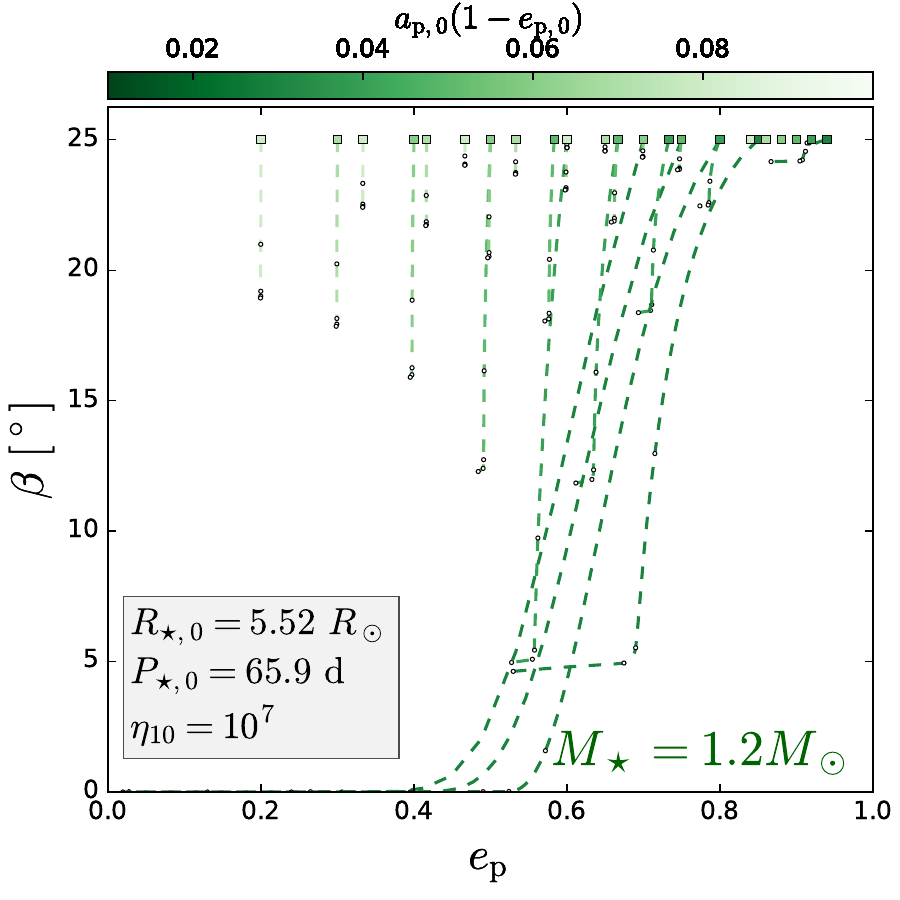}
\end{center}
\caption{Evolutionary tracks in the $\beta$--$e_{\rm p}$ plane for rapid rotators (top row) and intermediate rotators (bottom row). Stellar masses are $0.8\,M_\odot$ (blue), $1.0\,M_\odot$ (orange), and $1.2\,M_\odot$ (green), from left to right. Each panel shows all 30 combinations of $r_{{\rm peri},0}=0.03$, $0.04$, $0.05$, $0.06$, $0.07$, and $0.08$~au with $a_{{\rm p},0}=0.10$, $0.12$, $0.15$, $0.20$, and $0.50$~au. Color scale denotes the initial pericenter distance.
\label{fig:pms_obliquity_tracks}
}
\end{figure*}

For completeness, we also include the effect of conservative short-range forces (CSRF; e.g., \citealp{Liu:2015}),
although these do not directly contribute to changes in obliquity or eccentricity:
\begin{align}
\dot{\bf L}_{\rm p}\Big|_{\rm CSRF}=& 
\dot{\bf L}_{\rm p}\Big|_{\rm Q,\star}
  \\
  \frac{\dot{\bf e}_{\rm p}}{e_{\rm p}}\bigg|_{\rm CSRF}=&
    \frac{\dot{\bf e}_{\rm p}}{e_{\rm p}}\bigg|_{\rm Q,\star}
    +  
  \frac{\dot{\bf e}_{\rm p}}{e_{\rm p}}\bigg|_{\rm Q,p}  
  +
  \frac{\dot{\bf e}_{\rm p}}{e_{\rm p}}\bigg|_{\rm GR}
\end{align}
where we include the precessional effects of the rotationally-induced stellar and planetary quadrupoles, and of 1-PN general relativistic
corrections. These are given by
\begin{align}
\frac{1}{\Lambda_{\rm p}}\dot{\bf L}_{\rm p}\Big|_{\rm Q,\star}=&
-\frac{3k_{q,\star}M_{\rm p}R_\star^5\Omega_\star^2}{2\mu_{\rm p}a_{\rm p}^5\Omega^2}
\frac{}{}\frac{\Omega}{(1-e_{\rm p}^2)^{3/2}}
(\hat{\bf s}\cdot\hat{\bf n})
\hat{\bf s}\times\hat{\bf n} 
\\
  \frac{\dot{\bf e}_{\rm p}}{e_{\rm p}}\bigg|_{\rm Q,\star} =&
 \frac{3}{2}k_{q,\star}\frac{M_{\rm p}}{\mu_{\rm p}}\frac{R_\star^5}{a_{\rm p}^5} 
 \frac{\Omega_\star^2 }{\Omega^2}
  \frac{\Omega}{(1-e_{\rm p}^2)^2}
  \label{eq:stellar_quadrupole_precession}
  \\
  &~~~~~~~
  \times
  \Big\{
    \tfrac{1}{2}\left[1-5  (\hat{\bf s}\cdot\hat{\bf n})^2\right]\hat{\bf u}\times\hat{\bf n}
    \\
      &~~~~~~~~~~
  -(\hat{\bf s}\cdot\hat{\bf n})\hat{\bf s}\times\hat{\bf u}
  \Big\}
\\
  \frac{\dot{\bf e}_{\rm p}}{e_{\rm p}}\bigg|_{\rm Q,p} =&
 \frac{3}{2}k_{q,{\rm p}}\frac{M_\star}{\mu_{\rm p}}\frac{R_{\rm p}^5}{a_{\rm p}^5} 
 \frac{\Omega_{\rm p}^2 }{\Omega^2}
  \frac{\Omega}{(1-e_{\rm p}^2)^2}
\hat{\bf n}\times\hat{\bf u}
\\
  \frac{\dot{\bf e}_{\rm p}}{e_{\rm p}}\bigg|_{\rm GR}=&
 - \frac{3{\cal G}(M_\star+M_{\rm p})}{c^2a_{\rm p}(1-e_{\rm p}^2)}\Omega
~\hat{\bf u}\times\hat{\bf n}
\label{eq:gr_1pn_precession}
\end{align}
where $\Lambda_{\rm p}=\mu_{\rm p}\sqrt{{\cal G}(M_\star+ M_p)a_{\rm p}}$

The structural and tidal properties of the star stem from the structure evolution model of Section~\ref{sec:toy_model}.
For the planet, we set the default parameters to be
$M_{\rm p}=1M_{\rm J}$, $R_{\rm p}=1R_{\rm J}$,
$k_{\rm p}=0.25$, $k_{2,{\rm p}} = 0.37$, $k_{q,{\rm p}}=\tfrac{1}{3}k_{2,{\rm p}}=0.123$
and $\tau_{\rm eq,p}=0.1$~s. Unless stated otherwise, these choices will remain our default values for the remainder of this work.

We integrate a grid of planetary orbits with initial semimajor axis $a_{{\rm p},0}$, pericenter distance $r_{{\rm peri},0}=a_{{\rm p},0}(1-e_{{\rm p},0})$, and obliquity $\beta_0=25^\circ$
while also varying the stellar mass ($0.8M_\odot$, $1M_\odot$ and $1.2M_\odot$). For each stellar mass, we consider rapid and intermediate rotators with nominal ZAMS rotation periods of $P_{\star,{\rm ZAMS}}=0.2$~d and $1$~d, respectively. We obtain the initial periods by evolving backward from the ZAMS at fixed stellar spin angular momentum, $S_\star=k_\star(t)M_\star R_\star^2(t)\Omega_\star$, neglecting magnetic braking in this mapping (see Section~\ref{sec:ms_evolution}). 
We adopt initial radii as$R_{\star,0}/R_\odot=5(M_\star/M_\odot)^{0.55}$, in a scenario we call `early start', meaning that the PMS stars are still substantially inflated and fully convective. The initial rotation periods for early start models are thus $P_{\star,0}\simeq13$--$16$~d for rapid rotators and $70$--$80$~d for intermediate rotators.
All integrations are run with  $\eta_{10}=10^7$.

In Figure~\ref{fig:pms_obliquity_tracks}, we show that the tracks for rapid rotators are nearly vertical across all three stellar masses,
 indicating that obliquity damps with little change in eccentricity. Intermediate rotators show a wider range of outcomes. Obliquity damping is weak for orbits with large initial pericenters and low eccentricities. At small initial pericenters, some highly eccentric orbits may partially circularize and migrate inward, outcompeting realignment (see Section~\ref{sec:timescales}).

\subsection{Long-Term Evolution and Population Synthesis}
In order to provide a meaningful comparison to observational trends, 
we need to provide a population-level assessment of obliquity evolution across a varied range of orbital configurations and stellar ages. We thus resort
to Monte Carlo experiments. For this, we must take stellar evolution beyond the PMS phase into account.

\subsubsection{Long Term Evolution During the Main Sequence}\label{sec:ms_evolution}
We minimally expand our toy stellar model by including radius expansion during the MS and spin down 
due to magnetic braking (MB).

\paragraph{Long-term Radius and Spin Evolution of the Star}
Following PMS contraction, Solar type stars slowly expand during the MS. To minimally capture this radius evolution, we 
adopt a simplified version of the phenomenological law found by \citet{Hurley:2002}
\begin{equation}\label{eq:radius_expansion}
R_\star(t)=R_{\star,\mathrm{\scriptscriptstyle PMS}}(t)
\left\{
1+\left[1.6\left(\frac{M_\star}{M_\odot}\right)^{0.3}-1\right]\frac{t}{t_\mathrm{\scriptscriptstyle MS}}
\right\}
\end{equation}
with $t_\mathrm{\scriptscriptstyle MS}=10(M_\star/M_\odot)^{-7/2}{\rm Gyr}$.

Over $\sim$Gyr time scales, we must also include the effects of magnetic braking \citep{Skumanich:1972} in the evolution of the stellar spin,
as the spin frequency $\Omega_\star$ plays a crucial role in our parametrization of IW dissipation.
The evolution equation for the stellar spin is thus modified from $\dot{\bf S}_\star=-\dot{\bf L}_{\rm p}$ to
\begin{equation}
\dot{\bf S}_\star=-\dot{\bf L}_{\rm p} -T_{\rm wind}\hat{\bf s}
\end{equation}
where $T_{\rm wind}$ is rate of angular momentum carried away by a magnetic wind. Using the empirically calibrated
spin down law of \citet{Matt:2015}, we write the wind torque as
\begin{equation}\label{eq:wind_torque}
T_{\rm wind}=S_0\Omega_\star
\left\{
\begin{array}{lrr}
\tau_{\rm conv}^2\Omega_\star^2 & \Omega_\star\leq  29\tau_{\rm conv}^{-1} & \text{(unsat.)} \\
840& \Omega_\star>  29\tau_{\rm conv}^{-1} & \text{(sat.)}
\end{array}
\right.
\end{equation}
which captures both ``saturated'' and ``unsaturated'' regimes. The scaling coefficient
$S_0$ has units of angular momentum and is given by 
\begin{equation}
S_0\equiv 7.2\times10^{-17}M_\odot \sqrt{{\cal G}M_\odot R_\odot }
\left(\frac{R_\star}{R_\odot}\right)^{3.1}
\left(\frac{M_\star}{M_\odot}\right)^{0.5}
\end{equation}
(see equation 8 in \citealp{Matt:2015}). In Equation~(\ref{eq:wind_torque})
$\tau_{\rm conv}$ is also changing during the MS. \citet{Sadeghi:2017} identify the 
empirical relation $\tau_{\rm conv}\propto R_\star^{-1.2}$ during the MS, and therefore, in accordance
to Equation~(\ref{eq:radius_expansion}), we write
\begin{equation}
\tau_{\rm conv}=\tau_{{\rm conv},\mathrm{\scriptscriptstyle PMS}}(t)
\left\{
1+\left[1.6\left(\frac{M_\star}{M_\odot}\right)^{0.3}-1\right]\frac{t}{t_\mathrm{\scriptscriptstyle MS}}
\right\}^{-1.2}
\end{equation}
where $\tau_{{\rm conv},\mathrm{\scriptscriptstyle PMS}}(t)$ is the PMS convective
time of Equation~(\ref{eq:pms_convective_time}).

\begin{figure}
\begin{center}
\includegraphics[width=0.49\textwidth]{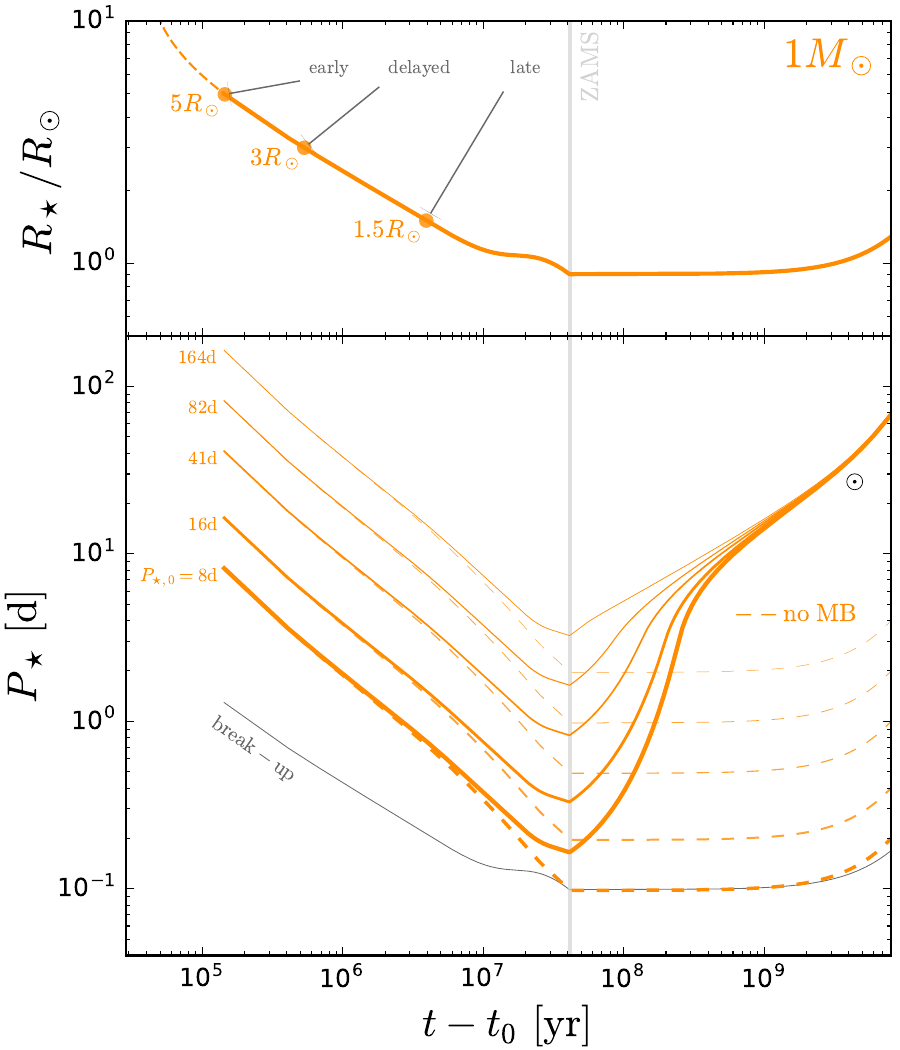}
\end{center}
\caption{Evolution of stellar radius (top) and rotational period (bottom)  for a $1M_\odot$ star as dictated by the toy model of PMS  followed by MS expansion
alongside magnetic braking (MB). Spin evolution is shown for 5 values of $P_{\star,0}$, where the initial time is defined as the moment at which the stellar radius is $5R_\odot$ ('early start'). For reference, the top panel shows the times and radii for `delayed start' ($R_{\star,0}=3R_\odot$) and `late start' ($R_{\star,0}=1.5R_\odot$) scenarios.
For comparison, the bottom panel shows the spin evolution without the effect of MB (dashed lines) as well as the  curve depicting the break-up spin rate
of $P_{\star,{\rm bu}}\equiv 2\pi ({\cal G}M_\star/R_\star^3)^{-1/2}$.
\label{fig:spin_evolution}
}
\end{figure}

In Figure~\ref{fig:spin_evolution}, we show the radius (top panel) and spin period evolution (bottom panel)
over long time scales for a $1M_\odot$ star evolving in isolation for different initial spins periods ($P_{\star,0}=$8--160~d).
The figure also depicts the current Solar spin period of 27~d at an age of 4.5~Gyr,  showing that the spin-down model implemented here generally
produces somewhat slower rotators than the average for G stars.
While this quantitative discrepancy is not troubling given the crudeness of our stellar evolution model, the qualitative trend of Figure~\ref{fig:spin_evolution}, 
is worth pointing out: after $\sim2$Gyr, rapid rotators have caught up to the slow ones due to the exponential spin down during the saturated phase. This means that,
while our proposed mechanism required rapid rotators to operate efficiently, once-rapid rotators are difficult to identify observationally unless they are relatively young.

\subsubsection{Monte Carlo Experiments}\label{sec:monte_carlo}

\paragraph{Initial Conditions}
To further compare to observational trends,
we carry out a numerical exploration of the evolution of a synthetic population of WJs.
We assume that these systems are initially eccentric and inclined relative to the stellar equator.
For concreteness, we adopt Rayleigh distributions of parameters $\sigma_e$ and $\sigma_\beta$ for the initial values of $e_{\rm p}$ and $\beta$,
respectively, motivated by results  from $N$-body experiments of planet-planet scattering \citep[e.g.,][]{Juric:2008,Petrovich:2014}. 
For simplicity, and unless stated otherwise, we will assume
\begin{equation}
\sigma_e=2\sigma_\beta
\end{equation}
\citep[e.g.,][]{Lissauer:1993,Ida:1992}
and that random draws of $e_{\rm p}$ and $\beta$  are not mutually correlated. We truncate the drawn samples at
$e_{\rm p}=0.95$ and $\beta=\pi/2$, i.e., we only consider prograde systems.
Finally, initial semi-major axes are drawn from a log-uniform
distribution between 0.1~au and 0.6~au. 


\paragraph{Fiducial Parameters}
After generating a planet sample of size $N$ by
randomly drawing the orbital elements, $a_{\rm p}$, $e_{\rm p}$ and $\beta$,
we evolve each system for $8$~Gyr
according to the equations of motion of Section~\ref{sec:equations_of_motion}. Unless stated otherwise,
we assume the following set of `fiducial' parameters:
$M_{\rm p}=1M_{\rm J}$, $M_\star=1M_\odot$, $R_{\rm p}=1R_{\rm J}$,
$R_{\star,0}=5R_\odot$, $P_{\star,0}=8.2$~d.  
The  tidal and structural properties of the star are evolved according to the toy model of Section~\ref{sec:toy_model}
plus MS expansion and MB torques, with a default (1,0)  enhancement parameter set to $\eta_{10}=10^7$.
For the planet, we use the same parameters as in Section~\ref{sec:short_term}.


\subsubsection{Results}\label{sec:montecarlo_results}
 
\begin{figure*}[htbp]
\centering
\includegraphics[width=0.88\textwidth]{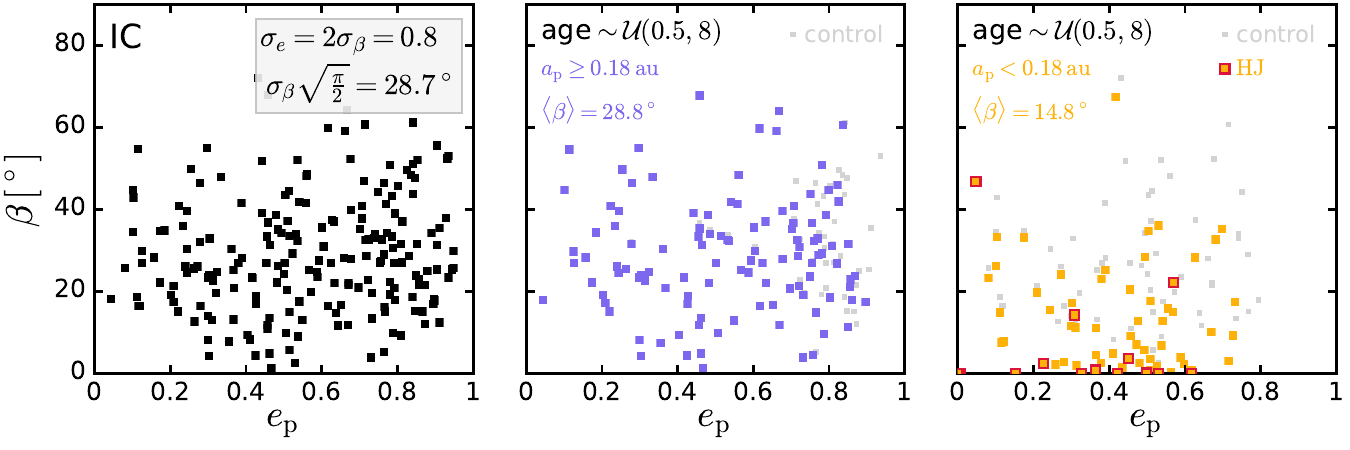} \\[-4pt]
\includegraphics[width=0.88\textwidth]{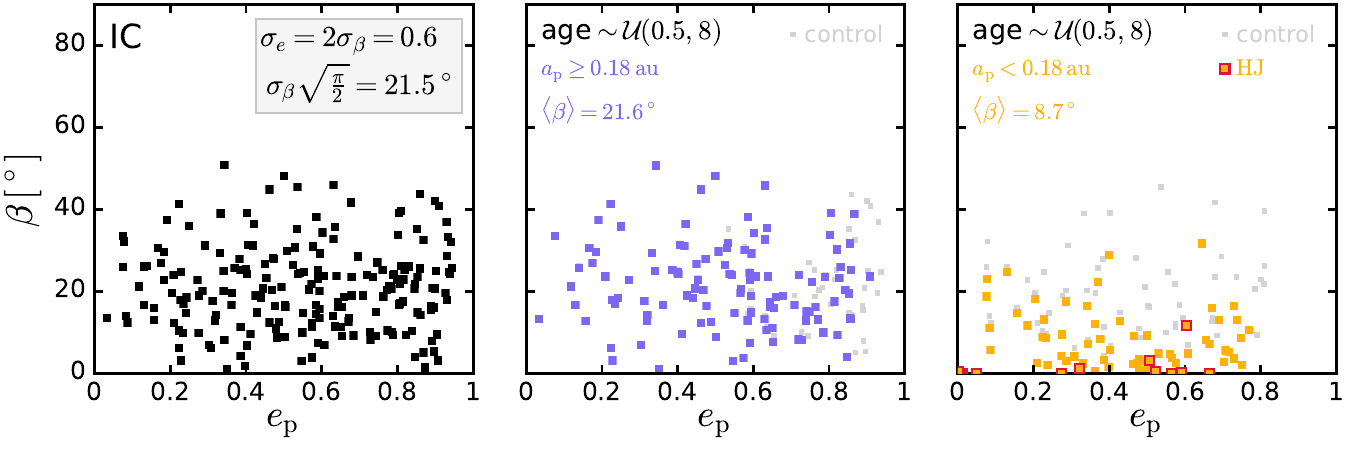} \\[-4pt]
\includegraphics[width=0.88\textwidth]{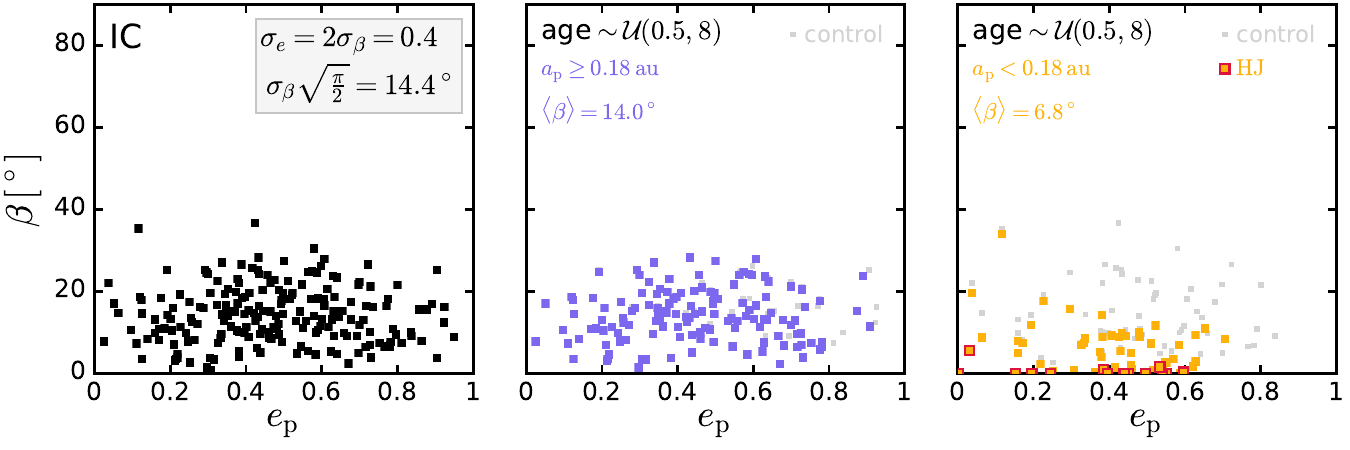} \\[-4pt]
\includegraphics[width=0.88\textwidth]{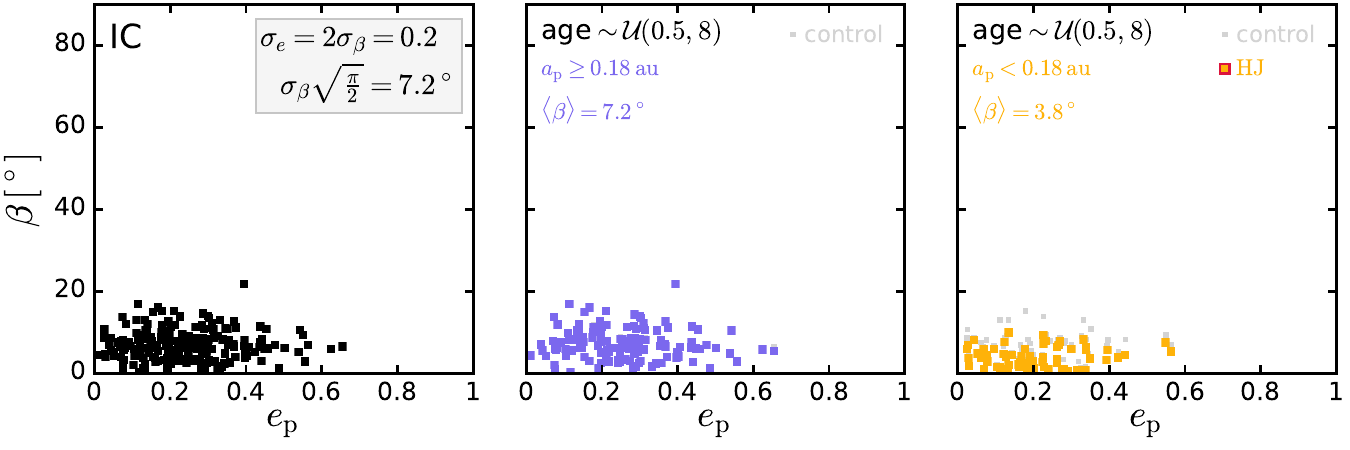} \\[-4pt]
\caption{Obliquities and eccentricities for a tidally evolved synthetic population of $N=200$ WJs initially placed
at $a_{\rm p}\in[0.1,0.6]~{\rm au}$ with $\beta$ and $e_{\rm p}$ drawn from Rayleigh distributions
 with parameters,  from top to bottom,  $\sigma_e=2\sigma_\beta=0.8$,0.6, 0.4 and 0.2  (Section~\ref{sec:monte_carlo}).
\textit{Left column:} initial conditions for which the obliquity averages are $\langle\beta\rangle=\sigma_\beta\sqrt{{\pi}/{2}}$.
\textit{Middle column:} tidally evolved systems with $a_{\rm p}\geq0.18$~au after a time 
$t$ where t is uniformly sampled between 0.5 and 8 Gyr.  
\textit{Right column:} tidally evolved systems with $a_{\rm p}<0.18$~au, where the
points highlighted in red denote nominal HJs ($a_{\rm p}<0.1$~au) that result from the orbital decay of the most eccentric WJs.
On the middle and right panels `control' denotes the sample of systems evolved assuming weak friction
tides only for the same set of initial conditions.
\label{fig:montecarlo_dispersion}
}
\end{figure*}

In Figure~\ref{fig:montecarlo_dispersion} (left panels, black squares), we show 4 different samples of size $N=200$
with $e_{\rm p}$  and $\beta$ drawn from uncorrelated  Rayleigh distributions with $\sigma_e=2\sigma_\beta$ for
$\sigma_e=0.8,~0.6,~0.4$ and $0.2$ (from top to bottom). 

We create an `observed' sample of WJs/HJs by uniformly
sampling a stellar age between 0.5~Gyr and 8~Gyr and extracting the orbital elements at that specific time. These results
are shown in the middle and right panels of Figure~\ref{fig:montecarlo_dispersion}, which separate a `cooler' population
of WJs (with $a_{\rm p}\geq0.18$~au in blue square markers) from a `hotter' one (with $a_{\rm p}<0.18$~au in orange square markers). 
In each panel, the gray markers depict a `control' sample evolved with weak friction theory alone (i.e., when $\tau_{10}=\tau_{{\rm eq},\star}$).
The subsample of more distant WJs  (middle panels) consistently show an empirical average obliquity 
$\langle\beta\rangle$ close to the initial mean $\langle\beta\rangle_0=\sigma_\beta\sqrt{\pi/2}$ evidencing little to no tidal processing of the orbital elements.
The closer-in subsample (right panels) show, on the other hand,  that the mean value in $\beta$ has been reduced by 
$50\%$ to $60\%$.

Some of these systems evolve into nominal HJs ($a_{\rm p}<0.1$~au; highlighted in red) owing to tidal dissipation within the planet. The number of these systems is sensitive to the choice of $\sigma_e$ and our truncation of the drawn sample at $e_{\rm p}=0.95$. In most cases, these systems undergo only mild migration down to $a_{\rm p}\lesssim0.1$~au, but some migrate down to $\sim0.05$~au. Note that many of these HJs retain high eccentricities as they become fully spin-orbit aligned.

\begin{figure*}[t]
\centering
\includegraphics[width=0.99\textwidth]{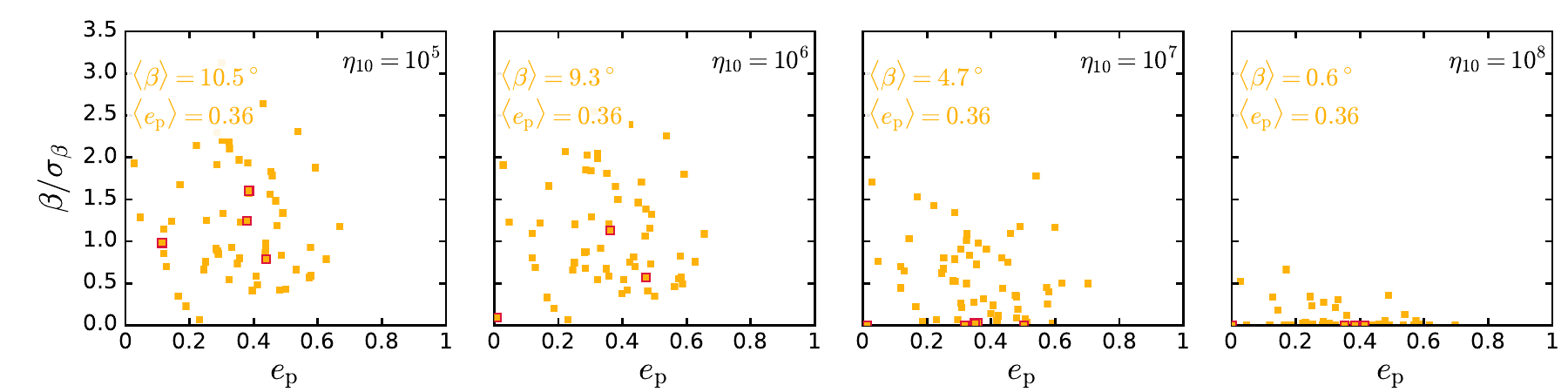}
\caption{Similar to the right panels in Figure~\ref{fig:montecarlo_dispersion}, showing tidally evolved WJs and HJs for
a randomly drawn age, but with $\beta$ normalized by $\sigma_\beta$ and  now for different values of the obliquity damping
strength $\eta_{10}$. From left to right, $\eta_{10}=10^5,10^6, 10^7$ and  $10^8$ (with other parameters fixed to
fiducial values). As in Figure~\ref{fig:dispersion_vs_planetmass}, initial conditions are generated for $\sigma_e=2\sigma_\beta=0.3$ for which the initial
eccentricity and obliquity means
are
$\langle e\rangle_0=\sigma_e\sqrt{\pi/2}=0.38$ and 
$\langle\beta\rangle_0=\sigma_\beta\sqrt{\pi/2}\approx11^\circ$.
\label{fig:dispersion_vs_strength}
}
\end{figure*}

Focusing on a single set of initial conditions with $\sigma_e=0.3$ \citep[e.g.,][]{Juric:2008,Dong:2021}, we repeat the Monte Carlo
experiment varying the (1,0) enhancement coefficient $\eta_{10}$, planet mass $M_{\rm p}$, the initial rotation period
$P_{\star,0}$, the stellar mass $M_\star$ and the initial stellar radius $R_{\star,0}$

As before, we generate a distribution of WJs by randomizing the stellar ages between 0.5~Gyr and 8~Gyr and
focus on systems for which $a_{\rm p}<0.18$~au. In 
Figure~\ref{fig:dispersion_vs_strength}, we show the distribution of obliquities and eccentricities for different values of $\eta_{10}$ (from left to right):
$10^5$, $10^6$,$10^7$ and $10^8$. These results show that, given fiducial parameters, a damping strength
of $10^5$ produces virtually no effect in $\beta$, as the initial distribution has
$\langle\beta\rangle_0=\sigma_\beta\sqrt{\pi/2}\approx11^\circ$. A value of $\eta_{10}=10^8$, on the other hand, produces a reduction of the mean
obliquity by a factor of 50, resulting in multiple WJs and HJs with substantial eccentricities and obliquities that are virtually zero. Whether realistic stellar models 
can yield values of $\eta_{10}$ in this range remains an open question (see Section~\ref{sec:discussion_spin-over}).

\begin{figure*}
\centering
\includegraphics[width=0.85\textwidth]{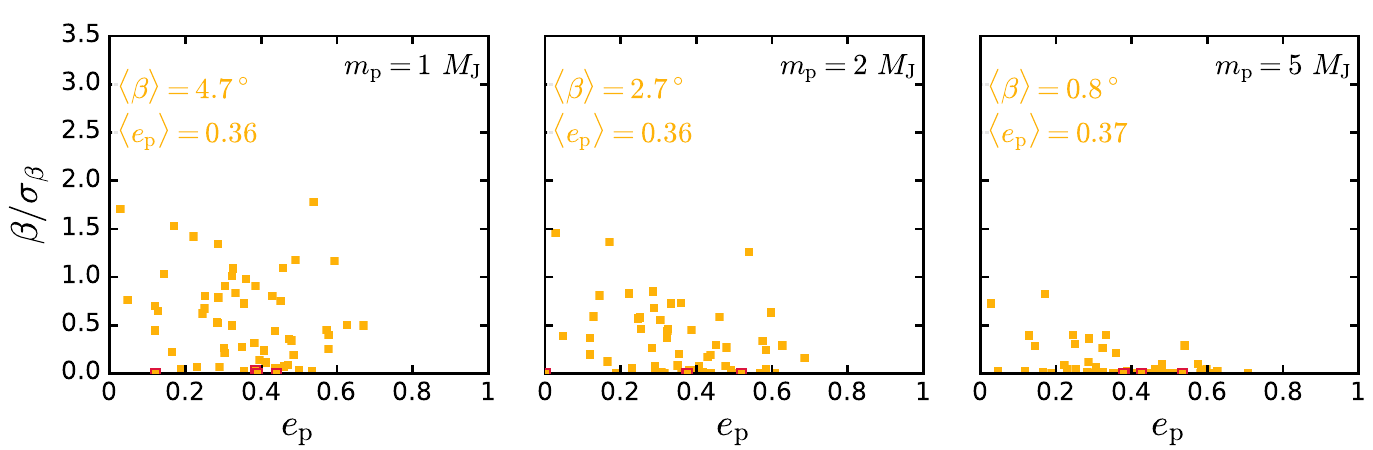}
\caption{
Similar to Figure~\ref{fig:dispersion_vs_strength},  but now exploring different values of the planet mass $M_{\rm p}$ and for the same set of initial conditions
($\langle e\rangle_0=0.38$ and $\langle\beta\rangle_{0}=11^\circ$).
From left to right, $M_{\rm p}=1M_{\rm J}$, $M_{\rm p}=2M_{\rm J}$ and  $M_{\rm p}=5M_{\rm J}$ (with other parameters fixed to
fiducial values). 
\label{fig:dispersion_vs_planetmass}
}
\end{figure*}

The obliquity damping time-scale due to the (10) response 
$t_{\beta}^{({\rm 10,\star})}$ (Equation~\ref{eq:realignment_timescale}), 
suggests that many of the physical parameters we are exploring here can be entirely mutually degenerate. This, however, is not necessarily the case,
as the orbital decay and circularization timescales due to weak friction tides in the planet also depend on quantities such as  $M_{\rm p}$, which add nonlinearity
to the total evolution. To address these partial degeneracies, we explore dependence on $M_{\rm p}$, $P_{\star,0}$, $M_\star$ and $R_{\star,0}$, one parameter at a time.

We show in Figure~\ref{fig:dispersion_vs_planetmass} the dependence of the WJ population's obliquity
on planet mass. Increasing
the planet mass to $2M_{\rm J}$ and $5M_{\rm J}$ reduces  the $\langle \beta\rangle$ by a factor of $\sim2$ and $\sim6$, respectively, relative to the fiducial
case with $M_{\rm p}=1M_{\rm J}$. This trend could help explain why some of the most eccentric WJs tend to be massive planets in the super-Jupiter regime 
(e.g.,\citealp{Gupta:2024}; see discussion in Section~\ref{sec:discussion_warm_jupiters}).

\begin{figure*}
\centering
\includegraphics[width=\textwidth]{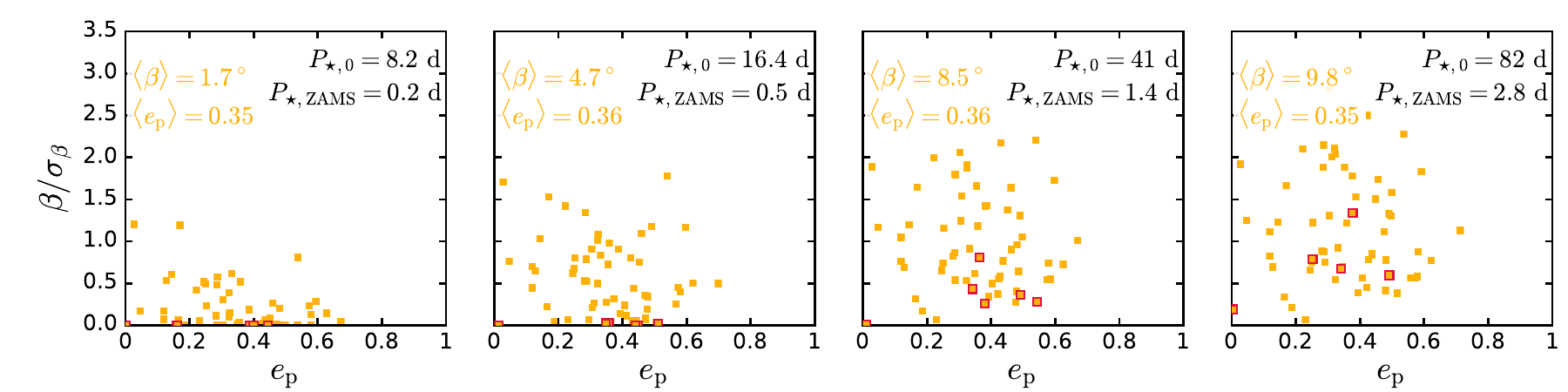}
\caption{Similar to Figure~\ref{fig:dispersion_vs_planetmass}, but now exploring different values of the initial rotation period $P_{\star,0}$ for an `early start'
system ($R_{\star,0}=5R_\odot$) and for the same set of initial conditions
($\langle e\rangle_0=0.38$ and $\langle\beta\rangle_{0}=11^\circ$).
From left to right, $P_{\star,0}=8.2,16.4, 41$ and  $82$ d (with other parameters fixed to
fiducial values). For these four values of $P_{\star,0}$, the rotation periods at the ZAMS are, respectively,
$P_{\star,{\rm ZAMS}}=0.2, 0.5, 1.4$ and  $2.8$ d.
\label{fig:dispersion_vs_rotation}
}
\end{figure*}

In Figure~\ref{fig:dispersion_vs_rotation}, we show how sensitive obliquity damping is to stellar rotation (see Equation~\ref{eq:realignment_timescale}).
As before, the figure depicts the obliquity
and eccentricity sample, this time for different values of the stellar rotation period at time zero (defined, in the fiducial `early start' scenario as the time
at which $R_\star=5R_\odot$). Each of these rotation periods can be mapped onto a period at the ZAMS which results from a combination of contraction-driven spin-up and MB.
As the rotation periods of young stars can be measured (e.g., \citealp{Stassun:1999,Herbst:2002}; see also Section~\ref{sec:discussion_likelihood}),
$P_{\star,{\rm ZAMS}}$ is a useful label to compare to observations. If we define rapid rotators as those with $P_{\star,{\rm ZAMS}}<1$~d, from the figure we
infer that only rapid rotators can efficiently damp obliquities at the population level, while rotators with  $P_{\star,{\rm ZAMS}}>1$~d will only produce a small effect
for the rest of the fiducial parameters kept unchanged.

\begin{figure*}
\centering
\includegraphics[width=0.85\textwidth]{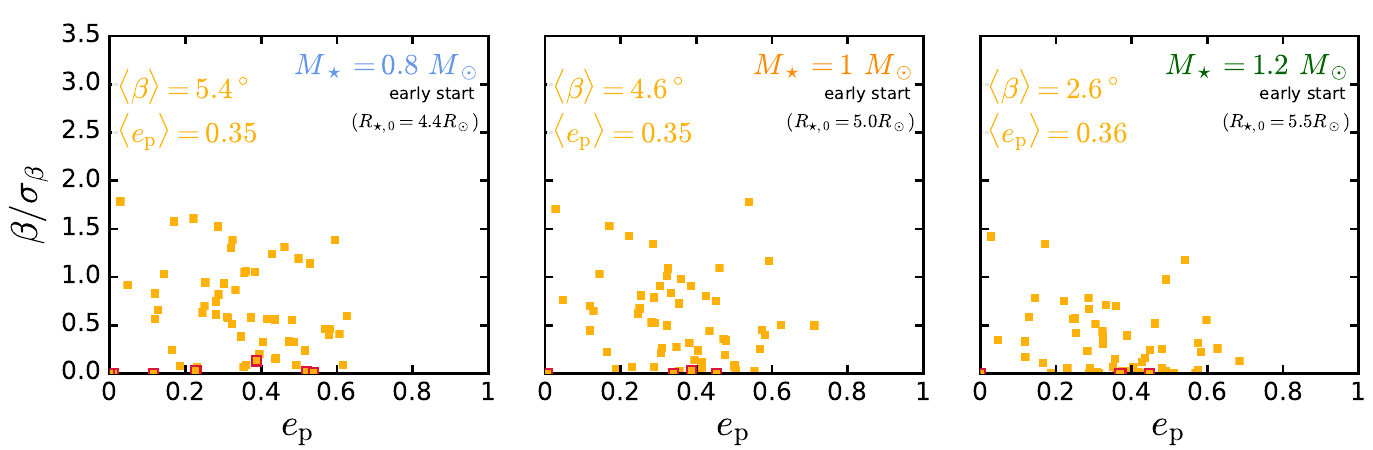}
\includegraphics[width=0.85\textwidth]{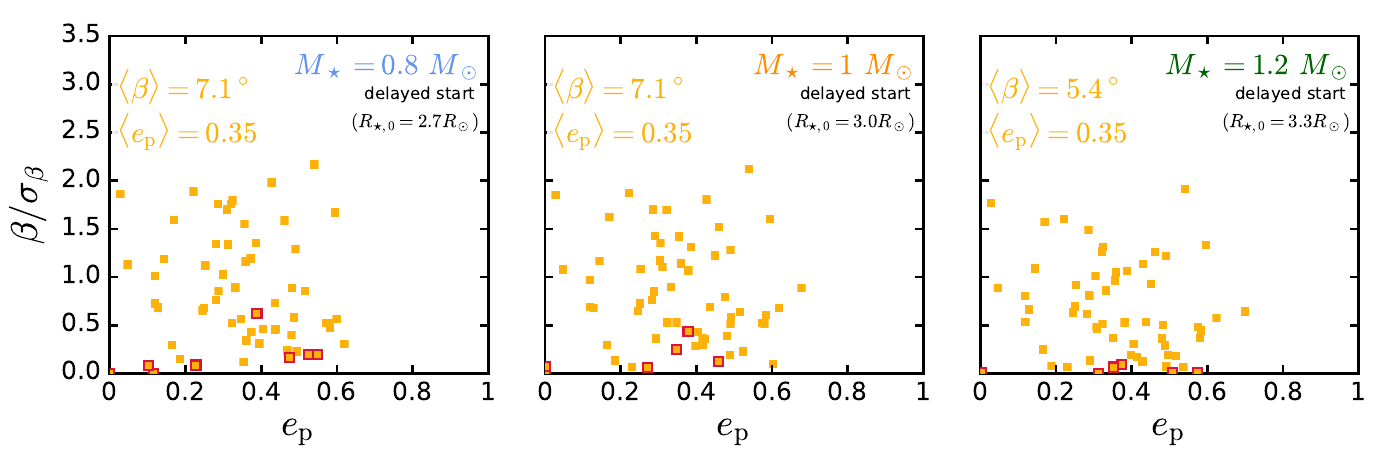}
\includegraphics[width=0.85\textwidth]{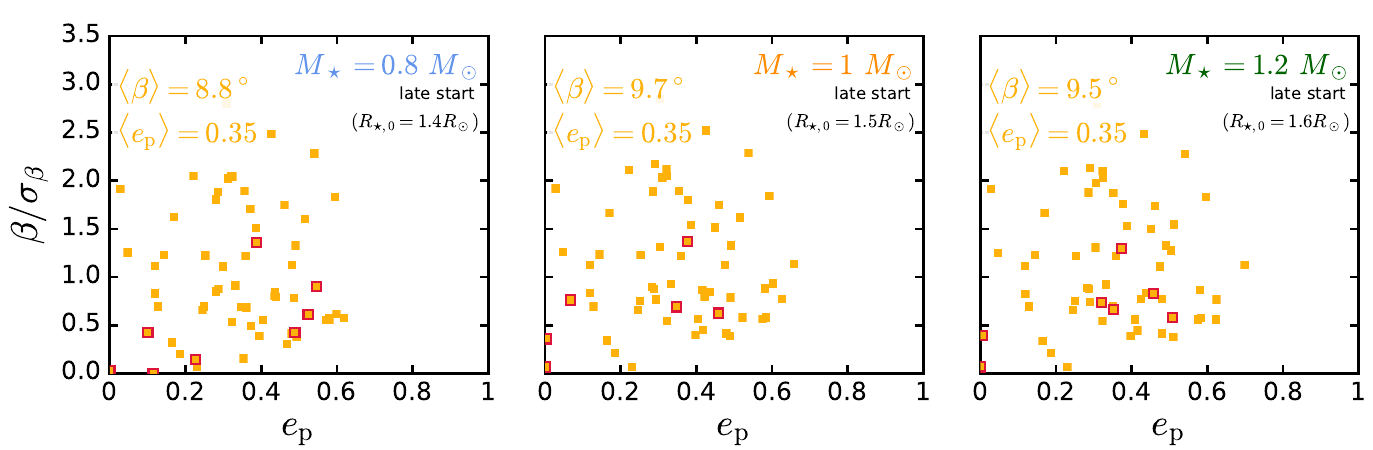}
\caption{
Similar to Figure~\ref{fig:dispersion_vs_planetmass}, but now exploring different values of the stellar mass (columns) and start time (rows).
From left to right, columns depict stellar masses of $0.8M_\odot$, $1M_\odot$ and $1.2M_\odot$ (see Figure~\ref{fig:stellar_contraction}). 
From top to bottom, start times represent `early', `delayed' and `late', which correspond to different initial stellar radii depending on the stellar mass: 
$R_{\star,0}=4.4,~2.7$ and $1.4R_\odot$ for $M_\star=0.8M_\odot$;
$R_{\star,0}=5,~3.0$ and $1.5R_\odot$ for $M_\star=1M_\odot$;
and
$R_{\star,0}=5.5,~3.3$ and $1.6R_\odot$ for $M_\star=1.2M_\odot$.
\label{fig:dispersion_vs_stellarmass}
}
\end{figure*}

Finally, in Figure~\ref{fig:dispersion_vs_stellarmass}, we show how stellar mass $M_\star$ and initial stellar radius
$R_{\star,0}$ (or equivalently, the choice of `start time') affect the measured distribution of obliquities. For a fiducial
`early start' scenario, the increase of mass provides a somewhat surprising finding: a star of $1.2M_\odot$, with a convective envelope
of less than $\sim10\%$ of its total radius at the ZAMS (Figure~\ref{fig:stellar_contraction})  is able to damp obliquities more efficiently than a star of $0.8M_\odot$ for which the convective envelope is at $\sim50\%$ at the ZAMS.  Moreover, a $1.2M_\odot$ star arrives to the ZAMS faster than a low mass star, and has thus less time to effectuate obliquity damping.
These two properties are nonetheless subdominant when compared to
the effect of an increased mass and radius Equations~\ref{eq:realignment_timescale}, which allow the star to cause realignment faster, regardless of the reduced time spent in the PMS.
To test this explanation, we repeat the experiment with a `delayed start' and a `late start', and show the results in the middle and bottom rows of the figure. For the  `delayed start' case,
all systems experience a less effective orbital realignment than in the `early' case, with the more massive star still exhibiting a more aggressive damping effect than their less massive counterparts. For the `late' case, the mean obliquity is almost indistinguishable from that of the initial condition ($\langle\beta\rangle_0=11^\circ$), although in this case the trend appears to reverse, with the lower mass star exhibiting a slightly more efficient damping effect, presumably because of its substantial convective envelope and extended PMS lifetime. These results imply that for the proposed mechanism to operate effectively, planet formation must take place early, including during a period in which the host star, regardless of its mass, is fully convective.  The implications for this scenario are potentially important for the dependence of stellar obliquity as a function of stellar effective temperature, which exhibits a strong correlation for HJ systems \citep{Winn:2010,Albrecht:2012}, but which has been reported to not apply for eccentric HJs \citep{Rice:2022a}, which could form as the low-perihelion tail-end of the WJ population we model in this work (see discussion in Section~\ref{sec:discussion_warm_jupiters}).

\section{Obliquity Damping in Stellar Binaries}\label{sec:stellar_binaries}
Our work thus far has focused on the spin-orbit evolution of isolated planet-star systems, with the inherent assumption
that these planets formed nearly in situ, with minimal to no migration within the protoplanetary disk.
However, an alternative scenario is that these planets do not form close to the star, but instead they form as cold Jupiters that get
relocated inwardly via some version of HEM.  This hypothesis has found support in the finding
of EWJs that lie on the tidal migration track \citep[e.g.,][]{Dong:2014}, defined as the 
curve with $a_{\rm p}(1-e_{\rm p}^2)\approx$constant, implying that over tidal time-scales, WJs with sufficiently high eccentricities
may end up with final semi-major axes $a_{\rm f}=a_{\rm p}(1-e_{\rm p}^2)\lesssim0.1$~au. These so-called `proto hot Jupiters' (PHJs) 
\citep[e.g.,][]{Dawson:2012,Dong:2021,Espinoza:2023,Gupta:2024} may thus represent crucial test-beds for the diverse variants of HEM, such as
Zeipel-Lidov Kozai (ZLK) cycles in stellar binaries \citep{Wu:2003,Fabrycky:2007}.

If EWJs are the result of incomplete or failed HEM, then we need to study how the stellar obliquity evolves when coupling conventional HEM evolution
with our proposed model of enhanced IW dissipation. To fully explore this possibility, one needs a separate Monte Carlo experiment for 3-body dynamics, which
is beyond the scope of this work. However, we still offer some insights and preliminary results below.

We conduct a  3-body experiment that produces HJs via standard ZLK oscillations driven by a distant, stellar-mass binary companion \citep[e.g.][]{Wu:2003,Fabrycky:2007,Naoz:2012,Petrovich:2015a,Anderson:2016,Munoz:2016a}. To this end, 
we extend our equations of motion to account for 3-body effects. In Milankovitch vector form, we now evolve
\begin{equation}
\begin{split}
\dot{\bf L}_{\rm p}=& \dot{\bf L}_{\rm p}\Big|_{\rm 2b}
  +\dot{\bf L}_{\rm p}\Big|_{\rm 3b}
  \\
  \dot{\bf e}_{\rm p}=& 
  \dot{\bf e}_{\rm p}\Big|_{\rm 2b}
  +\dot{\bf e}_{\rm p}\Big|_{\rm 3b}
\end{split}
\end{equation}
where $\dot{\bf L}_{\rm p}\Big|_{\rm 2b}$ and $ \dot{\bf e}_{\rm p}\Big|_{\rm 2b}$ are given by
Equations~(\ref{eq:2body_system_L}) and~(\ref{eq:2body_system_e}), respectively, and where
the three-body coupling is given by
\begin{equation}
\begin{split}
\dot{\bf L}_{\rm p}\Big|_{\rm 3b}=&
-{\bf L}_{\rm p}\times\nabla_{{\bf L}_{\rm p}}{\cal H}_{\rm 3b}
-{\bf e}_{\rm p}\times\nabla_{{\bf e}_{\rm p}}{\cal H}_{\rm 3b}
\\
\dot{\bf e}_{\rm p}\Big|_{\rm 3b}=&
-\Lambda_{\rm p}^{-2}~{\bf L}_{\rm p}\times\nabla_{{\bf e}_{\rm p}}{\cal H}_{\rm 3b}
-{\bf e}_{\rm p}\times\nabla_{{\bf L}_{\rm p}}{\cal H}_{\rm 3b}
\end{split}
\end{equation}
\citep[e.g.,][]{Milankovitch:1939,Dullin:2004,Breiter:2005}, where
${\cal H}_{\rm 3b}$ is the three-body interaction Hamiltonian \citep[e.g.,][]{Harrington:1968,Naoz:2013}, which we truncate
to octupole order  \citep[e.g.,][]{Liu:2015,Petrovich:2015a}. We then set up the initial system by choosing a companion of mass $m_{\rm out}=1M_\odot$, semi-major axis $a_{\rm out}=200$~au, eccentricity $e_{\rm b}=0$, and inclined by $85^{\circ }$, and a planet of mass $m_{\rm p}=5M_{\rm J}$ and initial semi-major axis $a_{\rm p}=1.5$~au. Structural and tidal parameters for the planet are the same as elsewhere in this paper.

We carry out our calculations retaining our toy model of stellar evolution, but for easier comparison to other works, we initialize the system at the ZAMS, i.e., when the host star has reached a radius
of 0.9~$R_\odot$ and has begun to expand slowly. We compare three models: one rapid rotator with $P_{\star,{\rm ZAMS}}=0.35$~d (the second-fastest model shown in Figure~\ref{fig:spin_evolution}), an intermediate rotator with $P_{\star,{\rm ZAMS}}=2$~d, and a slow rotator with $P_{\star,{\rm ZAMS}}=5$~d. For the IW dissipation model,
we retain $\eta_{10}=10^7$ across all numerical experiments. 

\begin{figure*}
\centering
\includegraphics[width=0.99\textwidth]{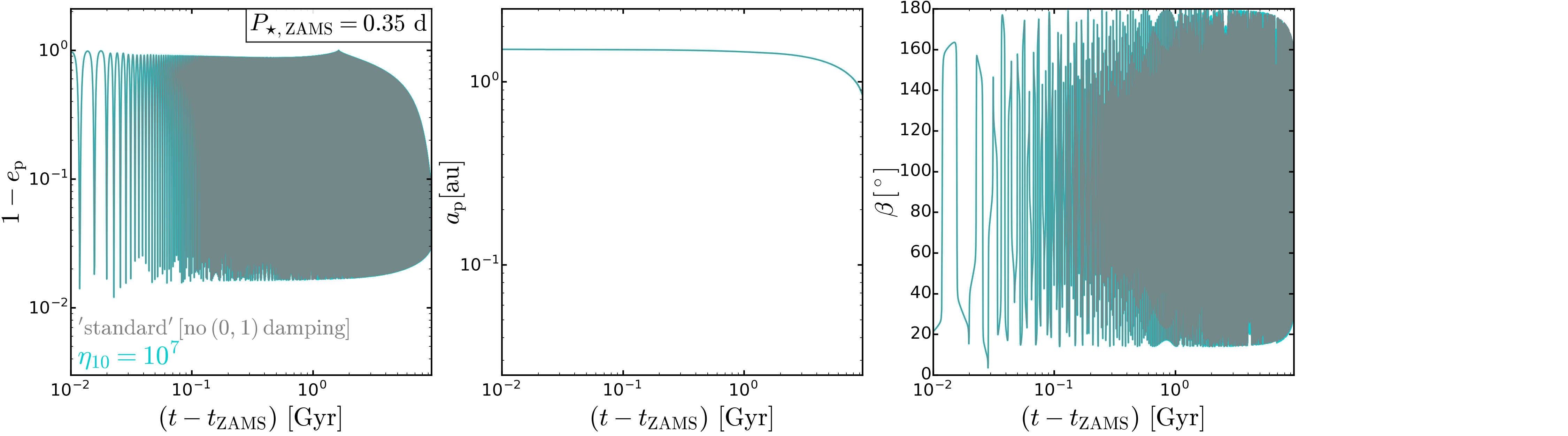}
\includegraphics[width=0.99\textwidth]{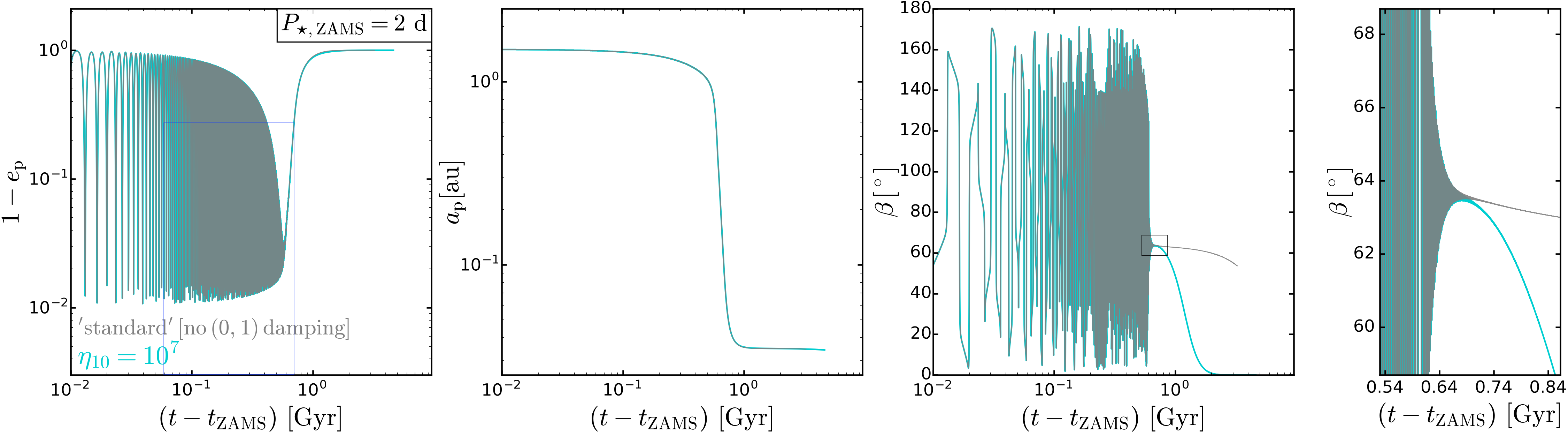}
\includegraphics[width=0.99\textwidth]{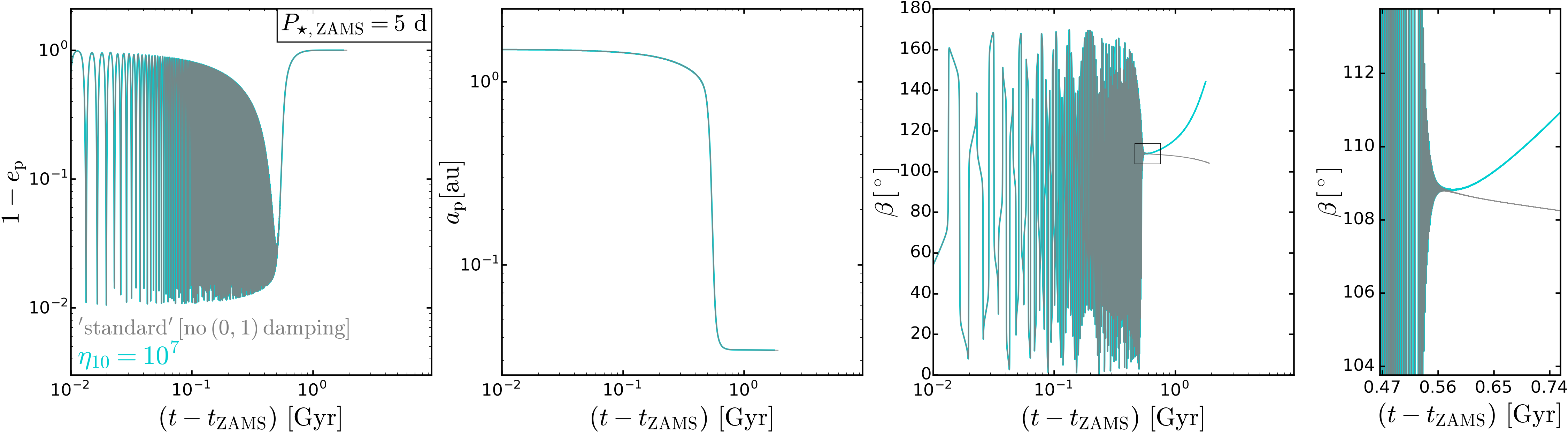}
\caption{Binary-driven high-eccentricity migration with (blue) and without
(gray) enhanced $(1,0)$ tides for initial stellar rotation periods of
$P_{\star,{\rm ZAMS}}=0.35$~d (top row), $2$~d (middle row), and $5$~d
(bottom row). Each row shows the evolution over 10~Gyr of eccentricity
(plotted as $1-e_{\rm p}$), semimajor axis $a_{\rm p}$, and stellar
obliquity $\beta$.
The stellar, planetary, and binary companion parameters are
$M_\star=1\,M_\odot$, $m_{\rm p}=5\,M_{\rm J}$, $a_{\rm p,0}=1.5$~au,
$e_{\rm p,0}=0.01$, $m_{\rm out}=1\,M_\odot$, $a_{\rm out}=200$~au,
$e_{\rm b}=0$, and $I_{\rm mut}=85^\circ$.
In all cases, the remaining orbital angles are initialized to
$\omega_{\rm in,0}=\omega_{\rm out,0}=\Omega_{\rm in,0}=0$ and
$\Omega_{\rm out,0}=180^\circ$, with an initial stellar obliquity of
$\beta_{\rm in,0}=0.5^\circ$.
The planetary structure and tidal parameters are the same as elsewhere
in this paper.
\label{fig:hem_examples}
}
\end{figure*}

We show the result of these numerical integrations in Figure~\ref{fig:hem_examples}, which can be compared to fig. 1 in \citet{Anderson:2016}. From top to bottom, the figure
shows the eccentricity and obliquity evolution in order of increasing $P_{\star,{\rm ZAMS}}$.  The fastest rotator caps eccentricity oscillations at $e_{\rm p}\approx0.98$
due to the star's rotational quadrupole (Equation~\ref{eq:stellar_quadrupole_precession}). This curbing effect  can be quantified in terms of the parameter
 \begin{equation}\label{eq:epsilon_rot}
\varepsilon_{\rm Rot}=
\frac{3}{2}k_{q,\star}\frac{R_\star^5}{a_{\rm p}^5}\frac{M_\star+m_{\rm p}}{m_{\rm out}}\frac{\Omega_\star^2}{{\cal G}M_\star}
a_{\rm out}^3(1-e_{\rm out}^2)^{3/2}
\end{equation}
 \citep{Liu:2015}, which becomes greater than 1 when ZLK oscillations are substantially suppressed in amplitude. As the maximum eccentricity is suppressed,
 more ZLK cycles are needed to extract substantial orbital energy during the high-eccentricity phases \citep[e.g.,][]{Anderson:2016}, which can delay the tidal migration
 time-scale, potentially even past the MS lifetimes of the star. Equation~(\ref{eq:epsilon_rot}) shows that, were we to initialize the system in the PMS, the suppression effect of the stellar quadrupole could become extreme, potentially suppressing the ZLK effect entirely. These inflated stellar radii could, in addition,  cause some unsuppressed
high-eccentricity orbits to plunge directly into the star.

Most importantly, in all three cases, the `standard' and the  `enhanced' models are virtually indistinguishable
in $a_{\rm p}$ and $e_{\rm p}$, with only $\beta$ showing substantial differences after the planet has become a HJ.
For  $P_{\star,{\rm ZAMS}}=2$~d, and 5~d (middle and bottom panels in Figure~\ref{fig:hem_examples}),  the models begin to diverge only after the ZLK cycles have been arrested (see inset panels) and  both $e_{\rm p}$ and $a_{\rm p}$ have decayed substantially.  At this point, the time-scale ratio ${t_{\beta}^{(10,\star)}}/{t_{e}^{({\rm eq,p})}}$
(Equation~\ref{eq:damping_timescale_ratio}) is too large, and orbits circularize and migrate much faster than they can realign. After the planet has fully circularized, we see
differences in $\beta$ over long time-scales. 

These preliminary calculations suggest that enhanced IW dissipation cannot help explain low-obliquity EWJs in the context of HEM, although
it might still play a role in the realignment of circular HJ orbits over long time-scales \citep[e.g., see][for an alternative scenario]{Zanazzi:2024}.
Similarly compelling is the study of the combined effect of enhanced IW dissipation and tidal migration under more general scenarios and variants of HEM.
This is especially timely in light of recent evidence that octupole-level (i.e., ``coplanar")  HEM  \citealp{Lee:2003,Petrovich:2015b}) not only produces systematically lower stellar obliquities than quadrupole-level HEM, but also HJs at larger separations \citep{Bhaskar:2026}.

\section{Discussion}
\subsection{Implications of this Work}

\subsubsection{Obliquity as a Function of Stellar Effective Temperature}
An important consequence of the hypothesis we have put forward is that, by virtue of realignment happening mainly during the PMS, the final obliquity is
not a strong function of the effective temperature the star attains during the MS. A similar claim was already reported by \citet{Rice:2022a}, who
found that the sub-sample of eccentric HJs shows no clear break in $T_{\rm eff}$ as do circular HJs \citep{Winn:2010,Albrecht:2012}.
There might already be observational evidence that this trend extends to EWJs.
Some of the most eccentric WJs with low obliquities orbit stars with $T_{\rm eff}>6000$~K. Some of these include
TOI-677~b \citep[$e_{\rm p}=0.435$, $a_{\rm p}=0.1038\,\mathrm{au}$, $T_{\rm eff}=6295\,\mathrm{K}$][]{Jordan:2020,Sedaghati:2023},
HD~17156~b \citep[$e_{\rm p}=0.6772$, $a_{\rm p}=0.1632\,\mathrm{au}$, $T_{\rm eff}=6046\,\mathrm{K}$][]{Fischer:2007,Narita:2009,Kane:2023},
TOI-4127~b \citep[$e_{\rm p}=0.7484$, $a_{\rm p}=0.303\,\mathrm{au}$, $T_{\rm eff}=6096\,\mathrm{K}$][]{Gupta:2023,Mireles:2025},
TOI-3362~b \citep[$e_{\rm p}=0.720$, $a_{\rm p}=0.155\,\mathrm{au}$, $T_{\rm eff}=6532\,\mathrm{K}$][]{Dong:2021,Espinoza:2023},
and TOI-558~b \citep[$e_{\rm p}=0.31$, $a_{\rm p}=0.128\,\mathrm{au}$, $T_{\rm eff}=6466\,\mathrm{K}$][]{Ikwut-Ukwa:2022,Espinoza:2025}.
While this sample is still small, it could help support the hypothesis that tidal realignment happens during the PMS, when all Solar type stars go through a period in which they are completely convective.

\subsubsection{Implications for Warm Jupiter Formation}\label{sec:discussion_warm_jupiters}
The dynamical scenario that we have favored here relies on a `fast and early' formation of WJs {\it in situ} or with only minimal disk-driven migration. 
In principle, a high enough pebble flux might be able to support this scenario (see Section~\ref{sec:discussion_likelihood}), and consequently imply that
WJs in general, and EWJs in particular, could be compositionally different than close-in gas giants that formed as cold Jupiters that build their envelope farther out before undergoing high-eccentricity migration. If a growing planet undergoes runaway accretion in the inner disk, acquiring most of its envelope inside the water snow line while inward-drifting icy pebbles are still supplying vapor, the resulting atmosphere can have elevated O/H and, if water dominates the volatile enrichment, lower C/O \citep{Schneider:2021}. 

Our preliminary binary-driven HEM calculations did not produce low-obliquity EWJs (Section~\ref{sec:stellar_binaries}). This supports the idea that EWJs could be a distinct population from HJs, even as their appearing to lie on the tidal migration track may suggest otherwise.

Moreover, locally formed WJs could also be more massive on average if their cores reach runaway earlier in the disk’s lifetime than the cores of the cold Jupiters that later become HJs. A high early pebble flux can accelerate core assembly, allowing runaway to begin while a substantial gas reservoir remains. If the disk can sustain gas delivery, the longer accretion window would favor more massive envelopes \citep{Savvidou:2023}. In principle, such favorable conditions could also form multiple critical cores. For a nascent planet pair, if an inner planet forms first, the outer planet still has direct access to incoming pebbles and gas, and if the outer planet forms first, enough gas can still cross its gap to feed runaway accretion onto an already assembled inner core. The formation of more than one WJ under these conditions would favor the
dynamical evolution via planet-planet scattering after disk dispersal, and thus justify our choice of initial orbital elements in equipartition (Section~\ref{sec:monte_carlo}).

Observational evidence already appears to support both that planet-planet scattering is behind the measured eccentricities of WJs \citet{Dong:2026,Morgan:2026} and that EWJs can be very massive. Planets like TOI-2145b \citep{Dong:2025}, TOI-4127 b \citep{Gupta:2023,Mireles:2025} and TOI-3362 b \citep{Espinoza:2023} are super Jupiters  on eccentric and well-aligned orbits located at the transition between WJs and HJs. We consequently posit that EWJs that are the focus of this study might represent the most massive tail end of in situ WJ formation.

\subsubsection{Polar and Anti-aligned Orbits}
Equation~(\ref{eq:obliquity_damping_10}) implies that there exist four fixed points at 
$0$, $\pi/2$, $\beta_*{\equiv}\cos^{-1}\left(-{S_\star}/{L_{\rm p}}\right)$, and $\pi$. 
Linearization around these equilibria indicates that $0$ and $\pi$ are stable points, $\pi/2$ is semi-stable (repelling for $\beta<\pi/2$ and attracting for $\beta>\pi/2$)
and $\beta_*$ is unstable. Thus, according to tidal evolution given by
the (1,0) component alone, systems that begin with retrograde orbits might evolve toward polar orbits ($\beta=\pi/2$)
or anti-aligned orbits ($\beta=\pi$) depending on whether the initial obliquity is smaller or greater than $\beta_*$. Initially prograde orbits, on the other hand, always evolve toward
$\beta=0$.

 Figure~\ref{fig:hem_examples} already illustrates this type of behavior. Due to the chaotic nature of ZLK with spin-orbit back-reaction \citep{Storch:2014b,Storch:2017},
 HJ systems can exit the ZLK cycling phase with a broad range of possible obliquities. In the middle panel of the figure ($P_{\star,{\rm ZAMS}}=2$~d), ZLK cycles end with
 $\beta\approx64^\circ$, which proceeds to decay afterward, with the model with (1,0) enhancement approaching realignment more rapidly than the standard case.  In the bottom panel of the figure ($P_{\star,{\rm ZAMS}}=5$~d), ZLK cycles end with $\beta\approx109^\circ$. In this case, the model with (1,0) enhancement evolves toward anti-alignment. 
 In both these cases,
 stellar spin-down has brought the angular momentum ratio down to $S_\star/L_{\rm p}\approx 0.04$--$0.05$, and thus 
$ \beta_*\approx92^\circ$, by the time ZLK cycles have been arrested. This means that nearly all retrograde orbits ($\beta\gtrsim90^\circ$) will evolve
toward anti-alignment. In principle, for very dissipative planets, ``fast'' tidal migration could be completed in fewer ZLK cycles \citep[e.g.,][]{Petrovich:2015a},
and thus $\beta_*$ would be larger by the time tidal migration has been completed. With larger $\beta_*$, there is a window within which evolution toward $\pi/2$ is possible. This behavior, nonetheless, might be short-lived, as further spin down would reduce $\beta_*$  until it caught up to $\beta$ before the orbit became fully polar. In such a case, the evolution of obliquity would revert sign and progress toward anti-alignment. Future work will address this behavior in further detail.

\subsection{Caveats and Extensions}
Our work consists of a rigorous  derivation, followed by an applied model with several assumptions and uncertainties. In the following, we 
address the extent to which our approximations are reasonable.

\subsubsection{Realistic Stellar  Models and the True Nature of the Spin-over Mode in Stars}\label{sec:discussion_spin-over} 
A major simplification in our work is the implementation of a toy model for PMS and MS evolution. While such a parametric model 
helps with the reproducibility of our work, it is nonetheless true that realistic stellar evolution models are readily accessible to researchers
and these models have proven important in the quantitative study of tidal dissipation in stars \citep[e.g.,][]{Ma:2021,Zanazzi:2024}. Future extensions
of this work will implement realistic stellar evolution models using public numerical tools. 

A more consequential assumption of our model is that the obliquity tide dissipates efficiently at $\tilde{\omega}_{10}=-\Omega_\star$, the frequency of the spin-over mode. In the spherical limit, this large-scale inertial mode corresponds to a rigid tilt of the rotation axis \citep{Poincare:1910,Greenspan:1968}. The rigid tilt itself is strain-free and produces no internal viscous dissipation (Section~\ref{sec:iw_dissipation}, see also \citealp{Ogilvie:2013}).

The forced response at this frequency can include wavelike motions with finite shear. After accounting for mutual spin--orbit precession, \citet{Lin:2017} found such motions in spherical models with a rigid core. The resulting dissipation depended on the core size and density profile.

In a modal description, several modes can contribute to this type of response even when the forcing is detuned from their eigenfrequencies. The dissipative part of $K_{lm}(\tilde{\omega})$ within the inertial range can include contributions from the tails of f-mode resonances \citep{Wu:2024,Dewberry:2024,Dewberry:2026}. Since $\mathcal{D}$ is quadratic in the total strain, viscous cross-terms between modes can also be important away from individual resonances (Equations~\ref{eq:viscous_dissipation} and \ref{eq:specific_dissipation}). Calculations of the $l=m=2$ sectoral tide in aligned systems illustrate these effects.

The existence of a dissipative response leaves open whether it is strong enough to account for the observed trends. For a solar-type star with a 10-day spin period, \citet{Lin:2017} estimated a modified quality factor $Q'_{10}\sim10^7$ based on their rigid-core calculations. This is equivalent to a modified lag time of $\tau_{10}=\tfrac{2}{3}(Q'_{10}\Omega_\star)^{-1}\sim10^{-2}$~s, orders of magnitude below the values favored by our evolutionary models. The estimate depends on the rotation rate and internal structure, so the size of this discrepancy for rapidly rotating, inflated stars is uncertain. Since existing calculations offer little guidance on the value of $\eta_{10}$ for rapidly rotating, inflated stars, 
we continue to treat $\eta_{10}$ as a free parameter to be constrained by observations.

A near-resonant response\textemdash if real\textemdash is one way to reach the required values of $\eta_{10}$. Resonances in the $Y_{22}$ response can routinely produce peak-to-background ratios comparable to or larger than the values of $\eta_{10}$ considered here \citep{Wu:2024,Dewberry:2024,Dewberry:2026}. In centrifugally distorted models, \citet{Dewberry:2022} found that mode mixing can strengthen sectoral tidal overlaps and that the spin-over mode retains its zero inertial-frame frequency. Our conjecture is that centrifugal deformation could bring a separate, shearing mode close to $-\Omega_\star$, with mode mixing increasing its $Y_{21}$ overlap enough to enhance dissipation at the obliquity-tide frequency.

\subsubsection{Inflated Planetary Radii}
We have shown that the model presented here is most efficient at early phases of stellar evolution.
Assuming young ages naturally begs the question: if planets are young, are they inflated as well? Indeed, known close-in
giant planets around young stars such as HIP 67522 and V1298 Tau \citep{Rizzuto:2020,James:2026,David:2019}
do exhibit signs of substantially inflated radii. A larger planetary radius would directly shorten the circularization time scale
due to dissipation in the planet $t_e^{({\rm eq,p})}$ (see Section~\ref{sec:timescales}), and thus make it more difficult
to realign before circularizing (Equation~\ref{eq:damping_timescale_ratio}). This extra factor, however, can be absorbed
into the uncertainty of the weak friction lag time $\tau_{\rm eq,p}$, and thus would only change our result quantitatively.

\subsubsection{Is this Mechanism Likely in Practice?}\label{sec:discussion_likelihood}
As discussed above, the mechanism proposed in this work relies critically on the unconstrained
parameter  $\eta_{10}$ to be high.  The mechanism proposed is most effective when  stars are rapid rotators, and when the planets have already formed when the star is still in the PMS. To an extent, these requirements are mutually degenerate, and thus they might not all need to be satisfied for the mechanism to operate.

Ultimately, the large uncertainties in the value of $\eta_{10}$ preclude us from making a statistical statement as to how often we expect
EWJs or PHJs to be well aligned with their host stars. Future work on IW damping in realistic, tidally distorted models of stars will be instrumental in
assessing the viability of the dynamical processes described in this work as an explanation for the observations. Nonetheless, we can still address whether the values of $P_{\star,0}$ and the planet formation time scale that we have assumed here are broadly reasonable.

\paragraph{Planet Formation Time Scale}In this work, we assumed that planets are already present at some nominal `start time',
which we label as `early', `delayed', and `late,' which are defined relative to the initial epoch $t_0$ of our adopted PMS evolution (Section~\ref{sec:toy_model}). We favor early planet formation scenarios, when the host star remains inflated and obliquity damping is more efficient. Specifically, we adopt three reference stellar radii, $R_*=5R_\odot$, $3R_\odot$, and $1.5R_\odot$, corresponding to our `early', `delayed,' and `late' scenarios, respectively (e.g., see Fig.~\ref{fig:spin_evolution}). 

Planet formation may begin well before the zero-point $t_0$. If this epoch is loosely associated with the classical stellar ``birthline'' \citep{Stahler:1983,Baraffe:2012,Baraffe:2015}, the preceding embedded protostellar phase provides an additional $\sim10^5$--$10^6$~yr for planetary growth. Indeed, the early formation of calcium--aluminum-rich inclusions (CAIs) in the Solar System is consistent with the formation and processing of solids during the embedded Class~0/I phase \citep{Wood:2004,Krot:2009,Ciesla:2010}.

Once planetary embryos exist, pebble accretion can produce massive cores on $\sim10^5$--$10^6$~yr timescales under favorable disk conditions \citep{Ormel:2010,Lambrechts:2012,Lambrechts:2014b}. Numerical models further demonstrate that sufficiently high pebble fluxes can enable rapid core growth, gas accretion, and inward migration, producing hot and warm Jupiters \citep{Bitsch:2015,Bitsch:2019}. Although these models assume pre-existing embryos, they support the plausibility of WJs already being present while their host stars remain substantially inflated ($R_*\sim5R_\odot$).

\paragraph{Rapid Rotators}
Although young rapid rotators ($P_{\star,{\rm ZAMS}}<1$ day)
 constitute a minority of young stellar populations, they still represent a non-negligible fraction of stars with measured rotation periods. 
In the ~13 Myr h Persei cluster, at least $13\%$ of the measured-period sample exhibits sub-day periods \citep{Moraux:2013}.
In the younger Orion complex, the corresponding fraction ranges from approximately 2--6$\%$ among classical T Tauri stars to 12--15$\%$ among weak-line T Tauri stars \citep{Kounkel:2023}. Other Orion surveys report sub-day fractions of approximately 5--20$\%$, depending on sample selection and observational cadence 
\citep[e.g.,][]{Stassun:1999,Herbst:2002}. These estimates include candidate binary systems, which may contribute significantly (\citealp{Kounkel:2023}) to the rapid-rotator population, although it has not been established how many rapid rotators are unambiguously single.

On a more speculative note,
we point out that early-and-fast planet formation and rapid stellar rotation might not be entirely uncorrelated. A tumultuous, early planet formation process triggered
by solid-rich disk and high pebble flux could lead to the formation of more than one gas giant, which in turn could accelerate the dispersal of the inner disk. This dispersal would then be followed by orbital evolution via planet-planet scattering, without the damping effect usually provided by gas, thus motivating  our dynamically hot initial conditions.
More importantly, early
disk dispersal would disrupt the magnetic disk-star coupling that contributes to PMS spin-down \citep[e.g.,][]{Armitage:1996}.  Indeed, the fact that many non-accreting WTTSs
that are rapid rotators also reside in binaries, supports the hypothesis that early disk dispersal hampers the ability of these stars to spin down \citep{Stauffer:2018,Messina:2017,Kounkel:2023}.

\subsubsection{Additional Planets}
If planet--planet scattering produced the dynamically hot initial conditions adopted here \citep[e.g.,][]{Dong:2026}, some WJs may retain eccentric, mutually inclined outer companions. 

WJs are known to host nearby planetary companions \citep{Huang:2016} and outer Jovian companions at au-scale separations \citep{Dawson:2014,Rosenthal:2024}. Sensitivity to outer companions depends in part on the RV baseline (for instance, about 45 days in the RV follow-up observations of TOI-3362~b; \citealp{Dong:2021}). For a 90-day baseline with a short-period 50\% recovery threshold of $K_0=2\,\mathrm{m\,s^{-1}}$, the corresponding thresholds for circular orbits around a solar-mass star are $9.6$ and $14.7\,\mathrm{m\,s^{-1}}$ at $1.5$ and $2\,\mathrm{au}$, respectively \citep[Eq.~27]{Cumming:2004}. At these separations, a $0.3\,M_{\rm J}$ companion inclined by $30^\circ$--$50^\circ$ relative to a transiting WJ gives $K\simeq4$--$7\,\mathrm{m\,s^{-1}}$ and could potentially escape detection. A $1\,M_{\rm J}$ companion gives $K\simeq13$--$23\,\mathrm{m\,s^{-1}}$ and can fall below the 50\% recovery threshold at $2\,\mathrm{au}$ for some viewing geometries when the mutual inclination exceeds approximately $43^\circ$.

Detectable or not, the dynamical role of these additional planets may be limited. While, in principle, an eccentric outer planet can drive substantial eccentricity oscillations through octupole-order secular coupling even in a coplanar system \citep[e.g.,][]{Lee:2003,Petrovich:2015b}, apsidal precession induced by the stellar quadrupole can limit these oscillations (Section~\ref{sec:stellar_binaries}). Even as the stellar quadrupole weakens, general relativistic (GR) precession can continue to limit eccentricity excitation. In analogy with Equation~(\ref{eq:epsilon_rot}), the ratio of the circular-orbit GR precession rate to the characteristic quadrupole secular frequency is \citep[e.g.,][]{Fabrycky:2007,Liu:2015}
\begin{equation}
\varepsilon_{\rm GR}
=
\frac{3{\cal G}(M_\star+m_{\rm p})^2
a_{\rm out}^3(1-e_{\rm out}^2)^{3/2}}
{a_{\rm p}^4 c^2 m_{\rm out}} .
\end{equation}
For $M_\star=M_\odot$, $a_{\rm p}=0.15\,\mathrm{au}$, and $e_{\rm out}=0.2$,
$\varepsilon_{\rm GR}\simeq
1.5(m_{\rm out}/0.3\,M_{\rm J})^{-1}(a_{\rm out}/2\,\mathrm{au})^3$.
GR precession is therefore able to suppress secular forcing by a Saturn-mass companion at $2\,\mathrm{au}$, although closer or more massive companions may exert stronger forcing.

We expect the results of Section~\ref{sec:monte_carlo} to remain approximately valid when the companion induces only small eccentricity oscillations and the stellar spin and WJ orbit remain strongly coupled. In that regime, the outer torque causes the spin and orbit to precess nearly together. As the star spins down, their coupling can weaken, allowing the outer planet to drive obliquity evolution even if GR continues to limit eccentricity oscillations \citep{Anderson:2018}. The changing balance of nodal torques is analogous to a Laplace-plane transition as the central quadrupole declines \citep{Munoz:2015b,Schultz:2021}.

Alternatively, scattering during the PMS phase could also leave an EWJ as the sole survivor if the other planets collide with the inflated star.

\subsubsection{Other Applications of the Generalized  Tides Ansatz}
The generalized effective tides model can be used for components of the tidal response 
other than $(1,0)$. One  example is that of tidal phase locking \citep[][]{Witte:1999,Fuller:2016,Fuller:2017},
which describes the resonant coupling between a forcing component and an internal mode when the latter increases
in frequency over time. This model has been applied
to HJ migration \citep{Ma:2021} and HJ obliquity damping \citep{Zanazzi:2024}. For instance, \citet{Zanazzi:2024} focus on the resonance locking between stellar g-modes and
 the $(m,n)=(0,2)$ component  of a tidal potential. Following \citet{Zanazzi:2024}, one can estimate the value
 of $\tau_{02}$ from the stellar evolutionary time-scale, and then use the same methodology of this work to develop equations of motion in which the weak friction of tides
 and a $(m,n)=(0,2)$ component enhanced by resonance locks can operate simultaneously.


\section{Summary and Conclusions}
In this work, we have investigated whether stellar tides can account for the low projected obliquities of eccentric warm Jupiters (EWJs).

\begin{enumerate}[label=(\alph*)]
\item We derived secular tidal evolution equations in Milankovitch state-vector form for arbitrary eccentricities and inclinations, allowing a separate effective lag time $\tau_{mn}$ for each tidal harmonic. In particular, we focus on the $(1,0)$ component of the response, which leaves the semi-major axis and eccentricity unchanged. At fixed semi-major axis, its contribution to stellar obliquity evolution strengthens with eccentricity.

\item We applied these equations to initially misaligned EWJs. With an enhanced $(1,0)$ response, stellar realignment can outpace orbital circularization if the planet is already on an eccentric warm orbit while its host is still inflated and rapidly rotating during the pre-main sequence (PMS).

\item In our fiducial $1\,M_{\rm J}$ models, $\eta_{10}\sim10^7$ produces
appreciable obliquity damping. The stronger damping found for more massive
planets suggests that $\eta_{10}\sim10^6$ may suffice for super-Jupiters. The $(1,0)$ forcing frequency in the stellar rotating frame, $\tilde{\omega}_{10}=-\Omega_\star$, lies within the inertial-wave range and coincides with the frequency of the global spin-over mode. The rigid spin-over mode is strain-free and produces no internal viscous dissipation. Other motions excited at the same frequency can dissipate energy through shear. Whether these motions can supply the inferred enhancement in rapidly rotating, centrifugally distorted PMS stars remains uncertain.

\item We tested whether this realignment mechanism can operate during binary-driven high-eccentricity migration (HEM) from large initial separations. In the configurations tested, conditions favorable to fast realignment (namely, short rotation periods) can suppress secular eccentricity excitation. In configurations that sustained eccentricity forcing on the main sequence, orbital circularization proceeded faster than stellar realignment. These results cast doubt on the interpretation of aligned EWJs as relics of secularly driven HEM, even when they appear to lie along the tidal migration track.
\end{enumerate}

If PMS tidal realignment accounts for a substantial fraction of aligned EWJs, we expect the following trends:

\begin{enumerate}[label=(\alph*)]
\item Among sufficiently young EWJ systems, hosts with shorter rotation periods $P_\star$ should preferentially have smaller projected obliquities $|\lambda|$.

\item Among EWJs with comparable orbital properties, the aligned fraction should increase with planet mass.

\item Present-day stellar effective temperature should be a weak predictor of EWJ obliquity if most of the damping occurred during the PMS.

\item At comparable eccentricities and planet masses, the aligned fraction should decrease at larger semi-major axes. In our models, realignment becomes inefficient beyond roughly $a_{\rm p}\sim0.2$~au.
\end{enumerate}

If planet--planet scattering excited the eccentricities of some EWJs before realignment, those systems may retain eccentric Jovian companions at au-scale separations. Mutual inclinations could make these companions less likely to transit, and short-baseline RV observations may provide limited constraints on their orbits.

\section*{acknowledgements}
DJM is grateful to C. Petrovich, J.~I. Espinoza-Retamal, H. Bhaskar, JJ Zanazzi, J. Jackson, A. Polanksi, J. Fuller, Janosz Dewberry and H. Perets for constructive discussions.
  This work was supported by the National Science Foundation under grant AST-2511258.


\begin{thebibliography}{}
\expandafter\ifx\csname natexlab\endcsname\relax\def\natexlab#1{#1}\fi
\providecommand{\url}[1]{\href{#1}{#1}}
\providecommand{\dodoi}[1]{doi:~\href{http://doi.org/#1}{\nolinkurl{#1}}}
\providecommand{\doeprint}[1]{\href{http://ascl.net/#1}{\nolinkurl{http://ascl.net/#1}}}
\providecommand{\doarXiv}[1]{\href{https://arxiv.org/abs/#1}{\nolinkurl{https://arxiv.org/abs/#1}}}

\bibitem[{Abramowitz \& Stegun(1964)}]{abramowitz+stegun}
Abramowitz, M., \& Stegun, I.~A. 1964, Handbook of Mathematical Functions with
  Formulas, Graphs, and Mathematical Tables (New York: Dover)

\bibitem[{{Albrecht} {et~al.}(2012){Albrecht}, {Winn}, {Johnson}, {Howard},
  {Marcy}, {Butler}, {Arriagada}, {Crane}, {Shectman}, {Thompson}, {Hirano},
  {Bakos}, \& {Hartman}}]{Albrecht:2012}
{Albrecht}, S., {Winn}, J.~N., {Johnson}, J.~A., {et~al.} 2012, \apj, 757, 18,
  \dodoi{10.1088/0004-637X/757/1/18}

\bibitem[{{Albrecht} {et~al.}(2022){Albrecht}, {Dawson}, \&
  {Winn}}]{Albrecht:2022}
{Albrecht}, S.~H., {Dawson}, R.~I., \& {Winn}, J.~N. 2022, \pasp, 134, 082001,
  \dodoi{10.1088/1538-3873/ac6c09}

\bibitem[{{Alexander}(1973)}]{Alexander:1973}
{Alexander}, M.~E. 1973, \apss, 23, 459, \dodoi{10.1007/BF00645172}

\bibitem[{{Anderson} \& {Lai}(2018)}]{Anderson:2018}
{Anderson}, K.~R., \& {Lai}, D. 2018, \mnras, 480, 1402,
  \dodoi{10.1093/mnras/sty1937}

\bibitem[{{Anderson} {et~al.}(2016){Anderson}, {Storch}, \&
  {Lai}}]{Anderson:2016}
{Anderson}, K.~R., {Storch}, N.~I., \& {Lai}, D. 2016, \mnras, 456, 3671,
  \dodoi{10.1093/mnras/stv2906}

\bibitem[{{Armitage} \& {Clarke}(1996)}]{Armitage:1996}
{Armitage}, P.~J., \& {Clarke}, C.~J. 1996, \mnras, 280, 458,
  \dodoi{10.1093/mnras/280.2.458}

\bibitem[{{Baraffe} {et~al.}(2015){Baraffe}, {Homeier}, {Allard}, \&
  {Chabrier}}]{Baraffe:2015}
{Baraffe}, I., {Homeier}, D., {Allard}, F., \& {Chabrier}, G. 2015, \aap, 577,
  A42, \dodoi{10.1051/0004-6361/201425481}

\bibitem[{{Baraffe} {et~al.}(2012){Baraffe}, {Vorobyov}, \&
  {Chabrier}}]{Baraffe:2012}
{Baraffe}, I., {Vorobyov}, E., \& {Chabrier}, G. 2012, \apj, 756, 118,
  \dodoi{10.1088/0004-637X/756/2/118}

\bibitem[{{Barker}(2020)}]{Barker:2020}
{Barker}, A.~J. 2020, \mnras, 498, 2270, \dodoi{10.1093/mnras/staa2405}

\bibitem[{{Barker}(2022)}]{Barker:2022}
---. 2022, \apjl, 927, L36, \dodoi{10.3847/2041-8213/ac5b63}

\bibitem[{{Barker} {et~al.}(2016){Barker}, {Braviner}, \&
  {Ogilvie}}]{Barker:2016}
{Barker}, A.~J., {Braviner}, H.~J., \& {Ogilvie}, G.~I. 2016, \mnras, 459, 924,
  \dodoi{10.1093/mnras/stw701}

\bibitem[{{Barker} {et~al.}(2024){Barker}, {Efroimsky}, {Makarov}, \&
  {Veras}}]{Barker:2024}
{Barker}, A.~J., {Efroimsky}, M., {Makarov}, V.~V., \& {Veras}, D. 2024,
  \mnras, 527, 5131, \dodoi{10.1093/mnras/stad3530}

\bibitem[{{Barker} \& {Ogilvie}(2009)}]{Barker:2009}
{Barker}, A.~J., \& {Ogilvie}, G.~I. 2009, \mnras, 395, 2268,
  \dodoi{10.1111/j.1365-2966.2009.14694.x}

\bibitem[{{Belczynski} {et~al.}(2008){Belczynski}, {Kalogera}, {Rasio}, {Taam},
  {Zezas}, {Bulik}, {Maccarone}, \& {Ivanova}}]{Belczynski:2008}
{Belczynski}, K., {Kalogera}, V., {Rasio}, F.~A., {et~al.} 2008, \apjs, 174,
  223, \dodoi{10.1086/521026}

\bibitem[{{Bhaskar} {et~al.}(2026){Bhaskar}, {Petrovich}, \&
  {Mu{\~n}oz}}]{Bhaskar:2026}
{Bhaskar}, H.~G., {Petrovich}, C., \& {Mu{\~n}oz}, D.~J. 2026, \apjl, 1007,
  L12, \dodoi{10.3847/2041-8213/ae8dbc}

\bibitem[{{Bitsch} {et~al.}(2019){Bitsch}, {Izidoro}, {Johansen}, {Raymond},
  {Morbidelli}, {Lambrechts}, \& {Jacobson}}]{Bitsch:2019}
{Bitsch}, B., {Izidoro}, A., {Johansen}, A., {et~al.} 2019, \aap, 623, A88,
  \dodoi{10.1051/0004-6361/201834489}

\bibitem[{{Bitsch} {et~al.}(2015){Bitsch}, {Lambrechts}, \&
  {Johansen}}]{Bitsch:2015}
{Bitsch}, B., {Lambrechts}, M., \& {Johansen}, A. 2015, \aap, 582, A112,
  \dodoi{10.1051/0004-6361/201526463}

\bibitem[{{Breiter} \& {Ratajczak}(2005)}]{Breiter:2005}
{Breiter}, S., \& {Ratajczak}, R. 2005, \mnras, 364, 1222,
  \dodoi{10.1111/j.1365-2966.2005.09658.x}

\bibitem[{{Brumberg}(2012)}]{Brumberg:2012}
{Brumberg}, V.~A. 2012, {Analytical Techniques of Celestial Mechanics}
  (Springer Berlin Heidelberg)

\bibitem[{{Ciesla}(2010)}]{Ciesla:2010}
{Ciesla}, F.~J. 2010, \icarus, 208, 455, \dodoi{10.1016/j.icarus.2010.02.010}

\bibitem[{{Correia}(2009)}]{Correia:2009}
{Correia}, A. C.~M. 2009, \apjl, 704, L1, \dodoi{10.1088/0004-637X/704/1/L1}

\bibitem[{{Correia} {et~al.}(2011){Correia}, {Laskar}, {Farago}, \&
  {Bou{\'e}}}]{Correia:2011}
{Correia}, A. C.~M., {Laskar}, J., {Farago}, F., \& {Bou{\'e}}, G. 2011,
  Celestial Mechanics and Dynamical Astronomy, 111, 105,
  \dodoi{10.1007/s10569-011-9368-9}

\bibitem[{{Correia} \& {Valente}(2022)}]{Correia:2022}
{Correia}, A. C.~M., \& {Valente}, E. F.~S. 2022, Celestial Mechanics and
  Dynamical Astronomy, 134, 24, \dodoi{10.1007/s10569-022-10079-3}

\bibitem[{{Cumming}(2004)}]{Cumming:2004}
{Cumming}, A. 2004, \mnras, 354, 1165, \dodoi{10.1111/j.1365-2966.2004.08275.x}

\bibitem[{{Darwin}(1879)}]{Darwin:1879}
{Darwin}, G.~H. 1879, Philosophical Transactions of the Royal Society of London
  Series I, 170, 447

\bibitem[{{David} {et~al.}(2019){David}, {Petigura}, {Luger}, {Foreman-Mackey},
  {Livingston}, {Mamajek}, \& {Hillenbrand}}]{David:2019}
{David}, T.~J., {Petigura}, E.~A., {Luger}, R., {et~al.} 2019, \apjl, 885, L12,
  \dodoi{10.3847/2041-8213/ab4c99}

\bibitem[{{Dawson} \& {Johnson}(2012)}]{Dawson:2012}
{Dawson}, R.~I., \& {Johnson}, J.~A. 2012, \apj, 756, 122,
  \dodoi{10.1088/0004-637X/756/2/122}

\bibitem[{{Dawson} \& {Johnson}(2018)}]{Dawson:2018}
---. 2018, \araa, 56, 175, \dodoi{10.1146/annurev-astro-081817-051853}

\bibitem[{{Dawson} {et~al.}(2015){Dawson}, {Murray-Clay}, \&
  {Johnson}}]{Dawson:2015}
{Dawson}, R.~I., {Murray-Clay}, R.~A., \& {Johnson}, J.~A. 2015, \apj, 798, 66,
  \dodoi{10.1088/0004-637X/798/2/66}

\bibitem[{{Dawson} {et~al.}(2014){Dawson}, {Johnson}, {Fabrycky},
  {Foreman-Mackey}, {Murray-Clay}, {Buchhave}, {Cargile}, {Clubb}, {Fulton},
  {Hebb}, \& et~al.}]{Dawson:2014}
{Dawson}, R.~I., {Johnson}, J.~A., {Fabrycky}, D.~C., {et~al.} 2014, \apj, 791,
  89, \dodoi{10.1088/0004-637X/791/2/89}

\bibitem[{{Dewberry}(2023)}]{Dewberry:2023}
{Dewberry}, J.~W. 2023, \mnras, 521, 5991, \dodoi{10.1093/mnras/stad546}

\bibitem[{{Dewberry}(2026)}]{Dewberry:2026}
---. 2026, \apj, 997, 27, \dodoi{10.3847/1538-4357/ae18ca}

\bibitem[{{Dewberry} \& {Lai}(2022)}]{Dewberry:2022}
{Dewberry}, J.~W., \& {Lai}, D. 2022, \apj, 925, 124,
  \dodoi{10.3847/1538-4357/ac3ede}

\bibitem[{{Dewberry} \& {Wu}(2024)}]{Dewberry:2024}
{Dewberry}, J.~W., \& {Wu}, S.~C. 2024, \mnras, 527, 2288,
  \dodoi{10.1093/mnras/stad3164}

\bibitem[{{Dong} {et~al.}(2026){Dong}, {Lee}, {Kokubo}, {Murray-Clay}, {Gupta},
  \& {Turner}}]{Dong:2026}
{Dong}, J., {Lee}, E.~J., {Kokubo}, E., {et~al.} 2026, \apjl, 1007, L29,
  \dodoi{10.3847/2041-8213/ae930e}

\bibitem[{{Dong} {et~al.}(2021){Dong}, {Huang}, {Zhou}, {Dawson}, {Rodriguez},
  {Eastman}, {Collins}, {Quinn}, {Shporer}, {Triaud}, {Wang}, {Beatty},
  {Jackson}, {Collins}, {Abe}, {Suarez}, {Crouzet}, {M{\'e}karnia},
  {Dransfield}, {Jensen}, {Stockdale}, {Barkaoui}, {Heitzmann}, {Wright},
  {Addison}, {Wittenmyer}, {Okumura}, {Bowler}, {Horner}, {Kane}, {Kielkopf},
  {Liu}, {Plavchan}, {Mengel}, {Ricker}, {Vanderspek}, {Latham}, {Seager},
  {Winn}, {Jenkins}, {Christiansen}, \& {Paegert}}]{Dong:2021}
{Dong}, J., {Huang}, C.~X., {Zhou}, G., {et~al.} 2021, \apjl, 920, L16,
  \dodoi{10.3847/2041-8213/ac2600}

\bibitem[{{Dong} {et~al.}(2025){Dong}, {Chontos}, {Zhou}, {Stefansson}, {Wang},
  {Huang}, {Gupta}, {Halverson}, {Kanodia}, {Luhn}, {Mahadevan}, {Monson},
  {Alvarado-Montes}, {Ninan}, {Robertson}, {Roy}, {Schwab}, \&
  {Wright}}]{Dong:2025}
{Dong}, J., {Chontos}, A., {Zhou}, G., {et~al.} 2025, \aj, 169, 4,
  \dodoi{10.3847/1538-3881/ad8ec1}

\bibitem[{{Dong} {et~al.}(2014){Dong}, {Katz}, \& {Socrates}}]{Dong:2014}
{Dong}, S., {Katz}, B., \& {Socrates}, A. 2014, \apjl, 781, L5,
  \dodoi{10.1088/2041-8205/781/1/L5}

\bibitem[{Dullin(2004)}]{Dullin:2004}
Dullin, H.~R. 2004, Regular and chaotic dynamics, 9, 255

\bibitem[{{Efroimsky}(2012)}]{Efroimsky:2012}
{Efroimsky}, M. 2012, Celestial Mechanics and Dynamical Astronomy, 112, 283,
  \dodoi{10.1007/s10569-011-9397-4}

\bibitem[{{Efroimsky}(2015)}]{Efroimsky:2015}
---. 2015, \aj, 150, 98, \dodoi{10.1088/0004-6256/150/4/98}

\bibitem[{{Eggleton} {et~al.}(1998){Eggleton}, {Kiseleva}, \&
  {Hut}}]{Eggleton:1998}
{Eggleton}, P.~P., {Kiseleva}, L.~G., \& {Hut}, P. 1998, \apj, 499, 853

\bibitem[{{Eggleton} \& {Kiseleva-Eggleton}(2001)}]{Eggleton:2001}
{Eggleton}, P.~P., \& {Kiseleva-Eggleton}, L. 2001, \apj, 562, 1012,
  \dodoi{10.1086/323843}

\bibitem[{{Espinoza-Retamal} {et~al.}(2023){Espinoza-Retamal}, {Brahm},
  {Petrovich}, {Jord{\'a}n}, {Stef{\'a}nsson}, {Sedaghati}, {Hobson},
  {Mu{\~n}oz}, {Boyle}, {Leiva}, \& {Suc}}]{Espinoza:2023}
{Espinoza-Retamal}, J.~I., {Brahm}, R., {Petrovich}, C., {et~al.} 2023, \apjl,
  958, L20, \dodoi{10.3847/2041-8213/ad096d}

\bibitem[{{Espinoza-Retamal} {et~al.}(2025){Espinoza-Retamal}, {Jord{\'a}n},
  {Brahm}, {Petrovich}, {Sedaghati}, {Stef{\'a}nsson}, {Hobson}, {Tala Pinto},
  {Mu{\~n}oz}, {Boyle}, \& et~al.}]{Espinoza:2025}
{Espinoza-Retamal}, J.~I., {Jord{\'a}n}, A., {Brahm}, R., {et~al.} 2025, \aj,
  170, 70, \dodoi{10.3847/1538-3881/ade22e}

\bibitem[{{Fabrycky} \& {Tremaine}(2007)}]{Fabrycky:2007}
{Fabrycky}, D., \& {Tremaine}, S. 2007, \apj, 669, 1298, \dodoi{10.1086/521702}

\bibitem[{{Fabrycky} \& {Winn}(2009)}]{Fabrycky:2009}
{Fabrycky}, D.~C., \& {Winn}, J.~N. 2009, \apj, 696, 1230,
  \dodoi{10.1088/0004-637X/696/2/1230}

\bibitem[{{Fischer} {et~al.}(2007){Fischer}, {Vogt}, {Marcy}, {Butler}, {Sato},
  {Henry}, {Robinson}, {Laughlin}, {Ida}, {Toyota}, \& et~al.}]{Fischer:2007}
{Fischer}, D.~A., {Vogt}, S.~S., {Marcy}, G.~W., {et~al.} 2007, \apj, 669,
  1336, \dodoi{10.1086/521869}

\bibitem[{{Fuller}(2017)}]{Fuller:2017}
{Fuller}, J. 2017, \mnras, 472, 1538, \dodoi{10.1093/mnras/stx2135}

\bibitem[{{Fuller} {et~al.}(2016){Fuller}, {Luan}, \& {Quataert}}]{Fuller:2016}
{Fuller}, J., {Luan}, J., \& {Quataert}, E. 2016, \mnras, 458, 3867,
  \dodoi{10.1093/mnras/stw609}

\bibitem[{{Giacaglia}(1976)}]{Giacaglia:1976}
{Giacaglia}, G.~E.~O. 1976, Celestial Mechanics, 14, 515,
  \dodoi{10.1007/BF01229062}

\bibitem[{{Goldreich} \& {Keeley}(1977)}]{Goldreich:1977b}
{Goldreich}, P., \& {Keeley}, D.~A. 1977, \apj, 211, 934,
  \dodoi{10.1086/155005}

\bibitem[{{Goldreich} \& {Nicholson}(1977)}]{Goldreich:1977a}
{Goldreich}, P., \& {Nicholson}, P.~D. 1977, \icarus, 30, 301,
  \dodoi{10.1016/0019-1035(77)90163-4}

\bibitem[{{Goldreich} \& {Nicholson}(1989)}]{Goldreich:1989}
---. 1989, \apj, 342, 1079, \dodoi{10.1086/167665}

\bibitem[{{Goodman} \& {Dickson}(1998)}]{Goodman:1998}
{Goodman}, J., \& {Dickson}, E.~S. 1998, \apj, 507, 938, \dodoi{10.1086/306348}

\bibitem[{{Goodman} \& {Oh}(1997)}]{Goodman:1997}
{Goodman}, J., \& {Oh}, S.~P. 1997, \apj, 486, 403, \dodoi{10.1086/304505}

\bibitem[{Greenspan(1968)}]{Greenspan:1968}
Greenspan, H. 1968, The Theory of Rotating Fluids, Cambridge Monographs on
  Mechanics and Applied Mathematics (Cambridge University Press)

\bibitem[{{Gupta} {et~al.}(2023){Gupta}, {Jackson}, {H{\'e}brard}, {Lin},
  {Stassun}, {Dong}, {Villanueva}, {Dragomir}, {Mahadevan}, {Wright},
  {Almenara}, {Blake}, {Boisse}, {Cort{\'e}s-Zuleta}, {Dalba}, {D{\'\i}az},
  {Ford}, {Forveille}, {Gagliano}, {Halverson}, {Heidari}, {Kanodia}, {Kiefer},
  {Latham}, {McElwain}, {Mireles}, {Moutou}, {Pepper}, {Ricker}, {Robertson},
  {Roy}, {Schlecker}, {Schwab}, {Seager}, {Shporer}, {Stef{\'a}nsson},
  {Terrien}, {Ting}, {Winn}, \& {Youngblood}}]{Gupta:2023}
{Gupta}, A.~F., {Jackson}, J.~M., {H{\'e}brard}, G., {et~al.} 2023, \aj, 165,
  234, \dodoi{10.3847/1538-3881/accb9b}

\bibitem[{{Gupta} {et~al.}(2024){Gupta}, {Millholland}, {Im}, {Dong},
  {Jackson}, {Carleo}, {Libby-Roberts}, {Delamer}, {Giovinazzi}, {Lin},
  {Kanodia}, {Wang}, {Stassun}, {Masseron}, {Dragomir}, {Mahadevan}, {Wright},
  {Alvarado-Montes}, {Bender}, {Blake}, {Caldwell}, {Ca{\~n}as}, {Cochran},
  {Dalba}, {Everett}, {Fernandez}, {Golub}, {Guillet}, {Halverson}, {Hebb},
  {Higuera}, {Huang}, {Klusmeyer}, {Knight}, {Leroux}, {Logsdon}, {Loose},
  {McElwain}, {Monson}, {Ninan}, {Nowak}, {Palle}, {Patel}, {Pepper}, {Primm},
  {Rajagopal}, {Robertson}, {Roy}, {Schneider}, {Schwab}, {Schweiker}, {Sgro},
  {Shimizu}, {Simard}, {Stef{\'a}nsson}, {Stevens}, {Villanueva}, {Wisniewski},
  {Will}, \& {Ziegler}}]{Gupta:2024}
{Gupta}, A.~F., {Millholland}, S.~C., {Im}, H., {et~al.} 2024, \nat, 632, 50,
  \dodoi{10.1038/s41586-024-07688-3}

\bibitem[{Hansen {et~al.}(1994)Hansen, Kawaler, \& Trimble}]{Hansen:2004}
Hansen, C.~J., Kawaler, S.~D., \& Trimble, V. 1994, Stellar Interiors: Physical
  Principles, Structure, and Evolution, 1st edn. (New York: Springer-Verlag)

\bibitem[{{Harrington}(1968)}]{Harrington:1968}
{Harrington}, R.~S. 1968, \aj, 73, 190, \dodoi{10.1086/110614}

\bibitem[{{Herbst} {et~al.}(2002){Herbst}, {Bailer-Jones}, {Mundt},
  {Meisenheimer}, \& {Wackermann}}]{Herbst:2002}
{Herbst}, W., {Bailer-Jones}, C.~A.~L., {Mundt}, R., {Meisenheimer}, K., \&
  {Wackermann}, R. 2002, \aap, 396, 513, \dodoi{10.1051/0004-6361:20021362}

\bibitem[{{Huang} {et~al.}(2016){Huang}, {Wu}, \& {Triaud}}]{Huang:2016}
{Huang}, C., {Wu}, Y., \& {Triaud}, A. H.~M.~J. 2016, \apj, 825, 98,
  \dodoi{10.3847/0004-637X/825/2/98}

\bibitem[{{Hughes}(1981)}]{Hughes:1981}
{Hughes}, S. 1981, Celestial Mechanics, 25, 101, \dodoi{10.1007/BF01301812}

\bibitem[{{Hurley} {et~al.}(2002){Hurley}, {Tout}, \& {Pols}}]{Hurley:2002}
{Hurley}, J.~R., {Tout}, C.~A., \& {Pols}, O.~R. 2002, \mnras, 329, 897,
  \dodoi{10.1046/j.1365-8711.2002.05038.x}

\bibitem[{{Hut}(1981)}]{Hut:1981}
{Hut}, P. 1981, \aap, 99, 126

\bibitem[{{Ida} \& {Makino}(1992)}]{Ida:1992}
{Ida}, S., \& {Makino}, J. 1992, \icarus, 96, 107,
  \dodoi{10.1016/0019-1035(92)90008-U}

\bibitem[{{Ikwut-Ukwa} {et~al.}(2022){Ikwut-Ukwa}, {Rodriguez}, {Quinn},
  {Zhou}, {Vanderburg}, {Ali}, {Bunten}, {Gaudi}, {Latham}, {Howell}, \&
  et~al.}]{Ikwut-Ukwa:2022}
{Ikwut-Ukwa}, M., {Rodriguez}, J.~E., {Quinn}, S.~N., {et~al.} 2022, \aj, 163,
  9, \dodoi{10.3847/1538-3881/ac2ee1}

\bibitem[{{James} {et~al.}(2026){James}, {Barber}, {Mann}, {Thao}, {Kraus},
  {Tofflemire}, {Vanderburg}, {Abe}, {Guillot}, {Suarez}, {Mekarnia}, {Triaud},
  {Deloupy}, {Mori}, {de Leon}, {Narita}, {Fukui}, {Livingston}, {Zhou},
  {Guen{\'e}e}, {Becker}, {Lopez Murillo}, {Collins}, {Schwarz}, {Stockdale},
  {Tan}, {Horne}, {Boff}, {Wood}, {Morrell}, \& {Bouma}}]{James:2026}
{James}, H.~R., {Barber}, M.~G., {Mann}, A.~W., {et~al.} 2026, arXiv e-prints,
  arXiv:2609.35979, \dodoi{10.48550/arXiv.2609.35979}

\bibitem[{{Jord{\'a}n} {et~al.}(2020){Jord{\'a}n}, {Brahm}, {Espinoza},
  {Henning}, {Jones}, {Kossakowski}, {Sarkis}, {Trifonov}, {Rojas}, {Torres},
  {Drass}, {Nandakumar}, {Barbieri}, {Davis}, {Wang}, {Bayliss}, {Bouma},
  {Dragomir}, {Eastman}, {Daylan}, {Guerrero}, {Barclay}, {Ting}, {Henze},
  {Ricker}, {Vanderspek}, {Latham}, {Seager}, {Winn}, {Jenkins}, {Wittenmyer},
  {Bowler}, {Crossfield}, {Horner}, {Kane}, {Kielkopf}, {Morton}, {Plavchan},
  {Tinney}, {Addison}, {Mengel}, {Okumura}, {Shahaf}, {Mazeh}, {Rabus},
  {Shporer}, {Ziegler}, {Mann}, \& {Hart}}]{Jordan:2020}
{Jord{\'a}n}, A., {Brahm}, R., {Espinoza}, N., {et~al.} 2020, \aj, 159, 145,
  \dodoi{10.3847/1538-3881/ab6f67}

\bibitem[{{Juri{\'c}} \& {Tremaine}(2008)}]{Juric:2008}
{Juri{\'c}}, M., \& {Tremaine}, S. 2008, \apj, 686, 603, \dodoi{10.1086/590047}

\bibitem[{{Kane} {et~al.}(2023){Kane}, {Hill}, {Dalba}, {Fetherolf}, {Henry},
  {Fajardo-Acosta}, {Gnilka}, {Howard}, {Howell}, \& {Isaacson}}]{Kane:2023}
{Kane}, S.~R., {Hill}, M.~L., {Dalba}, P.~A., {et~al.} 2023, \aj, 165, 252,
  \dodoi{10.3847/1538-3881/acd17a}

\bibitem[{Kerswell(2002)}]{Kerswell:2002}
Kerswell, R.~R. 2002, Annual Review of Fluid Mechanics, 34, 83,
  \dodoi{10.1146/annurev.fluid.34.081701.171829}

\bibitem[{{Kippenhahn} {et~al.}(2012){Kippenhahn}, {Weigert}, \&
  {Weiss}}]{Kippenhahn:2012}
{Kippenhahn}, R., {Weigert}, A., \& {Weiss}, A. 2012, {Stellar Structure and
  Evolution} (Berlin: Springer), \dodoi{10.1007/978-3-642-30304-3}

\bibitem[{{Kounkel} {et~al.}(2023){Kounkel}, {Stassun}, {Hillenbrand},
  {Hern{\'a}ndez}, {Serna}, \& {Curtis}}]{Kounkel:2023}
{Kounkel}, M., {Stassun}, K.~G., {Hillenbrand}, L.~A., {et~al.} 2023, \aj, 165,
  182, \dodoi{10.3847/1538-3881/acc2bd}

\bibitem[{{Kozai}(1973)}]{Kozai:1973}
{Kozai}, Y. 1973, SAO Special Report, 349

\bibitem[{{Krot} {et~al.}(2009){Krot}, {Amelin}, {Bland}, {Ciesla}, {Connelly},
  {Davis}, {Huss}, {Hutcheon}, {Makide}, {Nagashima}, {Nyquist}, {Russell},
  {Scott}, {Thrane}, {Yurimoto}, \& {Yin}}]{Krot:2009}
{Krot}, A.~N., {Amelin}, Y., {Bland}, P., {et~al.} 2009, \gca, 73, 4963,
  \dodoi{10.1016/j.gca.2008.09.039}

\bibitem[{Kundu {et~al.}(2001)Kundu, Cohen, \& Hu}]{KunduCohen:2001}
Kundu, P., Cohen, I., \& Hu, H. 2001, Fluid Mechanics (Academic Press).
\newblock \url{https://books.google.com/books?id=AFVnbQqCJSQC}

\bibitem[{{Lacaze} {et~al.}(2004){Lacaze}, {Le Gal}, \& {Le
  Diz{\`e}s}}]{Lacaze:2004}
{Lacaze}, L., {Le Gal}, P., \& {Le Diz{\`e}s}, S. 2004, Journal of Fluid
  Mechanics, 505, 1, \dodoi{10.1017/S0022112004008328}

\bibitem[{{Lai}(2012)}]{Lai:2012}
{Lai}, D. 2012, \mnras, 423, 486, \dodoi{10.1111/j.1365-2966.2012.20893.x}

\bibitem[{{Lai} {et~al.}(1993){Lai}, {Rasio}, \& {Shapiro}}]{Lai:1993}
{Lai}, D., {Rasio}, F.~A., \& {Shapiro}, S.~L. 1993, \apjs, 88, 205,
  \dodoi{10.1086/191822}

\bibitem[{{Lai} {et~al.}(1994){Lai}, {Rasio}, \& {Shapiro}}]{Lai:1994a}
---. 1994, \apj, 420, 811, \dodoi{10.1086/173606}

\bibitem[{{Lambrechts} \& {Johansen}(2012)}]{Lambrechts:2012}
{Lambrechts}, M., \& {Johansen}, A. 2012, \aap, 544, A32,
  \dodoi{10.1051/0004-6361/201219127}

\bibitem[{{Lambrechts} \& {Johansen}(2014)}]{Lambrechts:2014b}
---. 2014, \aap, 572, A107, \dodoi{10.1051/0004-6361/201424343}

\bibitem[{{Landau} \& {Lifshitz}(1959)}]{Landau_Fluids}
{Landau}, L.~D., \& {Lifshitz}, E.~M. 1959, {Fluid mechanics} (Oxford: Pergamon
  Press)

\bibitem[{{Lee} \& {Peale}(2003)}]{Lee:2003}
{Lee}, M.~H., \& {Peale}, S.~J. 2003, \apj, 592, 1201, \dodoi{10.1086/375857}

\bibitem[{{Lin} {et~al.}(1996){Lin}, {Bodenheimer}, \& {Richardson}}]{Lin:1996}
{Lin}, D.~N.~C., {Bodenheimer}, P., \& {Richardson}, D.~C. 1996, \nat, 380,
  606, \dodoi{10.1038/380606a0}

\bibitem[{{Lin} \& {Ogilvie}(2017)}]{Lin:2017}
{Lin}, Y., \& {Ogilvie}, G.~I. 2017, \mnras, 468, 1387,
  \dodoi{10.1093/mnras/stx540}

\bibitem[{{Lissauer}(1993)}]{Lissauer:1993}
{Lissauer}, J.~J. 1993, \araa, 31, 129,
  \dodoi{10.1146/annurev.aa.31.090193.001021}

\bibitem[{{Liu} {et~al.}(2015){Liu}, {Mu{\~n}oz}, \& {Lai}}]{Liu:2015}
{Liu}, B., {Mu{\~n}oz}, D.~J., \& {Lai}, D. 2015, \mnras, 447, 751,
  \dodoi{10.1093/mnras/stu2396}

\bibitem[{{Ma} \& {Fuller}(2021)}]{Ma:2021}
{Ma}, L., \& {Fuller}, J. 2021, \apj, 918, 16, \dodoi{10.3847/1538-4357/ac088e}

\bibitem[{{Mathews} {et~al.}(2002){Mathews}, {Herring}, \&
  {Buffett}}]{Mathews:2002}
{Mathews}, P.~M., {Herring}, T.~A., \& {Buffett}, B.~A. 2002, Journal of
  Geophysical Research (Solid Earth), 107, 2068, \dodoi{10.1029/2001JB000390}

\bibitem[{{Matt} {et~al.}(2015){Matt}, {Brun}, {Baraffe}, {Bouvier}, \&
  {Chabrier}}]{Matt:2015}
{Matt}, S.~P., {Brun}, A.~S., {Baraffe}, I., {Bouvier}, J., \& {Chabrier}, G.
  2015, \apjl, 799, L23, \dodoi{10.1088/2041-8205/799/2/L23}

\bibitem[{{Messina} {et~al.}(2017){Messina}, {Lanzafame}, {Malo}, {Desidera},
  {Buccino}, {Zhang}, {Artemenko}, {Millward}, \& {Hambsch}}]{Messina:2017}
{Messina}, S., {Lanzafame}, A.~C., {Malo}, L., {et~al.} 2017, \aap, 607, A3,
  \dodoi{10.1051/0004-6361/201730444}

\bibitem[{{Mignard}(1979)}]{Mignard:1979}
{Mignard}, F. 1979, Moon and Planets, 20, 301, \dodoi{10.1007/BF00907581}

\bibitem[{{Milankovitch}(1939)}]{Milankovitch:1939}
{Milankovitch}, M. 1939, Bull. Acad. Sci. Math. Nat. A, 6, 1

\bibitem[{{Miranda} \& {Lai}(2015)}]{Miranda:2015}
{Miranda}, R., \& {Lai}, D. 2015, \mnras, 452, 2396,
  \dodoi{10.1093/mnras/stv1450}

\bibitem[{{Mireles} {et~al.}(2025){Mireles}, {Murgas}, {Dragomir}, {Palle},
  {Dong}, {Carleo}, \& {Esparza-Borges}}]{Mireles:2025}
{Mireles}, I., {Murgas}, F., {Dragomir}, D., {et~al.} 2025, \aap, 704, A194,
  \dodoi{10.1051/0004-6361/202554335}

\bibitem[{{Moe} \& {Kratter}(2018)}]{Moe:2018}
{Moe}, M., \& {Kratter}, K.~M. 2018, \apj, 854, 44,
  \dodoi{10.3847/1538-4357/aaa6d2}

\bibitem[{{Moraux} {et~al.}(2013){Moraux}, {Artemenko}, {Bouvier}, {Irwin},
  {Ibrahimov}, {Magakian}, {Grankin}, {Nikogossian}, {Cardoso}, {Hodgkin},
  {Aigrain}, \& {Movsessian}}]{Moraux:2013}
{Moraux}, E., {Artemenko}, S., {Bouvier}, J., {et~al.} 2013, \aap, 560, A13,
  \dodoi{10.1051/0004-6361/201321508}

\bibitem[{{Morgan} {et~al.}(2026){Morgan}, {Bowler}, \& {Tran}}]{Morgan:2026}
{Morgan}, M., {Bowler}, B.~P., \& {Tran}, Q.~H. 2026, \aj, 171, 19,
  \dodoi{10.3847/1538-3881/ae0e16}

\bibitem[{{Morton} \& {Johnson}(2011)}]{Morton:2011}
{Morton}, T.~D., \& {Johnson}, J.~A. 2011, \apj, 729, 138,
  \dodoi{10.1088/0004-637X/729/2/138}

\bibitem[{{Mu{\~n}oz} \& {Lai}(2015)}]{Munoz:2015b}
{Mu{\~n}oz}, D.~J., \& {Lai}, D. 2015, Proceedings of the National Academy of
  Science, 112, 9264, \dodoi{10.1073/pnas.1505671112}

\bibitem[{{Mu{\~n}oz} {et~al.}(2016){Mu{\~n}oz}, {Lai}, \& {Liu}}]{Munoz:2016a}
{Mu{\~n}oz}, D.~J., {Lai}, D., \& {Liu}, B. 2016, \mnras, 460, 1086,
  \dodoi{10.1093/mnras/stw983}

\bibitem[{{Mu{\~n}oz} \& {Perets}(2018)}]{Munoz:2018}
{Mu{\~n}oz}, D.~J., \& {Perets}, H.~B. 2018, \aj, 156, 253,
  \dodoi{10.3847/1538-3881/aae7d0}

\bibitem[{{Naoz} {et~al.}(2013){Naoz}, {Farr}, {Lithwick}, {Rasio}, \&
  {Teyssandier}}]{Naoz:2013}
{Naoz}, S., {Farr}, W.~M., {Lithwick}, Y., {Rasio}, F.~A., \& {Teyssandier}, J.
  2013, \mnras, 431, 2155, \dodoi{10.1093/mnras/stt302}

\bibitem[{{Naoz} {et~al.}(2012){Naoz}, {Farr}, \& {Rasio}}]{Naoz:2012}
{Naoz}, S., {Farr}, W.~M., \& {Rasio}, F.~A. 2012, \apjl, 754, L36,
  \dodoi{10.1088/2041-8205/754/2/L36}

\bibitem[{{Narita} {et~al.}(2009){Narita}, {Hirano}, {Sato}, {Winn}, {Suto},
  {Turner}, {Aoki}, {Tamura}, \& {Yamada}}]{Narita:2009}
{Narita}, N., {Hirano}, T., {Sato}, B., {et~al.} 2009, \pasj, 61, 991,
  \dodoi{10.1093/pasj/61.5.991}

\bibitem[{{Newcomb}(1895)}]{newc1895}
{Newcomb}, S. 1895, United States. Nautical Almanac Office. Astronomical paper
  ; v.5, 5, 1

\bibitem[{{Ogilvie}(2009)}]{Ogilvie:2009}
{Ogilvie}, G.~I. 2009, \mnras, 396, 794,
  \dodoi{10.1111/j.1365-2966.2009.14814.x}

\bibitem[{{Ogilvie}(2013)}]{Ogilvie:2013}
---. 2013, \mnras, 429, 613, \dodoi{10.1093/mnras/sts362}

\bibitem[{{Ogilvie}(2014)}]{Ogilvie:2014}
---. 2014, \araa, 52, 171, \dodoi{10.1146/annurev-astro-081913-035941}

\bibitem[{{Ogilvie} \& {Lin}(2004)}]{Ogilvie:2004}
{Ogilvie}, G.~I., \& {Lin}, D.~N.~C. 2004, \apj, 610, 477,
  \dodoi{10.1086/421454}

\bibitem[{{Ogilvie} \& {Lin}(2007)}]{Ogilvie:2007}
---. 2007, \apj, 661, 1180, \dodoi{10.1086/515435}

\bibitem[{Olver {et~al.}(2010)Olver, of~Standards, \& (U.S.)}]{Olver:2010}
Olver, F., of~Standards, N.~I., \& (U.S.), T. 2010, NIST Handbook of
  Mathematical Functions Paperback and CD-ROM (Cambridge University Press).
\newblock \url{https://books.google.com/books?id=7EZ6PwDy4qEC}

\bibitem[{{Ormel} \& {Klahr}(2010)}]{Ormel:2010}
{Ormel}, C.~W., \& {Klahr}, H.~H. 2010, \aap, 520, A43,
  \dodoi{10.1051/0004-6361/201014903}

\bibitem[{{Papaloizou} \& {Savonije}(1997)}]{Papaloizou:1997}
{Papaloizou}, J.~C.~B., \& {Savonije}, G.~J. 1997, \mnras, 291, 651,
  \dodoi{10.1093/mnras/291.4.651}

\bibitem[{{Penev} \& {Sasselov}(2011)}]{Penev:2011}
{Penev}, K., \& {Sasselov}, D. 2011, \apj, 731, 67,
  \dodoi{10.1088/0004-637X/731/1/67}

\bibitem[{{Petrovich}(2015{\natexlab{a}})}]{Petrovich:2015b}
{Petrovich}, C. 2015{\natexlab{a}}, \apj, 805, 75,
  \dodoi{10.1088/0004-637X/805/1/75}

\bibitem[{{Petrovich}(2015{\natexlab{b}})}]{Petrovich:2015a}
---. 2015{\natexlab{b}}, \apj, 799, 27, \dodoi{10.1088/0004-637X/799/1/27}

\bibitem[{{Petrovich} {et~al.}(2014){Petrovich}, {Tremaine}, \&
  {Rafikov}}]{Petrovich:2014}
{Petrovich}, C., {Tremaine}, S., \& {Rafikov}, R. 2014, \apj, 786, 101,
  \dodoi{10.1088/0004-637X/786/2/101}

\bibitem[{{Plummer}(1918)}]{plummer1918}
{Plummer}, H.~C.~K. 1918, {An introductory treatise on dynamical astronomy}
  (Cambridge, UK: Cambridge University Press)

\bibitem[{{Poincar{\'e}}(1910)}]{Poincare:1910}
{Poincar{\'e}}, H. 1910, Bulletin Astronomique, Serie I, 27, 321

\bibitem[{{Poisson} \& {Will}(2014)}]{Poisson:2014}
{Poisson}, E., \& {Will}, C.~M. 2014, {Gravity} (Cambridge, UK: Cambridge
  University Press)

\bibitem[{{Press} \& {Teukolsky}(1977)}]{Press:1977}
{Press}, W.~H., \& {Teukolsky}, S.~A. 1977, \apj, 213, 183,
  \dodoi{10.1086/155143}

\bibitem[{{Rasio} {et~al.}(1996){Rasio}, {Tout}, {Lubow}, \&
  {Livio}}]{Rasio:1996a}
{Rasio}, F.~A., {Tout}, C.~A., {Lubow}, S.~H., \& {Livio}, M. 1996, \apj, 470,
  1187, \dodoi{10.1086/177941}

\bibitem[{{Rice} {et~al.}(2022{\natexlab{a}}){Rice}, {Wang}, \&
  {Laughlin}}]{Rice:2022a}
{Rice}, M., {Wang}, S., \& {Laughlin}, G. 2022{\natexlab{a}}, \apjl, 926, L17,
  \dodoi{10.3847/2041-8213/ac502d}

\bibitem[{{Rice} {et~al.}(2022{\natexlab{b}}){Rice}, {Wang}, {Wang},
  {Stef{\'a}nsson}, {Isaacson}, {Howard}, {Logsdon}, {Schweiker}, {Dai},
  {Brinkman}, {Giacalone}, \& {Holcomb}}]{Rice:2022b}
{Rice}, M., {Wang}, S., {Wang}, X.-Y., {et~al.} 2022{\natexlab{b}}, \aj, 164,
  104, \dodoi{10.3847/1538-3881/ac8153}

\bibitem[{{Rizzuto} {et~al.}(2020){Rizzuto}, {Newton}, {Mann}, {Tofflemire},
  {Vanderburg}, {Kraus}, {Wood}, {Quinn}, {Zhou}, {Thao}, {Law}, {Ziegler}, \&
  {Brice{\~n}o}}]{Rizzuto:2020}
{Rizzuto}, A.~C., {Newton}, E.~R., {Mann}, A.~W., {et~al.} 2020, \aj, 160, 33,
  \dodoi{10.3847/1538-3881/ab94b7}

\bibitem[{Rose(1957)}]{Rose:1957}
Rose, M. 1957, Elementary Theory of Angular Momentum, Structure of matter
  series (New York: Wiley)

\bibitem[{{Rosenthal} {et~al.}(2024){Rosenthal}, {Howard}, {Knutson}, \&
  {Fulton}}]{Rosenthal:2024}
{Rosenthal}, L.~J., {Howard}, A.~W., {Knutson}, H.~A., \& {Fulton}, B.~J. 2024,
  \apjs, 270, 1, \dodoi{10.3847/1538-4365/acffc0}

\bibitem[{{Sadeghi Ardestani} {et~al.}(2017){Sadeghi Ardestani}, {Guillot}, \&
  {Morel}}]{Sadeghi:2017}
{Sadeghi Ardestani}, L., {Guillot}, T., \& {Morel}, P. 2017, \mnras, 472, 2590,
  \dodoi{10.1093/mnras/stx2039}

\bibitem[{{Savonije} \& {Papaloizou}(1997)}]{Savonije:1997}
{Savonije}, G.~J., \& {Papaloizou}, J.~C.~B. 1997, \mnras, 291, 633,
  \dodoi{10.1093/mnras/291.4.633}

\bibitem[{{Savvidou} \& {Bitsch}(2023)}]{Savvidou:2023}
{Savvidou}, S., \& {Bitsch}, B. 2023, \aap, 679, A42,
  \dodoi{10.1051/0004-6361/202245793}

\bibitem[{{Schlaufman}(2010)}]{Schlaufman:2010}
{Schlaufman}, K.~C. 2010, \apj, 719, 602, \dodoi{10.1088/0004-637X/719/1/602}

\bibitem[{{Schneider} \& {Bitsch}(2021)}]{Schneider:2021}
{Schneider}, A.~D., \& {Bitsch}, B. 2021, \aap, 654, A71,
  \dodoi{10.1051/0004-6361/202039640}

\bibitem[{{Schultz} {et~al.}(2021){Schultz}, {Spalding}, \&
  {Batygin}}]{Schultz:2021}
{Schultz}, K., {Spalding}, C., \& {Batygin}, K. 2021, \mnras, 506, 2999,
  \dodoi{10.1093/mnras/stab1899}

\bibitem[{{Sedaghati} {et~al.}(2023){Sedaghati}, {Jord{\'a}n}, {Brahm},
  {Mu{\~n}oz}, {Petrovich}, \& {Hobson}}]{Sedaghati:2023}
{Sedaghati}, E., {Jord{\'a}n}, A., {Brahm}, R., {et~al.} 2023, \aj, 166, 130,
  \dodoi{10.3847/1538-3881/acea84}

\bibitem[{{Skumanich}(1972)}]{Skumanich:1972}
{Skumanich}, A. 1972, \apj, 171, 565, \dodoi{10.1086/151310}

\bibitem[{{Spalding} {et~al.}(2014){Spalding}, {Batygin}, \&
  {Adams}}]{Spalding:2014}
{Spalding}, C., {Batygin}, K., \& {Adams}, F.~C. 2014, \apjl, 797, L29,
  \dodoi{10.1088/2041-8205/797/2/L29}

\bibitem[{{Stahler}(1983)}]{Stahler:1983}
{Stahler}, S.~W. 1983, \apj, 274, 822, \dodoi{10.1086/161495}

\bibitem[{{Stassun} {et~al.}(1999){Stassun}, {Mathieu}, {Mazeh}, \&
  {Vrba}}]{Stassun:1999}
{Stassun}, K.~G., {Mathieu}, R.~D., {Mazeh}, T., \& {Vrba}, F.~J. 1999, \aj,
  117, 2941, \dodoi{10.1086/300881}

\bibitem[{{Stauffer} {et~al.}(2018){Stauffer}, {Rebull}, {Cody}, {Hillenbrand},
  {Pinsonneault}, {Barrado}, {Bouvier}, \& {David}}]{Stauffer:2018}
{Stauffer}, J., {Rebull}, L.~M., {Cody}, A.~M., {et~al.} 2018, \aj, 156, 275,
  \dodoi{10.3847/1538-3881/aae9ec}

\bibitem[{{Storch} {et~al.}(2014){Storch}, {Anderson}, \& {Lai}}]{Storch:2014b}
{Storch}, N.~I., {Anderson}, K.~R., \& {Lai}, D. 2014, Science, 345, 1317,
  \dodoi{10.1126/science.1254358}

\bibitem[{{Storch} \& {Lai}(2014)}]{Storch:2014}
{Storch}, N.~I., \& {Lai}, D. 2014, \mnras, 438, 1526,
  \dodoi{10.1093/mnras/stt2292}

\bibitem[{{Storch} {et~al.}(2017){Storch}, {Lai}, \& {Anderson}}]{Storch:2017}
{Storch}, N.~I., {Lai}, D., \& {Anderson}, K.~R. 2017, \mnras, 465, 3927,
  \dodoi{10.1093/mnras/stw3018}

\bibitem[{{Tremaine} {et~al.}(2009){Tremaine}, {Touma}, \&
  {Namouni}}]{Tremaine:2009}
{Tremaine}, S., {Touma}, J., \& {Namouni}, F. 2009, \aj, 137, 3706,
  \dodoi{10.1088/0004-6256/137/3/3706}

\bibitem[{{Triana} {et~al.}(2019){Triana}, {Rekier}, {Trinh}, \&
  {Dehant}}]{Triana:2019}
{Triana}, S.~A., {Rekier}, J., {Trinh}, A., \& {Dehant}, V. 2019, Geophysical
  Journal International, 218, 1071, \dodoi{10.1093/gji/ggz212}

\bibitem[{{Varshalovich} {et~al.}(1988){Varshalovich}, {Moskalev}, \&
  {Khersonskii}}]{Varshalovich:1988}
{Varshalovich}, D.~A., {Moskalev}, A.~N., \& {Khersonskii}, V.~K. 1988,
  {Quantum Theory of Angular Momentum} (World Scientific Publishing Company),
  \dodoi{10.1142/0270}

\bibitem[{{Vick} \& {Lai}(2020)}]{Vick:2020}
{Vick}, M., \& {Lai}, D. 2020, \mnras, 496, 3767,
  \dodoi{10.1093/mnras/staa1784}

\bibitem[{{Wahl} {et~al.}(2017){Wahl}, {Hubbard}, \& {Militzer}}]{Wahl:2017}
{Wahl}, S.~M., {Hubbard}, W.~B., \& {Militzer}, B. 2017, \icarus, 282, 183,
  \dodoi{10.1016/j.icarus.2016.09.011}

\bibitem[{{Wahl} {et~al.}(2020){Wahl}, {Parisi}, {Folkner}, {Hubbard}, \&
  {Militzer}}]{Wahl:2020}
{Wahl}, S.~M., {Parisi}, M., {Folkner}, W.~M., {Hubbard}, W.~B., \& {Militzer},
  B. 2020, \apj, 891, 42, \dodoi{10.3847/1538-4357/ab6cf9}

\bibitem[{{Wang} {et~al.}(2024){Wang}, {Rice}, {Wang}, {Kanodia}, {Dai},
  {Logsdon}, {Schweiker}, {Teske}, {Butler}, {Crane}, {Shectman}, {Quinn},
  {Kostov}, {Osborn}, {Goeke}, {Eastman}, {Shporer}, {Rapetti}, {Collins},
  {Watkins}, {Relles}, {Ricker}, {Seager}, {Winn}, \& {Jenkins}}]{Wang:2024}
{Wang}, X.-Y., {Rice}, M., {Wang}, S., {et~al.} 2024, \apjl, 973, L21,
  \dodoi{10.3847/2041-8213/ad7469}

\bibitem[{Wigner(1959)}]{Wigner:1959}
Wigner, E. 1959, Group Theory: And Its Application to the Quantum Mechanics of
  Atomic Spectra (New York: Academic Press)

\bibitem[{{Winn} {et~al.}(2010){Winn}, {Fabrycky}, {Albrecht}, \&
  {Johnson}}]{Winn:2010}
{Winn}, J.~N., {Fabrycky}, D., {Albrecht}, S., \& {Johnson}, J.~A. 2010, \apjl,
  718, L145, \dodoi{10.1088/2041-8205/718/2/L145}

\bibitem[{{Witte} \& {Savonije}(1999)}]{Witte:1999}
{Witte}, M.~G., \& {Savonije}, G.~J. 1999, \aap, 350, 129,
  \dodoi{10.48550/arXiv.astro-ph/9909073}

\bibitem[{{Wood}(2004)}]{Wood:2004}
{Wood}, J.~A. 2004, \gca, 68, 4007, \dodoi{10.1016/j.gca.2004.04.003}

\bibitem[{{Wu} {et~al.}(2024){Wu}, {Dewberry}, \& {Fuller}}]{Wu:2024}
{Wu}, S.~C., {Dewberry}, J.~W., \& {Fuller}, J. 2024, \apj, 963, 34,
  \dodoi{10.3847/1538-4357/ad1e54}

\bibitem[{{Wu} \& {Murray}(2003)}]{Wu:2003}
{Wu}, Y., \& {Murray}, N. 2003, \apj, 589, 605, \dodoi{10.1086/374598}

\bibitem[{{Zahn}(1975)}]{Zahn:1975}
{Zahn}, J.~P. 1975, \aap, 41, 329

\bibitem[{{Zahn}(1977)}]{Zahn:1977}
{Zahn}, J.-P. 1977, \aap, 57, 383

\bibitem[{{Zanazzi} {et~al.}(2024){Zanazzi}, {Dewberry}, \&
  {Chiang}}]{Zanazzi:2024}
{Zanazzi}, J.~J., {Dewberry}, J., \& {Chiang}, E. 2024, \apjl, 967, L29,
  \dodoi{10.3847/2041-8213/ad4644}

\bibitem[{Zhang \& Liao(2017)}]{Zhang-Liao:2017}
Zhang, K., \& Liao, X. 2017, Theory and Modeling of Rotating Fluids, Cambridge
  Monographs on Mechanics (Cambridge University Press).
\newblock \url{https://books.google.com/books?id=A9vFDgAAQBAJ}

\end{thebibliography}

\appendix
\section{Standard Weak Friction Limit}\label{app:weak_friction}
In the weak friction limit, we take $\tau_{mn}=\tau$ in Equations~(\ref{eq:torque_lag_u})-(\ref{eq:precession_lag_n}). This procedure  is straightforward
thanks to the identities of Hansen coefficients  (Appendix~\ref{app:hansen_coefficients}) and Wigner-D matrices (Appendix~\ref{app:wigner_matrices}). For succinctness,
we only lay out the derivation of $\langle{T_u}\rangle_{\!\rm \scriptscriptstyle eq}$ (Equation~\ref{eq:torque_u_equi}), but the derivation of the other equations of motion follows in a largely analogous way.

Starting from Equation~(\ref{eq:torque_lag_u}), we set $\tau_{mn}=\tau$ and replace $\tilde{\omega}_{mn}=n\Omega-m\Omega_\star$:
\begin{equation}
\begin{split}
\langle{T_u}\rangle_{\!\rm \scriptscriptstyle eq}
=&
-
\frac{{\cal G}M_{\rm p}^2R_\star^5}{a_{\rm p}^6}
\frac{5}{4\pi}\tau
\bigg\{
\Omega
\!\sum_{\scriptscriptstyle m n p r}
n
W_{2p}
W_{2r}
X_{n}^{-3,p}
X_{n}^{-3,r}
{\rm Re}\big\{
{D^{2^{\scriptstyle *}}_{p\mu}(\alpha\beta0)D^{2}_{r-1, m }}(\alpha\beta0)
b^{(-)}_{r}
+D^{2^{\scriptstyle *}}_{p\mu}(\alpha\beta0){D^{2}_{r+1, m }}(\alpha\beta0)
b^{(+)}_{r}
\big\}
\\
&~~~~~~
+
\Omega_\star
\!\sum_{\scriptscriptstyle m n p r}
m
W_{2p}
W_{2r}
X_{n}^{-3,p}
X_{n}^{-3,r}
{\rm Re}\big\{
{D^{2^{\scriptstyle *}}_{p\mu}(\alpha\beta0)D^{2}_{r-1, m }}(\alpha\beta0)
b^{(-)}_{r}
+D^{2^{\scriptstyle *}}_{p\mu}(\alpha\beta0){D^{2}_{r+1, m }}(\alpha\beta0)
b^{(+)}_{r}
\big\}
\bigg\}
\end{split}
\end{equation}
The uniform time-lag assumption allows us to carry the sum over $m$ explicitly. Using 
Equation~(\ref{eq:orthogonality}) and~(\ref{eq:orthogonality_first})
and noting that $W_{2p}W_{2,p\pm1}$ always vanishes, we have that
\begin{equation}
\begin{split}
\langle{T_u}\rangle_{\!\rm \scriptscriptstyle eq}
=
&
\frac{{\cal G}M_{\rm p}^2R_\star^5}{a_{\rm p}^6}
\frac{5}{4\pi}\tau
\sin\beta\cos\alpha
\Omega_\star
\!\sum_{\scriptscriptstyle n  p r}
W_{2p}
W_{2r}
X_{n}^{-3,p}
X_{n}^{-3,r}
\big\{
 b^{(-)}_{r}b^{(-)}_{r-1} \delta_{r-2,p}
+  \left[b^{(-)}_{r}b^{(+)}_{r-1}
 +
  b^{(+)}_{r}b^{(-)}_{r+1}\right] \delta_{r,p}
  \\
&~~~~~~~~~~~~~~~~~~~~~~~~~~~
~~~~~~~~~~~~~~~~~~~~~~~~~~~
~~~~~~~~~~~~
+ b^{(+)}_{r}b^{(+)}_{r+1}\delta_{r+2,p}
\big\}
\\
=
&
\frac{{\cal G}M_{\rm p}^2R_\star^5}{a_{\rm p}^6}
\frac{5}{4\pi}\tau
\sin\beta\cos\alpha
\Omega_\star
\!\sum_{\scriptscriptstyle n  p}
\left(W_{2p}\right)^2
\big\{
-
[b_p^{(+)}]^2
X_{n}^{-3,p}
X_{n}^{-3,p+2} 
+  
[X_{n}^{-3,p}]^2
\left[[b^{(-)}_{p}]^2+[b^{(+)}_{p}]^2\right]
  \\
&~~~~~~~~~~~~~~~~~~~~~~~~~~~
~~~~~~~~~~~~~~~~~~~~
- 
[b_p^{(-)}]^2
X_{n}^{-3,p}
X_{n}^{-3,p-2}
\big\}
\end{split}
\end{equation}
where we have used the identity
$W_{2,p\pm2}=- W_{2p}\left({b_p^{(\pm)}}/{b_{p\pm1}^{(\pm)}}\right)$ and the relations$b^{(-)}_{p}=b^{(+)}_{p-1}$. Next, we carry out
the sum over $n$ using identify~(\ref{eq:hansen_zeroth}) and then sum over $p$ using the definitions of $W_{2p}$ and $b_{p}^{(\pm)}$,
we finally get Equation~(\ref{eq:torque_u_equi}):
\begin{equation}
\begin{split}
\langle{T_u}\rangle_{\!\rm \scriptscriptstyle eq}
=
&
\frac{3{\cal G}M_{\rm p}^2R_\star^5}{a_{\rm p}^6}
\tau
\sin\beta\cos\alpha
\Omega_\star
\frac{1}{2}\left(X_{0}^{-6,0}-X_{0}^{-6,2}\right)~.
\end{split}
\end{equation}
The remaining Equations~(\ref{eq:torque_v_equi})-(\ref{eq:precession_n_equi}) are obtained similarly.

\section{Hansen Coefficients}\label{app:hansen_coefficients}
\subsection{Hansen Coefficients in Closed Form}
In Section~\ref{sec:orbital_perturbations}, the equations of motion
in Milankovitch form (\ref{eq:eom_angmom}-\ref{eq:eom_eccentricity})
are expressed as quadruple sums (\ref{eq:torque_lag_u}-\ref{eq:precession_lag_n})
involving Hansen coefficients. In the weak friction limit (Section~\ref{sec:weak_friction}),
all Hansen coefficients are of the form $X_0^{\alpha,m}$. Similarly, the Hansen coefficients
appearing in the equations of motion for the $(m,n)=(1,0)$ tidal component (Section~\ref{sec:inertial_waves})
are all in the form $X_0^{\alpha,m}$. These coefficients can be expressed in closed form \citep{Hughes:1981,Kozai:1973}
using the Gaussian hypergeometric function and its associated hypergeometric series:
\begin{align}
X_0^{\alpha,m}(e)
=&\left(-\frac{e}{2}\right)^m
 \binom{\alpha+m+1}{m}
~{_2}F_1\left(\frac{m-\alpha-1}{2},\frac{m-\alpha}{2},m+1;e^2\right)
\end{align}
which is valid regardless of the sign of  $\alpha\in\mathbb{Z}$.

Importantly, the Hansen coefficients with $\alpha<0$ that are relevant to this work end up containing a finite number of
terms in eccentricity. This is
because the series representation of ${_2}F_1(a,b,c;z)$ terminates whenever $a<0$ or $b<0$  \citep[e.g.,][\S15.2]{Olver:2010}. Using  
identity \S15.3.3 in \citet{abramowitz+stegun} and the series representation of ${_2}F_1(a,b,c;z)$ we obtain
\begin{align}\label{eq:hansen_closed_form}
 \begin{array}{lr}
\displaystyle
X_0^{-|\alpha|,m}(e)
=\frac{(-e/2)^m}{(1-e^2)^{|\alpha|-3/2}}
 \binom{m-|\alpha|+m}{m}
 \!\!
\sum_{j=0}^{\lfloor|\frac{m-|\alpha|+2}{2}|\rfloor}
\frac{\left(\frac{m-|\alpha|}{2}+\frac{3}{2}\right)_j\left(\frac{m-|\alpha|}{2}+1\right)_j}{\left(\frac{m+|\alpha|}{2}\right)_j}\frac{e^{2j}}{j!},
&~
\displaystyle
\left(\frac{m-|\alpha|+2}{2} \in \mathbb{Z}_0^- \right)
\end{array}
\end{align}
where $\lfloor \cdot \rfloor$ denotes the floor function and where
$(x)_j=\Gamma(x+j)/\Gamma(x)$ is the Pochhammer symbol. For negative arguments $x=-a$ with $a\in\mathbb{Z}$, the
Pochhammer symbol $(-|a|)_j$ in evaluates to zero for $j> |a|$, which explains the series terminating at finite $j$.
For $\alpha=-6$,  for example, $X_0^{-6,m}$ will be proportional to $(1-e^2)^{-9/2}$ multiplied by a polynomial in eccentricity
whose lowest-order and highest-order terms are, respectively,  ${\cal O}(e^m)$ and  ${\cal O}(e^{|m-4|+m})$ 

We use Equation~(\ref{eq:hansen_closed_form}) to compute $X_0^{-|\alpha|,m}$ in Sections~\ref{sec:weak_friction} and~\ref{sec:10_component}
and in Appendix~\ref{app:hut_functions} below. Note that we do not use eq. 11 in \citet{Hughes:1981}, as it appears to be incorrect.

\subsection{Moments of the Hansen Coefficients}
From the definition of the Hansen coefficients (Equation~\ref{eq:hansen_expansion}),
one can write
\begin{equation}
\left(\frac{r_{\rm p}}{a_{\rm p}}\right)^{2\alpha} {\rm e}^{{\rm i} (\mu-\nu) f}
=
\sum_{n=-\infty}^\infty \sum_{n'=-\infty}^\infty X_{n}^{\alpha,\mu} X_{n'}^{\alpha,-\nu}
{\rm e}^{{\rm i} {(n-n')\cal M}}
=\sum_{n=-\infty}^\infty X_{n}^{2\alpha,\mu-\nu}{\rm e}^{{\rm i} n{\cal M}}~,
\end{equation}
and thus, after averaging over the mean anomaly/time one obtains,
\begin{equation}\label{eq:hansen_zeroth}
\sum_{n=-\infty}^\infty  X_{n}^{\alpha,\mu} X_{n}^{\alpha,-\nu}
= X_{0}^{2\alpha,\mu-\nu}
\end{equation}
of which a special case is the `zeroth moment' of the Hansen coefficients:
\begin{equation}
\sum_{n=-\infty}^\infty
\left[X_n^{l,p}\right]^2
=X_0^{2l,0}
\end{equation}
Following a similar procedure, several identities can be proved for the higher order moments
of the Hansen coefficients \citep[see][]{Correia:2022}. In this work, we will use:
\begin{equation}\label{eq:hansen_first}
\sum_{n=-\infty}^\infty
n\left[X_n^{l,p}\right]^2
=p\sqrt{1-e^2} X_0^{2l-2,0}
\end{equation}
and
\begin{equation}\label{eq:hansen_second}
\sum_{n=-\infty}^\infty
n^2\left[X_n^{l,p}\right]^2=
l^2\left[
2X_0^{2l-3,0}
-X_0^{2l-2,0}
-(1-e^2)X_0^{2l-4,0}
\right]
+p^2(1-e^2)X_0^{2l-4,0}
\end{equation}
%

\subsection{Hansen Coefficients in terms of Hut's $f$-functions}\label{app:hut_functions}
In Section~{\ref{sec:weak_friction} we have recovered the weak-friction equations
in Milankovitch form of \citet{Correia:2011}. The closed-form Hansen coefficients 
$X_0^{k,m}$ can be written in terms of the
$f_i$ functions defined in \citet{Hut:1981}:
\begin{align}
X_0^{-6,0}(e)&=\frac{1+3e^2+\tfrac{3}{8}e^4}{(1-e^2)^{9/2}}=\frac{f_5(e^2)}{(1-e^2)^{9/2}}~,
\\
X_0^{-6,1}(e)&
=\frac{3}{2}e\frac{1+\tfrac{3}{2}e^2}{(1-e^2)^{9/2}}
=e\frac{3f_4(e^2)-f_5(e^2)}{(1-e^2)^{9/2}}
\\
X_0^{-6,2}(e)&
=\frac{3}{2}e^2\frac{1+\tfrac{1}{6}e^2}{(1-e^2)^{9/2}}
=\frac{f_5(e^2)-f_4(e^2)}{(1-e^2)^{9/2}}~,
\\
X_0^{-8,0}(e)&=\frac{1+\tfrac{15}{2}e^2+\tfrac{45}{8}e^4+\tfrac{5}{16}e^6}{(1-e^2)^{13/2}}=\frac{f_2(e^2)}{(1-e^2)^{13/2}}~,
\\
X_0^{-9,0}(e)&=\frac{1+\tfrac{21}{2}e^2+\tfrac{105}{8}e^4+\frac{35}{16}e^6}{(1-e^2)^{15/2}}~,
\\
X_0^{-10,0}(e)&=\frac{1+14 e^2+\tfrac{105}{4}e^4+\tfrac{35}{4}e^6+\tfrac{35}{128}e^8}{(1-e^2)^{17/2}}
=3\frac{X_0^{-9,0}}{1-e^2}-\frac{2(1-e^2)f_2(e^2)+\tfrac{9}{2}e^2f_3(e^2)}{(1-e^2)^{17/2}}
\end{align}
%

\section{Properties of  Wigner D-matrices}\label{app:wigner_matrices}
\subsection{Closed-Form Expressions and Symmetry Properties}
The Wigner D-matrices $D^l_{\mu m}(\alpha,\beta,\gamma)$ define the transformation of spherical
harmonics under a coordinate transformation given by the
$\alpha\beta\gamma$ Euler angle rotation in the $z-y-z$ convention \citep{Rose:1957}:
\begin{equation}
Y_{lm}(\theta',\phi')=
\!\sum_{\scriptscriptstyle m{=}-l}^{\scriptscriptstyle l} 
{D^{l}_{\mu m }}(\alpha\beta\gamma)
Y_{l \mu}(\theta,\phi)~
\end{equation}
The Euler angle dependence is separable such that the small-d Wigner matrix is introduced via
$D^l_{\mu m}(\alpha,\beta,\gamma)={\rm e}^{-{\rm i}\mu\alpha}d^l_{\mu m}(\beta){\rm e}^{-{\rm i}m\gamma}$
where $d^l_{\mu m}(\beta)\in\mathbb{R}$. 
These matrices exhibit the symmetries,
$d^l_{\mu m}(\beta)=d^l_{m \mu}(-\beta)$,
$d^l_{\mu m}(\beta)=(-1)^{\mu-m}d^l_{m \mu }(\beta)$
and $d^l_{\mu m}(\beta)=d^l_{-m,-\mu}(\beta)$, 
which also imply that 
 $d^l_{\mu m}(\beta)=(-1)^{\mu-m}d^l_{-\mu,-m}(\beta)$ \citep{Rose:1957}.
Equivalently,
${D^l}^*_{\mu m}(\alpha\beta\gamma)=(-1)^{\mu-m}D^l_{ -\mu,- m}(\alpha\beta\gamma)$
and
${D^l}^*_{\mu m}(\alpha\beta\gamma)={D^l}_{m \mu}(-\gamma,-\beta,-\alpha)$,
the latter of which is no other than the unitary property of rotational transformation.

The small d-matrices $d^l_{\mu m}(\beta)$ can be expressed 
in terms of  is Gauss' hypergeometric function ${_2}F_1$
\citep[][pp.50-54]{Rose:1957}. Focusing on $l=2$, we write
\begin{equation}
\begin{split}
d^{2}_{\mu m}(\beta)=&
\left[\frac{(2-m)!(2+\mu)!)}{(2+m)!(2-\mu)!}
\right]^{1/2}
\left[(\mu-m)!\right]^{-1}
\left(\cos\frac{\beta}{2}\right)^{4+m-\mu}
\left(-\sin\frac{\beta}{2}\right)^{\mu-m}
 {_2}F_1(\mu-2,-m-2;\mu-m+1;-\tan^2\frac{\beta}{2})
\end{split}
\end{equation}
which is only valid for $\mu\geq m$. If $m>\mu$, the values
of $d^l_{\mu m}$ can be obtained from the unitary property $d^{l}_{\mu m}(\beta)=d^{l}_{m\mu}(-\beta)$.

\subsection{Recursion and Orthogonality Relations}
The orthonormality properties of the Wigner matrices reads \citep{Rose:1957}:
\begin{equation}\label{eq:orthogonality}
 \!\sum_{\scriptscriptstyle m}
{D^{2^{\scriptscriptstyle *}}_{p m }}(\alpha\beta0)
{D^{2}_{rm }}(\alpha\beta0)
=\delta_{pr}
\in\mathbb{R}
\end{equation}
Using identity (4.8.17) in \citet{Varshalovich:1988}, we write
\begin{equation}\label{eq:varshalovich_identity_1}
\begin{split}
m D^2_{rm}(\alpha\beta0)=&r\cos\beta D^2_{rm}(\alpha\beta0)
+\sin\beta b^{(-)}_rD^2_{r-1,m}(\alpha\beta0){\rm e}^{-{\rm i}\alpha}
+\sin\beta b^{(+)}_rD^2_{r+1,m}(\alpha\beta0){\rm e}^{{\rm i}\alpha}
\end{split}
\end{equation}
from which we also write,
\begin{equation}
\begin{split}
m D^2_{r-1,m}(\alpha\beta0)=&(r-1)\cos\beta D^2_{r-1,m}(\alpha\beta0)
+\sin\beta b^{(-)}_{r-1}D^2_{r-2,m}(\alpha\beta0){\rm e}^{-{\rm i}\alpha}
+\sin\beta b^{(+)}_{r-1}D^2_{r,m}(\alpha\beta0){\rm e}^{{\rm i}\alpha}
\end{split}
\end{equation}
and
\begin{equation}
\begin{split}
m D^2_{r+1,m}(\alpha\beta0)=&(r+1)\cos\beta D^2_{r+1,m}(\alpha\beta0)
+\sin\beta b^{(-)}_{r+1}D^2_{r,m}(\alpha\beta0){\rm e}^{-{\rm i}\alpha}
+\sin\beta b^{(+)}_{r+1}D^2_{r+2,m}(\alpha\beta0){\rm e}^{{\rm i}\alpha}
\end{split}
\end{equation}
which means we can write the $m$ times first identity as
\begin{equation}
\begin{split}
m^2 D^2_{rm}(\alpha\beta0)=&
r\cos\beta mD^2_{rm}(\alpha\beta0)
+\sin\beta b^{(-)}_r {\rm e}^{-{\rm i}\alpha}m D^2_{r-1,m}(\alpha\beta0)
+\sin\beta b^{(+)}_r {\rm e}^{{\rm i}\alpha} mD^2_{r+1,m} (\alpha\beta0)
\\
=& r^2\cos^2\beta D^2_{rm}(\alpha\beta0)
+r\cos\beta \sin\beta b^{(-)}_r {\rm e}^{-{\rm i}\alpha} D^2_{r-1,m}(\alpha\beta0)
+r\cos\beta \sin\beta b^{(+)}_r {\rm e}^{{\rm i}\alpha} D^2_{r+1,m}(\alpha\beta0)
\\
+&\sin\beta b^{(-)}_r {\rm e}^{-{\rm i}\alpha}
\left[
(r-1)\cos\beta D^2_{r-1,m}(\alpha\beta0)
+\sin\beta b^{(-)}_{r-1}D^2_{r-2,m}(\alpha\beta0){\rm e}^{-{\rm i}\alpha}
+\sin\beta b^{(+)}_{r-1}D^2_{r,m}(\alpha\beta0){\rm e}^{{\rm i}\alpha}
\right]
\\
+&\sin\beta b^{(+)}_r {\rm e}^{{\rm i}\alpha} 
\left[
(r+1)\cos\beta D^2_{r+1,m}(\alpha\beta0)
+\sin\beta b^{(-)}_{r+1}D^2_{r,m}(\alpha\beta0){\rm e}^{-{\rm i}\alpha}
+\sin\beta b^{(+)}_{r+1}D^2_{r+2,m}(\alpha\beta0){\rm e}^{{\rm i}\alpha}
\right]
\\
=& \left[r^2\cos^2\beta 
+\sin^2\beta b^{(-)}_r  b^{(+)}_{r-1}
+\sin^2\beta b^{(+)}_r   b^{(-)}_{r+1}
\right]D^2_{r,m}(\alpha\beta0)
\\
+&
\cos\beta \sin\beta\left[
(2r-1) b^{(-)}_r {\rm e}^{-{\rm i}\alpha} D^2_{r-1,m}(\alpha\beta0)
+(2r+1)b^{(+)}_r {\rm e}^{{\rm i}\alpha} D^2_{r+1,m}(\alpha\beta0)
\right]
\\
+&
\sin^2\beta \left[b^{(-)}_{r-1} b^{(-)}_r {\rm e}^{-2{\rm i}\alpha}D^2_{r-2,m}(\alpha\beta0)
+ b^{(+)}_{r+1}  b^{(+)}_r {\rm e}^{2{\rm i}\alpha} D^2_{r+2,m}(\alpha\beta0)
\right]
\end{split}
\end{equation}

and thus
\begin{equation}
\begin{split}
m^2 D^2_{rm}(\alpha\beta0)=&
\left[r^2\cos^2\beta +\tfrac{1}{2}(6-r^2)\sin^2\beta\right]D^2_{rm}(\alpha\beta0)
\\
&+ \cos\beta\sin\beta\left[(2r-1) b^{(-)}_r D^2_{r-1,m}(\alpha\beta0){\rm e}^{-{\rm i}\alpha}
+(2r+1) b^{(+)}_r  D^2_{r+1,m}(\alpha\beta0){\rm e}^{{\rm i}\alpha}
\right]
\\
&
+\sin^2\beta \left[ b^{(-)}_r b^{(-)}_{r-1}D^2_{r-2,m}(\alpha\beta0){\rm e}^{-2{\rm i}\alpha}
+ b^{(+)}_r  b^{(+)}_{r+1}D^2_{r+2,m}(\alpha\beta0){\rm e}^{2{\rm i}\alpha}\right]
\end{split}
\end{equation}
which, along with the orthogonality condition, allows us to write
\begin{equation}\label{eq:orthogonality_first}
\begin{split}
\!\sum_{\scriptscriptstyle m }mD^{2^{\scriptstyle *}}_{pm } D^2_{rm}
=&
r\cos\beta \delta_{pr}
+\sin\beta b^{(-)}_r{\rm e}^{-{\rm i}\alpha} \delta_{p,r-1}
+\sin\beta  b^{(+)}_r{\rm e}^{{\rm i}\alpha}\delta_{p,r+1}
\end{split}
\end{equation}
and
\begin{equation}
\begin{split}
\!\sum_{\scriptscriptstyle m }m^2D^{2^{\scriptstyle *}}_{pm } D^2_{rm}
=&
\left[r^2\cos^2\beta +\tfrac{1}{2}(6-r^2)\sin^2\beta\right]\delta_{pr}
\\
&+ \cos\beta\sin\beta\left[(2r-1) b^{(-)}_r {\rm e}^{-{\rm i}\alpha}\delta_{p,r-1}
+(2r+1) b^{(+)}_r  {\rm e}^{{\rm i}\alpha}\delta_{p,r+1}
\right]
\\
&
+\sin^2\beta \left[ b^{(-)}_r b^{(-)}_{r-1}{\rm e}^{-2{\rm i}\alpha}
\delta_{p,r-2}
+ b^{(+)}_r  b^{(+)}_{r+1}{\rm e}^{2{\rm i}\alpha}\delta_{p,r+2}
\right]
\end{split}
\end{equation}
%

\section{Toy Model of Pre-Main Sequence Contraction}\label{app:pms_model}  
The contraction luminosity for an index $n$ polytrope is
$L_\star= -\tfrac{3}{5-n}{{\cal G}M_\star^2}/(2R_\star^2)\dot{R}_\star
=4\pi R_\star^2\sigma T_{\rm eff}^4
$,
or equivalently,
\begin{equation}\label{eq:polytrope_contraction}
\frac{dR_\star}{dt}
=- 
\left(\frac{5-n}{3}\right)
\frac{8\pi  \sigma T_{{\rm eff}}^4 }{{\cal G}M_\star^2}R_\star^4
\end{equation}
\citep[e.g.,][]{Hansen:2004}.
Assuming that the effective temperature is constant along a Hayashi track, and a polytropic index 
of $n=1/(1-\gamma)=3/2$ with $\gamma=5/3$ which is adequate for fully convective bodies,
 the solution to  
Equation~\ref{eq:polytrope_contraction} is
\begin{equation}\label{eq:hayashi_solution}
R_\star(t)
=R_{\star,0}\left[3\left(\frac{R_{\star,0}}{R_\odot}\right)^3
\frac{t-t_0}{\tau_{\rm Hay}}+1
\right]^{-1/3}~~
\text{with}
~~~
\tau_{\rm Hay}\equiv
\frac{1}{7}
\tau_{\rm KH}
\left(\frac{T_{{\rm eff},0}}{T_{{\rm eff},\odot}}\right)^{-4}
\left(\frac{M_\star}{M_\odot}\right)^{2}
\end{equation}
Assuming that $T_{\rm eff}\approx 4400~{\rm K} \left({M_\star}/{M_\odot}\right)^{7/51}$, we can also write
$
\tau_{\rm Hay}
=0.423
\tau_{\rm KH}
\left(\frac{M_\star}{M_\odot}\right)^{74/51} $
as shown in Equation~(\ref{eq:hayashi_timescale}).

A more general calculation can be carried out without assuming $T_{\rm eff}=$constant, but by adopting
a  mass-luminosity relationship $L_\star \propto M_\star^{s}T_{\rm eff}^{m}$. For $M_\star\sim 1M_\odot$, $s=4$ and $m=2$
are appropriate values for the Henyey track. In this case, the solution to  
Equation~(\ref{eq:polytrope_contraction}) is
\begin{equation}\label{eq:henyey_solution}
R_\star(t)
=
R_\odot
\left[
-\frac{t-t_{\rm min}}{\tau_{\rm Hen}}
+\left(\frac{R_{\star,{\rm min}}}{R_\odot}\right)
\right]
~~
\text{with}
~~~
\tau_{\rm Hen}\equiv
1.2348(5-n)
\tau_{\rm KH}
\left(\frac{M_\star}{M_\odot}\right)^{-3.5172}
\end{equation}
where  $n$ should take an intermediate value between that of fully convective bodies ($n=3/2$) and that of bodies with prominent
radiative cores ($n\approx3$), and where
$t_{\rm min}$ and  $R_{\star,{\rm min}}$  are integration constant, here identified as the time and radius at which the PMS star reaches its minimum
luminosity $L_{\rm min}$ along the Hayashi track. We can solve for $t_{\rm min}$ directly from Equation~(\ref{eq:hayashi_solution})
\begin{equation}
t_{\rm min}
=
\frac{\tau_{\rm Hay}}{3}
\left(\frac{R_{\star,0}}{R_\odot}\right)^{-3}\left[
\left(\frac{R_{\star,0}}{R_{\star,{\rm min}}}\right)^{3}
-1
\right]
+t_0
~~
\text{with}
~~~
R_{\star,{\rm min}} = R_{\star,0}\left(\frac{L_{\rm min}}{L_{\star,0}}\right)^{1/2}
\end{equation}
and where we empirically determine $L_{\rm min}/L_\odot \approx 0.52 (M_\star/M_\odot)^3$. In the same straightforward manner, we can estimate the contraction time
$t_{\rm contr}$ and the time for a radiative core to first appear $t_{\rm rc}$:
\begin{equation}
t_{\rm contr}
=
\frac{\tau_{\rm Hay}}{3}
\left(\frac{R_{\star,0}}{R_\odot}\right)^{-3}\left[
\left(\frac{R_{\star,0}}{R_{\star,{\rm final}}}\right)^{3}
-1
\right]
+t_0
~~
\text{and}
~~~
t_{\rm rc}
=
\frac{\tau_{\rm Hay}}{3}
\left(\frac{R_{\star,0}}{R_\odot}\right)^{-3}\left[
\left(\frac{R_{\star,0}}{R_{\rm rc}}\right)^{3}
-1
\right]
+t_0
\end{equation}

The first appearance of the radiative core is estimated following a standard calculation \cite{Kippenhahn:2012} in
which we solve for the condition$\nabla_{\rm rad}>\nabla_{\rm ad}$ at the core assuming a Kramers opacity:
\begin{equation}
\nabla_{\rm rad,c}={\cal F}_{\rm rad}\left(\frac{T_{\rm eff}}{900~{\rm K}}\right)^{-1}
\frac{Z(1+X)}{\mu^{7.5}}
\left(\frac{L_\star}{L_\odot}\right)^{5/4}
\left(\frac{M_\star}{M_\odot}\right)^{-5.5}>\nabla_{\rm ad,c}=\frac{2}{5}
\end{equation}
Using the compositions of \citet{Baraffe:2015}  ($Z=0.015, Y=0.270, X=0.715$), we can solve for the critical luminosity
$L_{\rm rc}$ above which a radiative core can exist:
\begin{equation}
L_{\star,\rm rc}\equiv 
 1.62 {\cal F}_{\rm rad}^{-4/5}~L_\odot 
\left(\frac{M_\star}{M_\odot}\right)^{22/5}
\left(\frac{T_{\rm eff}}{4400~{\rm K}}\right)^{4/5}
\end{equation}
which can be used to solve for the critical stellar radius $R_{\star,rc}$ assuming that $L_{\star,\rm rc} = 4\pi\sigma R_{\star,{}\rm rc }^2T_{\rm eff,0}^4$ and thus
\begin{equation}
\frac{R_{\star, {\rm rc}}}{R_\odot}\approx
2.19
{\cal F}_{\rm rad}^{-2/5}
\left(\frac{M_\star}{M_\odot}\right)^{11/5}
\left(\frac{T_{\rm eff}}{4400~{\rm K}}\right)^{-8/5}
\approx
2.1
\left(\frac{M_\star}{M_\odot}\right)^{1.38}
\end{equation}
where in the last step we assumed used that ${\cal F}_{\rm rad}\approx 1.12 (M_\star/M_\odot)^{3/2}$
and $T_{\rm eff,0}=4400{\rm K}\,(M_\star/M_\odot)^{7/51}$.

\end{document}